\documentstyle[12pt,epsf]{article}
\begin{document} 

\begin{titlepage}

\hrule 
\leftline{}
\leftline{Chiba Univ. Preprint
          \hfill   \hbox{\bf CHIBA-EP-123}}
\leftline{\hfill   \hbox{hep-th/0009152}}
\leftline{\hfill   \hbox{September 2000}}
\vskip 5pt
\hrule 
\vskip 1.0cm
\centerline{\large\bf 
  Dual superconductivity, 
} 
\vskip 0.5cm
\centerline{\large\bf  
monopole condensation and confining string
}
\vskip 0.5cm
\centerline{\large\bf  
in low-energy Yang-Mills theory.  Part I.
}

\vskip 1cm

\centerline{{\bf 
Kei-Ichi Kondo$^{1,}{}^{\dagger}{}$
}}  
\vskip 1cm
\begin{description}
\item[]{\it \centerline{ 
$^1$Department of Physics, Faculty of Science, 
Chiba University,  Chiba 263-8522, Japan}
  }
\end{description}

\centerline{{\bf Abstract}} 
We show that the QCD vacuum (without dynamical quarks) is a dual superconductor at least in the low-energy region 
in the sense that monopole condensation does really occur.  
In fact, we derive the dual Ginzburg-Landau theory (i.e., dual Abelian Higgs model) directly from the SU(2) Yang-Mills theory by adopting the maximal Abelian gauge. The dual superconductor can be on the border between type II and type I, excluding the London limit.  The masses of the dual Abelian gauge field is expressed by the Yang-Mills gauge coupling constant and the mass of the off-diagonal gluon of the original Yang-Mills theory.
Moreover, we can rewrite the Yang-Mills theory into an theory written in terms of the Abelian magnetic monopole alone at least in the low-energy region.  Magnetic monopole condensation originates in the non-zero mass of off-diagonal gluons.
Finally, we derive the confining string theory describing the low-energy Gluodynamics.  
Then the area law of the large Wilson loop is an immediate consequence of these constructions. 
Three low-energy effective theories give the same string tension.


\vskip 0.2cm
\hrule  
$^\dagger$ 
  E-mail:  kondo@cuphd.nd.chiba-u.ac.jp 
\par 
\par\noindent


\vskip 0.5cm  


\end{titlepage}

\pagenumbering{roman}
\tableofcontents
\pagenumbering{arabic}

\par
\section{Introduction}
\setcounter{equation}{0}

\par
The main aim of this paper is to discuss how the dual superconductor picture for explaining quark confinement is derived directly from  Quantum Chromodynamics (QCD).  It is believed that quark and gluon are confined into the inside of hadrons by the strong interaction described by QCD and that they can not be observed in isolation. 
 Quark confinement can be understood based on an idea of the electro-magnetic duality of ordinary superconductivity, i.e., the dual superconductor picture\cite{Nambu74}. 
In this picture, the color electric flux can be excluded from QCD vacuum as a dual superconductor (the dual Meissner effect), just as the magnetic field can not penetrate into the superconductor (the so-called Meissner effect).
In this context, the dual is used as implying electro-magnetic duality.  However, QCD is a non-Abelian gauge theory and its gluonic part is described by the Yang-Mills theory, whereas the usual superconductivity is described by the Abelian gauge theory. 
Therefore we must make clear the precise meaning of dual superconductivity in QCD. It is possible to assume that the diagonal component of color electric field can be identified with the dual of magnetic field in the ordinary superconductivity.   
If so, we must answer what is the role played by the off-diagonal component of gluons?  Finally, we must show that QCD vacuum has a tendency to exclude the (diagonal component of) color electric field.  
If a pair of heavy quark and anti-quark is immersed in the QCD vacuum, the color electric flux connecting a pair is squeezed into the tube-like region, leading to the formation of QCD (gluon) string between a quark and an anti-quark.  
As a result, the interquark potential $V(R)$ is proportional to the interquark distance $R$.  To separate a quark from an anti-quark, we need infinite energy.  In this sense,  quark confinement is achieved.    
\par
The dual superconductor picture of QCD vacuum is based on the following assumptions:
\begin{enumerate} 
\item  Existence of magnetic monopole: QCD has magnetic monopole.

\item  Monopole condensation: Magnetic monopole is condensed in QCD.

\item  Infrared Abelian dominance: Charged gluon (i.e., off-diagonal gluon) can be neglected at least in the low-energy region of QCD.

\item  Absence of quantum effect: Classical configuration, e.g., the magnetic monopole, is dominant.  Hence quantum effect can be neglected.
\end{enumerate}
Each of these statements should be explained based on QCD to really confirm the dual superconductor picture of QCD vacuum.
As a first step, a prescription of extracting the field configuration corresponding to magnetic monopole was proposed by 't Hooft \cite{tHooft81}.  This idea is called the Abelian projection. 
Immediately after the proposal of this idea, infrared Abelian dominance was suggested by Ezawa and Iwazaki \cite{EI82}.  However, recent simulations revealed that the Abelian dominance is not necessarily realized in all the Abelian gauges, see the review \cite{review}.
To author's knowledge, the best covariant gauge fixing condition realizing  Abelian dominance is given by the maximal Abelian (MA) gauge proposed by Kronfeld et al. \cite{KLSW87}. 
In fact, the infrared Abelian dominance is first confirmed in MA gauge based on Monte Carlo simulations by Suzuki and Yotsuyanagi \cite{SY90}. Subsequent simulations have also confirmed the monopole dominance in low-energy QCD\cite{SNW94}.
An analytical derivation of the dual Ginzburg-Landau (DGL) theory was tried by Suzuki \cite{Suzuki88} by way of Zwanziger formulation 
\cite{Zwanziger71} by neglecting the off-diagonal components of gluon fields by virtue of the Abelian dominance and by assuming condensation of magnetic monopoles. 
However, the DGL theory is not yet derived directly from the underlying theory, i.e. QCD, since the Abelian dominance and monopole condensation themselves must be derived in the same framework of the theory.
According to the recent Monte Carlo simulations \cite{dualsupertype}, the presumed dual Ginzburg-Landau theory is on the border of type II and type I.  

\par
In this paper we discuss how to derive the DGL theory as a low-energy effective theory (LEET) of QCD.
The LEET is not unique and we can derive various LEET's.  Of course, they should be equivalent to each other, if they are to be derived directly from QCD.
Even if a specific LEET of QCD is assumed, the parameters included in the LEET can be adjusted so as to meet the data of experiments or Monte Carlo simulation on a lattice.  
In the first paper \cite{KondoI} of a series of papers \cite{KondoII,KondoIII,KondoIV,KondoV,KondoVI} on the quark confinement in Yang-Mills theory in the Abelian gauge, the author has given a scenario of deriving the dual superconductivity in low-energy region of QCD and demonstrated that the DGL theory, i.e., the dual Abelian Higgs (DAH) theory in the London limit can be derived from QCD, provided that the condensation of magnetic monopole takes place.  In this scenario, the mass $m_b$ of the dual gluon (dual Abelian gauge field) is given directly by the non-zero condensate of the magnetic monopole current $k_\mu$, i.e., 
$\langle k_\mu k_\mu \rangle \not=0$.  Since the DGL theory is a LEET written in term of only the Abelian (diagonal) component extracted from the original non-Abelian field, the treatment of the off-diagonal component is crucial for deriving the LEET and also for giving reasonable interpretation of the result.  In the previous paper \cite{KondoI}, all the off-diagonal components are integrated out to write down the LEET in terms of the diagonal components alone.  
The resulting LEET is an Abelian gauge theory preserving a characteristic feature of non-Abelian gauge theory in the sense that the the $\beta$ function for the coupling constant has the same form as the original Yang-Mills theory, exhibiting the asymptotic freedom \cite{QR98,KondoI}.  For this procedure to be meaningful, the mass $M_A$ of the off-diagonal components must be heavier than the mass of the diagonal components.   
This procedure can be justified based on the Wilsonian renormalization group or the decoupling theorem \cite{AC75}. 
As a result, the LEET is valid in the energy region below $M_A$. 
\par
The aim of this paper is to show that the dual superconductivity can be derived at least in the low-energy region of Gluodynamics (i.e., the gluonic sector of QCD).  
The most important ingredient in deriving the dual superconductivity is the existence of non-zero mass of the off-diagonal gluons. 
Recently, it has been shown \cite{Schaden99,KS00a} that the off-diagonal gluons and off-diagonal ghosts (anti-ghosts) acquire non-zero masses in Yang-Mills theory in the {\it modified} MA gauge \cite{KS00a}. 
This result strongly supports the Abelian dominance in low-energy Gluodynamics.  
  The modified MA gauge was already proposed by the author from a different viewpoint in the paper \cite{KondoII}.  
A remarkable difference of MA gauge from the usual Lorentz type gauge lies in the fact that MA gauge is a nonlinear gauge.  In order to preserve the renormalizability of Yang-Mills theory in the MA gauge, it is indispensable to introduce quartic ghost interaction term as a piece of gauge fixing term \cite{MLP85}.  The modified MA gauge fixes the strength of the quartic ghost interaction by imposing the symmetry, i.e., orthosymplectic group $OSp(4|2)$.  The implications of the $OSp(4|2)$ symmetry have been discussed in the previous papers \cite{KondoII,KondoIII,KondoIV}.  The attractive quartic ghost interaction causes the ghost--anti-ghost condensation.  Consequently, the off-diagonal gluons become massive,%
\footnote{The off-diagonal gluon mass $M_A$ in the MA gauge has been calculated on a lattice by Amemiya and Suganuma \cite{AS99},
 $M_A=1.2$GeV for $SU(2)$.
}
 whereas the diagonal gluons remain massless in this gauge.  
Since the classical Yang-Mills theory is a scale invariant theory, the mass scale must be generated due to quantum effect.  
In this paper, we show that {\it the monopole condensation does really occur due to existence of non-zero mass of off-diagonal gluons}.   {\it The off-diagonal gluon mass also provides the mass of dual gauge field in the DGL theory.}
\par
It should be remarked that in the conventional approaches the off-diagonal components are completely neglected from the beginning in deriving the effective Abelian gauge theory by virtue of the Abelian dominance.  As a result, the conventional approach can not predict the physical quantities without some fitting of the parameters introduced by hand.
The  purpose of this paper is to bridge between the perturbative QCD in the high-energy region and the low-energy effective Abelian gauge theory.  Consequently, the undetermined parameters in the low-energy effective theory can in principle be expressed by the parameters of the original Yang-Mills theory, i.e., the gauge coupling constant $g$ and the renormalization group (RG) invariant scale $\Lambda_{QCD}$.
\par
In this paper we pay attention to the vacuum expectation value (VEV) of the Wilson loop operator $W_C[{\cal A}]$, which is written in the framework of the functional integration as
\begin{equation}
  \langle W_C[{\cal A}] \rangle_{YM} = Z_{YM}^{-1} 
\int d\mu \exp \left\{ i \int d^4x {\cal L}_{YM}^{tot} \right\} W_C[{\cal A}] ,
\end{equation}
where ${\cal L}_{YM}^{tot}$ is the total Yang-Mills Lagrangian, i.e.,  Yang-Mills Lagrangian ${\cal L}_{YM}$ plus the gauge fixing (GF) term ${\cal L}_{GF+FP}$ including the Faddeev-Popov (FP) ghost term in the modified MA gauge.  Our derivation is based on the Becchi-Rouet-Stora-Tyutin \cite{BRST}(BRST) formulation of the gauge theory.  Hence,  $d\mu$ is the integration measure which is BRST invariant,
$
  d\mu_{YM} := {\cal D}{\cal A}_\mu {\cal D}B {\cal D}C {\cal D}\bar C ,
$
where $B$ is the Nakanishi-Lautrap (NL) Lagrangian multiplier field.
We regard the Wilson loop as a special choice of the source term in general Yang-Mills theory.   
The Wilson loop is used as a probe to see the QCD vacuum. The shape of the Wilson loop is arbitrary at this stage.
\par
 We seek other theories (with the corresponding source term) which is equivalent to the original Yang-Mills theory at least in the low-energy region in the sense that it gives the same VEV of the Wilson loop operator as the original Yang-Mills theory for {\it large loop} $C$. 
 From this viewpoint, we obtain three LEET's, i.e., DGL theory (DAH theory), magnetic monopole theory and confining string theory.  For this derivation to be successful, the existence of quantum corrections coming from off-diagonal components of gluons and ghost (and anti-ghost) is indispensable.  
Without quantum corrections, we can not derive magnetic monopole condensation.  This is plausible, since it is the quantum correction that can introduce the scale into the gauge theory which is scale invariant at the classical level.

\par
This paper is organized as follows.
In section~\ref{sec:dualsuper}, we give a strategy (steps) of deriving the LEET of Gluodynamics.
In fact, we demonstrate that at least the London limit of the DGL theory (i.e., DAH model) of type II can be obtained as a very special limit of the resulting LEET of Gluodynamics.  However, our method is able to treat more general situation, not restricted to the London limit.  Rather, our result suggests that the DGL theory can not be in the London limit.  
\par
In section~\ref{sec:finalstep}, a more general case of the DGL theory is discussed.  
It is shown that the LEET of the supposed DGL theory agrees with the LEET derived from the Yang-Mills theory according to the above steps, in the energy region less than the mass $m_H$ of the dual Higgs mass. This way of showing equivalence is a little bit indirect.
The reason is as follows.
We know that the DAH model or the DGL theory is a renormalizable theory (within perturbation theory in the magnetic coupling constant $g_m=4\pi/g$) and hence is a meaningful theory at arbitrary energy scale.  On the other hand, we know that the high energy region of Gluodynamics is correctly described by the non-Abelian Yang-Mills theory with asymptotic freedom.  Therefore, the DGL theory is at best meaningful only in the low-energy region in this context, although it is a renormalizable theory.  
In this sense, we must be careful in saying that the DGL theory is regarded as a LEET of Gluodynamics.  
\par
In section~\ref{sec:estimate}, we try to estimate the neglected terms in the derivation of the LEET based on the large $N$ argument and the decoupling theorem.  It is shown that in the large $N$ expansion the higher-order terms are suppressed by $N^{-2}$ compared to the leading order term.
Thus the LEET obtained above is considered as the leading-order result of the large $N$ expansion.
\par
The LEET of Gluodynamics is not unique.  In fact, we can obtain various LEET's which are equivalent to each other. Other LEET's can be more convenient for the purpose of calculating some kinds of physical quantities.
\par
In section~\ref{sec:monopole}, we derive the magnetic monopole action as another LEET.  The magnetic monopole action $S_{MP}[k]$ is written entirely in terms of the magnetic monopole current $k_\mu$. 
\begin{equation}
  \langle W_C[{\cal A}] \rangle_{YM} = Z_{MP}^{-1} 
\int {\cal D}k_\mu \exp \left\{ i \int d^4x ({\cal L}_{MP}[k] + k_\mu \Xi^\mu ) \right\} ,
\end{equation}
by introducing the four-dimensional solid angle $\Xi_\mu$.
We show that the VEV of the Wilson loop exhibits area law decay for large loop $C$,
\begin{equation}
  \langle W_C[{\cal A}] \rangle_{YM} \cong \exp \{ - \sigma_{st} A(C) \} ,
\end{equation}
where $A(C)$ is the area of the minimal surface $S$ bounded by $C$.
The string tension $\sigma_{st}$ is explicitly obtained and it agrees with that predicted by the DGL theory..  
Then we show that the magnetic monopole condensation really occurs in the low-energy region in the sense that the current-current correlation at the same space-time point,
$\langle k_\mu k_\mu \rangle$, has non-vanishing expectation value,
\begin{equation}
  \langle k_\mu k^\mu \rangle_{YM} := Z_{MP}^{-1} 
\int {\cal D}k_\mu \exp \left\{ i \int d^4x  {\cal L}_{MP}[k] \right\} 
k_\mu k^\mu \not = 0 .
\end{equation}
The result is consistent with that of the monopole action on a lattice\cite{maKanazawa}.
\par
In section~\ref{sec:string}, we derive a string theory which is equivalent to the LEET's derived above,
\begin{equation}
  \langle W_C[{\cal A}] \rangle_{YM} = Z_{cs}^{-1} 
\int {\cal D}x_\mu(\sigma) \exp \left\{ i S_{cs}[x] \right\} 
 .
\end{equation}
The action $ S_{cs}[x]$ of the confining string is equal to the Nambu-Goto action for the world sheet $S$ with the Wilson loop $C$ as the boundary.
The string tension $\sigma_{cs}$ in the Nambu-Goto action is the same as $\sigma_{st}$ evaluated by the monopole action derived in section~\ref{sec:monopole}.
The Wilson loop exhibits area law with the string tension $\sigma_{st}$.
Therefore, the obtained string theory is regarded as a low-energy limit of the so-called confining string theory proposed by Polyakov \cite{Polyakov96}.
\par
In section~\ref{sec:numerical}, we discuss what values of the parameters in the LEET's should be chosen to reproduce the numerical results.

The final section~\ref{sec:conclusion} is devoted to summarizing the result obtained in this paper and discussing the future directions of our investigations.
The details of calculations are collected in Appendices, together with useful formulae.

\newpage
\section{\label{sec:dualsuper}Dual superconductivity in low-energy Gluodynamics}
\setcounter{equation}{0}

\subsection{Conventions}
\par
The gauge potential ${\cal A}_\mu$ is written as
\begin{equation}
 {\cal A}_\mu(x) := {\cal A}_\mu^A(x) T^A \quad (A=1, \cdots,N^2-1) ,
\end{equation}
where the generators $T^A(A=1, \cdots,N^2-1)$  of the Lie algebra ${\cal G}$ of the gauge group $G=SU(N)$
are taken to be Hermitian satisfying 
$
 [T^A, T^B] = i f^{ABC} T^C
$
and normalized as
$
 {\rm tr}(T^A T^B) =  {1 \over 2} \delta^{AB}.
$
 For a closed loop $C$, we define the Wilson loop operator $W(C)$ by
\begin{equation}
 W(C) = {\rm tr}\left[ P \exp \left\{ ig \oint_C dx^\mu {\cal A}_\mu(x) \right\} \right]/{\rm tr}(1) ,
\end{equation}
where $P$ denotes the path-ordered product and $g$ is the Yang-Mills coupling constant.
\par
We begin with the vacuum expectation value (VEV) of the Wilson loop operator 
$W(C)$ 
in the Yang-Mills theory with a gauge group $SU(N)$ defined by the functional integral,
\begin{equation}
  \langle W(C) \rangle_{YM} 
:= Z_{YM}^{-1} \int d\mu_{YM} \exp( iS_{YM}^{tot}) W(C) , 
\end{equation}
where $Z_{YM}$ is a normalization factor (or a partition function) to guarantee
$\langle 1 \rangle_{YM} \equiv 1$ 
and $S^{tot}_{YM}$ is the total action obtained by adding the gauge fixing (GF) and Faddeev-Popov (FP) ghost term $S_{GF+FP}$ to the Yang-Mills action $S_{YM}$, 
\begin{equation} 
 S_{YM}^{tot}=S_{YM}+S_{GF+FP} .
\end{equation}
The Yang-Mills action is of the usual form,
\begin{equation}
  S_{YM} = \int d^4x {-1 \over 4} {\cal F}_{\mu\nu}^A(x) {\cal F}^{\mu\nu}{}^A(x) ,
\end{equation}
where ${\cal F}_{\mu\nu}(x)$ is the field strength defined by
\begin{eqnarray}
 {\cal F}_{\mu\nu}(x) 
&:=&  {\cal F}_{\mu\nu}^A(x) T^A
:=   \partial_\mu {\cal A}_\nu(x) 
 -   \partial_\nu {\cal A}_\mu(x)
 - i g [{\cal A}_\mu(x), {\cal A}_\nu(x)] .
\end{eqnarray}
The GF+FP action $S_{GF+FP}$ is specified later (see section~\ref{sec:MAgauge}).
Finally, $d\mu_{YM}$ is the integration measure, 
\begin{equation}
 d\mu_{YM} := {\cal D}{\cal A}_\mu^A {\cal D}B^A {\cal D}C^A {\cal D}\bar C^A ,
\end{equation}
which is invariant under the Becchi-Rouet-Stora-Tyutin (BRST) transformation,
\begin{eqnarray}
   \delta_B {\cal A}_\mu(x)  &=&  {\cal D}_\mu[{\cal A}] {\cal C}(x)
   := \partial_\mu {\cal C}(x) - ig [{\cal A}_\mu(x), {\cal C}(x)],
    \nonumber\\
   \delta_B {\cal C}(x)  &=&  i{g \over 2} [{\cal C}(x), {\cal
C}(x)] ,
    \nonumber\\
   \delta_B \bar {\cal C}(x)  &=&   i B(x)  ,
    \nonumber\\
   \delta_B B(x)  &=&  0 ,
    \label{BRST0}
\end{eqnarray}
where $B$ is the Nakanishi-Lautrap (NL) field and ${\cal C}$ ($\bar {\cal C}$) is the ghost (anti-ghost) field.

\subsection{Step 1: Non-Abelian Stokes theorem for the Wilson loop}

 We make use of a version of the non-Abelian Stokes theorem (NAST) \cite{DP89,KondoIV,KT99} to rewrite the Wilson loop operator in terms of the diagonal components.  This version of NAST was first obtained by Diakonov and Petrov for $SU(2)$ \cite{DP89}.  It is possible to generalize their result to $SU(N)$ ($N \ge 3$).
\par
{\bf Theorem:}\cite{KT99}  {\it For a closed loop $C$, we define the non-Abelian 
Wilson loop operator by
\begin{equation}
  W_C[{\cal A}] = {\rm tr} \left\{ P \exp \left[ ig \oint_C dx^\mu {\cal A}_\mu(x) 
\right] \right\}/{\rm tr}(1) ,
\end{equation}
where $P$ is the path-ordered product.
Then it is rewritten as
\begin{eqnarray}
\mbox{\fboxsep=.1in \framebox{$\displaystyle
  W_C[{\cal A}] = \int d\mu_C(\xi) \exp \left[ ig \oint_C dx^\mu 
a_\mu^{\xi}(x) \right] 
=\int d\mu_C(\xi) \exp \left[ ig \int_S d\sigma^{\mu\nu} f_{\mu\nu}^{\xi}(x) 
\right] ,
$}}
\end{eqnarray}
where
\begin{equation}
\mbox{\fboxsep=.1in \framebox{$\displaystyle
 a_\mu^{\xi}(x) := \langle \Lambda | {\cal A}_\mu^\xi(x) |\Lambda \rangle , 
\quad
 {\cal A}_\mu^\xi(x) = \xi^\dagger(x) {\cal A}_\mu(x) \xi(x) + {i \over g} 
\xi^\dagger(x) \partial_\mu \xi(x) ,
$}}
\label{eq:acomp}
\end{equation}
and%
\footnote{
Note that $f_{\mu\nu}$ is not equal to the diagonal component of ${\cal F}_{\mu\nu}$.
}
\begin{equation}
\mbox{\fboxsep=.1in \framebox{$\displaystyle
  f_{\mu\nu}^{\xi}(x) := \partial_\mu a_\nu^{\xi}(x) - \partial_\nu a_\mu^{\xi}(x) .$}}
\end{equation}
Here $|\Lambda \rangle$ is the highest-weight state of the representation defining 
the Wilson loop and the measure $d\mu_C(\xi)$ is the product measure along the loop $C$,
$d\mu_C(\xi)=\prod_{x \in C}d\mu(\xi(x))$, 
where $d\mu(\xi)$ is the invariant Haar measure on $G/\tilde H$ with the maximal 
stability group $\tilde H$. The maximal stability group $\tilde H$ is the subgroup leaving the 
highest-weight state invariant (up to a phase factor), i.e.,
\begin{equation}
 g|\Lambda \rangle = \xi h |\Lambda \rangle = \xi |\Lambda \rangle e^{i\phi(h)} ,
\end{equation}
for $\xi \in G/\tilde H$ and $h \in \tilde H$.
It depends on the group $G$ and the 
representation in question.}
\par
For $G=SU(2)$, the $\tilde H$ is given by the maximal torus subgroup 
$H=U(1)$ irrespective of the representation. Hence $G/\tilde H=SU(2)/U(1)=CP^1=F_1$.  For 
$G=SU(N) (N \ge 3)$, however, $\tilde H$ does not necessarily agree with the maximal torus group 
$H=U(1)^{N-1}$ depending on the representation.  For $G=SU(3)$, all the 
representations can be classified by the Dynkin index $[m,n]$.  If $m=0$ or $n=0$,   
$\tilde H=U(2)$ and $G/\tilde H=CP^2$.  On the other hand, when $m\not=0$ and 
$n\not=0$, $\tilde H=U(1)^2=U(1)\times U(1)$ and $G/\tilde H=F_2$.  Here $CP^n$ is the complex 
projective space and $F_n$ the flag space.  This NAST is obtained by making use of the 
generalized coherent state.  For details,  see the reference\cite{KT99}.
\par
 For the fundamental representation, the expression (\ref{eq:acomp}) is greatly 
simplified as
\begin{equation}
 \langle \Lambda | (\cdots) |\Lambda \rangle = 2 {\rm tr}[{\cal H}(\cdots)] ,
\quad
 {\cal H}={1 \over 2}{\rm diag}\left({N-1 \over N},{-1 \over N}, \ldots, {-1 \over N} 
\right) .
\end{equation}
Therefore, 
\begin{eqnarray}
 a_\mu &=& {1 \over 2}{\cal A}_\mu^3 \quad  \mbox{for} \quad G=SU(2) 
\\
 a_\mu &=& {1 \over 3}\left[ {\cal A}_\mu^3+{1 \over \sqrt{3}} {\cal A}_\mu^8 \right]  \quad \mbox{for} \quad  G=SU(3).
\end{eqnarray}
This implies that the non-Abelian Wilson loop can be expressed by the diagonal 
(Abelian) components.  This is suggestive of the Abelian dominance in the expectation 
value of the Wilson loop.
\par
The monopole dominance in the Wilson loop is also expected to hold as shown follows.
We can rewrite $f_{\mu\nu}^{\xi}$ in the NAST as  
\begin{equation}
 f_{\mu\nu}^{\xi}(x) = \partial_\mu[n^A(x) {\cal A}_\nu^A(x)] - \partial_\nu[n^A(x) {\cal 
A}_\mu^A(x)] - {1 \over g} f^{ABC} n^A(x) \partial_\mu n^B(x) \partial_\nu n^C(x) ,
\end{equation}
where 
\begin{equation}
 n^A(x) T^A = \xi(x) {\cal H} \xi^\dagger(x) 
= U(x){\cal H}U^\dagger(x) .
\end{equation}
The  $f_{\mu\nu}^{\xi}$ is invariant under the full $G$ gauge transformation as well 
as the residual $H$ gauge transformation.  (Indeed, we can write a manifestly gauge 
invariant form, see ref.\cite{HU99}.)
The $f_{\mu\nu}^{\xi}$ is a generalization of the 't Hooft-Polyakov tensor of the non-Abelian magnetic monopole, 
if we identify $n^A(x)$ with the unit vector of the elementary Higgs scalar field in the 
gauge-Higgs theory:
\begin{equation}
 n^A(x) \leftrightarrow \hat \phi^A(x) := \phi^A(x)/|\phi(x)| .
\end{equation}
This implies that $n^A(x)$ is identified with the composite scalar field and plays the 
same role as the scalar field in the gauge-Higgs model, even though QCD has no 
elementary scalar field.  This fact could explain why the QCD vacuum can be dual 
superconductor due to magnetic monopole condensation.
\par
By introducing the magnetic monopole current $k$ by 
$k := \delta *f$ (see Appendix~\ref{sec:formula}), we have another expression,
\begin{eqnarray}
\mbox{\fboxsep=.1in \framebox{$\displaystyle
  W_C[{\cal A}] 
= \int d\mu_C(\xi) \exp \left[ ig (\Xi, k^{\xi}) \right] ,
\quad
 \Xi := \Delta^{-1}* d\Theta ,
$}}
\end{eqnarray}
where $\Delta$ is the Laplacian $\Delta:=d\delta+\delta d$ and $T$ is a two-form determined by the surface element $dS$ of the surface spanned 
by the Wilson loop $C$.
The derivation is given in Appendix~\ref{sec:NAST}.
Hence, the Wilson loop can also be expressed by the magnetic monopole current $k_\mu$.
The above results hold irrespective of which gauge theory we consider.
\par
In the case of $SU(2)$, the Wilson loop in an arbitrary representation is written in the form \cite{DP89,KondoIV},
\begin{eqnarray}
  W_C[{\cal A}] &=& \int d\mu_C(\xi) \exp \left[ ig J \oint_C dx^\mu 
a_\mu^{\xi}(x) \right] ,
\end{eqnarray}
where $a_\mu^\xi$ is the {\it Abelian} gauge field (or the diagonal component of ${\cal A}_\mu^\xi$) defined by
\begin{equation}
 a_\mu^\xi(x) := {\rm tr}\{ \sigma_3[\xi(x){\cal A}_\mu(x) \xi^\dagger(x) + ig^{-1}\xi(x) \partial_\mu \xi^\dagger(x)] \} ,
\end{equation}
for an element $\xi \in G/H=SU(2)/U(1) \cong S^2 \cong CP^1$.
Here $J$ is a character which distinguishes the different representation defining the Wilson loop,
$J={1 \over 2}, 1, {3 \over 2}, \ldots$.
Moreover, the use of the usual Stokes theorem leads to
\begin{equation}
\mbox{\fboxsep=.1in \framebox{$\displaystyle
 W(C) = \int d\mu_C(\xi) \exp \left[ig J \int_{S_C} dS^{\mu\nu} f_{\mu\nu}^\xi(x) \right] ,
$}}
\end{equation}
where $f_{\mu\nu}^\xi(x)$ is the {\it Abelian} field strength defined by 
$
 f_{\mu\nu}^\xi(x) := \partial_\mu a^\xi_\nu(x) - \partial_\nu a^\xi_\mu(x),
$
and $S_C$ is an arbitrary two-dimensional surface with the loop $C$ as the boundary.
Here it should be remarked that $f_{\mu\nu}^\xi(x)$ is invariant under the full $G=SU(2)$ gauge transformation as well as the residual $H=U(1)$ gauge transformation, since it has the same form as the usual 'tHooft--Polyakov tensor describing the magnetic monopole, see \cite{KondoIV}. 
\par

\par
\subsection{\label{sec:cumulant}Step 2: Cumulant expansion}

Now we specify the gauge theory in terms of which the VEV of the Wilson loop operator is evaluated.  We consider the Yang-Mills theory with gauge fixing term.  The gauge fixing is discussed in the next step.
\par
By applying the cumulant expansion to the VEV of the Wilson loop,
\begin{equation}
 \langle W(C) \rangle_{YM}  = \int d\mu_C(\xi) \Big\langle \exp \left[ig J \int_{S_C} dS^{\mu\nu} f_{\mu\nu}^\xi(x) \right] \Big\rangle_{YM} ,
\end{equation}
the VEV of the exponential is replaced by the exponential of the connected expectation as 
\begin{equation}
 \langle W(C) \rangle_{YM} = \int d\mu_C(\xi) \exp \left[ -{g^2 \over 2}J^2
  \int_{S_C} dS^{\mu\nu}(x) \int_{S_C} dS^{\rho\sigma}(y) \langle f_{\mu\nu}^\xi(x) f_{\rho\sigma}^\xi(y) \rangle_{YM} + \cdots \right] ,
\label{cumu}
\end{equation}
where we have used $\langle f_{\mu\nu}^\xi(x) \rangle_{YM}=0$
and $\cdots$ denotes the higher-order cumulants.
\par
The cumulant expansion is a well-known technique in statistical mechanics and quantum field theory.  
In what follows, we will neglect the higher-order cumulants in (\ref{cumu}) as in the analysis of the stochastic vacuum model.%
\footnote{In the non-perturbative study of QCD, the cumulant expansion is extensively utilized by the stochastic vacuum model (SVM) \cite{Dosch87} where the different version of the non-Abelian Stokes theorem is adopted.
In the SVM, the approximation of neglecting higher order cumulants is called the bilocal approximation. 
The validity of bilocal approximation in SVM was confirmed by Monte Carlo simulation on a lattice \cite{DDM97,BBDV98}.
The author would like to thank Dmitri Antonov for this information.
}
  The validity of this approximation, i.e, the truncation of the cumulant expansion, can be examined by Monte Carlo simulations on a lattice, as performed for the stochastic vacuum model, see \cite{DDM97}.  In the framework of our approach, this approximation can be justified in the sense that the approximation is self-consistent within the APEGT derived below. See section \ref{subsec:higher-order-cumulants} for more details.
\par
\subsection{\label{sec:MAgauge}Step 3: Maximal Abelian gauge fixing}

\par
First of all, we define the decomposition of the gauge potential into the diagonal and off-diagonal components,
\begin{equation}
  {\cal A}_\mu(x) = {\cal A}_\mu^A(x) T^A = a_\mu^i(x) T^i + A_\mu^a(x) T^a ,
\end{equation}
where $T^i \in {\cal H}$ and $T^a \in {\cal G}-{\cal H}$ with ${\cal H}$ being the Cartan subalgebra.
As a gauge fixing condition for the off-diagonal component, we adopt the modified version of the maximal Abelian (MA) gauge proposed by the author \cite{KondoII},
\begin{equation}
 S_{GF+FP} = \int d^4x i \delta_B \bar \delta_B \left[ {1 \over 2}A_\mu^a(x) A^{\mu}{}^a(x)
 - {\alpha \over 2}i C^a(x) \bar C^a(x) \right] ,
\label{MAGF}
\end{equation}
where $\alpha$ corresponds to the gauge fixing parameter for the off-diagonal components, since the explicit calculation of the anti-BRST transformation $\bar \delta_B$ yields
\begin{equation}
  S_{GF+FP} = - \int d^4x i \delta_B \left[ 
\bar C^a  \left\{ D_\mu[a]A^\mu + {\alpha \over 2} B \right\}^a
- i {\alpha \over 2} g f^{abi} \bar C^a \bar C^b  C^i
- i {\alpha \over 4} g f^{abc} C^a \bar C^b \bar C^c \right] .
\label{GF20}
\end{equation}
In order to see the effect of ghost self-interaction, we take
\begin{equation}
  S_{GF+FP} = - \int d^4x i \delta_B \left[ 
\bar C^a  \left\{ D_\mu[a]A^\mu + {\alpha \over 2} B \right\}^a
- i {\zeta \over 2} g f^{abi} \bar C^a \bar C^b  C^i
- i {\zeta \over 4} g f^{abc} C^a \bar C^b \bar C^c \right] ,
\label{GF2}
\end{equation}
where we must put $\zeta=\alpha$ to recover Eq.(\ref{MAGF}).  The most general form of the MA gauge was obtained by Hata and Niigata \cite{HN93}.

\par
By performing the BRST transformation explicitly, we obtain
\begin{eqnarray}
  S_{GF+FP} &=&   \int d^4x  \{ 
B^a D_\mu[a]^{ab}A^\mu{}^b+ {\alpha \over 2} B^a B^a
\nonumber\\
&&+ i \bar C^a D_\mu[a]^{ac} D^\mu[a]^{cb} C^b
- i g^2 f^{adi} f^{cbi} \bar C^a C^b A^\mu{}^c A_\mu^d 
\nonumber\\
&&+ i \bar C^a D_\mu[a]^{ac}(g f^{cdb}  A^\mu{}^d C^b)
+ i \bar C^a g  f^{abi} (D^\mu[a]^{bc}A_\mu^c) C^i 
\nonumber\\
&&+{\zeta \over 8} g^2 f^{abe}f^{cde} \bar C^a \bar C^b C^c C^d
+ {\zeta \over 4} g^2 f^{abc} f^{aid} \bar C^b \bar C^c C^i C^d
+ {\zeta \over 2} g f^{abc} i B^b C^a \bar C^c  
\nonumber\\
&&- \zeta  g f^{abi} i B^a \bar C^b C^i 
+ {\zeta \over 4} g^2 f^{abi} f^{cdi} \bar C^a \bar C^b C^c C^d \} .
\label{GF3}
\end{eqnarray}
In particular, the $SU(2)$ case is greatly simplified as
\begin{eqnarray}
  S_{GF+FP} &=&   \int d^4x  \{ 
B^a D_\mu[a]^{ab}A^\mu{}^b+ {\alpha \over 2} B^a B^a
\nonumber\\
&&+ i \bar C^a D_\mu[a]^{ac} D^\mu[a]^{cb} C^b
- i g^2 \epsilon^{ad} \epsilon^{cb} \bar C^a C^b A^\mu{}^c A_\mu^d 
\nonumber\\
&&
+ i \bar C^a g  \epsilon^{ab} (D_\mu[a]^{bc}A_\mu^c) C^3 
\nonumber\\
&&- \zeta  g \epsilon^{ab} i B^a \bar C^b C^3 
+ {\zeta \over 4} g^2 \epsilon^{ab} \epsilon^{cd} \bar C^a \bar C^b C^c C^d \} .
\label{GF4}
\end{eqnarray}
Integrating out the NL field $B^a$ leads to
\begin{eqnarray}
  S_{GF+FP} &=&   \int d^4x  \{ 
-{1 \over 2\alpha}(D_\mu[a]^{ab}A^\mu{}^b)^2  
+ ( 1-\zeta/\alpha ) i \bar C^a g  \epsilon^{ab} (D_\mu[a]^{bc}A_\mu^c) C^3 
\nonumber\\
&&+i \bar C^a D_\mu[a]^{ac} D^\mu[a]^{cb} C^b
- i g^2 \epsilon^{ad} \epsilon^{cb} \bar C^a C^b A^\mu{}^c A_\mu^d 
\nonumber\\
&&+ {\zeta \over 4} g^2 \epsilon^{ab} \epsilon^{cd} \bar C^a \bar C^b C^c C^d \} .
\label{GF5}
\end{eqnarray}

The advantages of the modified MA gauge (\ref{MAGF}) are as follows.
\begin{enumerate}
\item
 $S_{GF+FP}$ is BRST invariant, i.e., $\delta_B S_{GF+FP}=0$, due to nilpotency of the BRST transformation $\delta_B^2=0$.

\item
 $S_{GF+FP}$ is anti-BRST invariant, i.e., $\bar \delta_B S_{GF+FP}=0$, due to nilpotency of the anti-BRST transformation $\bar \delta_B^2=0$.

\item
 $S_{GF+FP}$ is supersymmetric, i.e., invariant under the $OSp(4|2)$ rotation among the component fields in the supermultiplet $({\cal A}_\mu, C, \bar C)$ defined on the superspace $(x_\mu, \theta, \bar \theta)$.  The hidden supersymmetry causes the dimensional reduction in the sense of Parisi-Sourlas.  Then the 4-dimensional GF+FP sector reduces to the 2-dimensional coset (G/H) nonlinear sigma model.  See ref.\cite{KondoII} for more details.

\item
 $S_{GF+FP}$ is invariant under the FP ghost conjugation,
\begin{equation}
 C^A \rightarrow \pm \bar C^A, \quad
 \bar C^A \rightarrow \mp C^A, \quad
 B^A \rightarrow - \bar B^A, \quad
 \bar B^A \rightarrow - B^A, \quad
 ({\cal A}_\mu^A \rightarrow {\cal A}_\mu^A) .
\end{equation}
  Therefore, we can treat $C$ and $\bar C$ on equal footing.  In other words, the theory is totally symmetric under the exchange of $C$ and $\bar C$.

\item
The Yang-Mills theory in the modified MA gauge (with the total action $S_{YM}+S_{GF+FP}$) is renormalizable.
The naive MA gauge 
\begin{eqnarray}
  S_{GF+FP} &=& - \int d^4x i \delta_B \left[ 
\bar C^a  \left\{ D_\mu[a]A^\mu + {\alpha \over 2} B \right\}^a \right] 
\\
&=&   \int d^4x  \{ 
B^a D_\mu[a]^{ab}A^\mu{}^b+ {\alpha \over 2} B^a B^a
\nonumber\\
&&+ i \bar C^a D_\mu[a]^{ac} D^\mu[a]^{cb} C^b
- i g^2 \epsilon^{ad} \epsilon^{cb} \bar C^a C^b A^\mu{}^c A_\mu^d 
\nonumber\\
&&
+ i \bar C^a g  \epsilon^{ab} (D_\mu[a]^{bc}A_\mu^c) C^3 
\end{eqnarray}
spoils the renormalizability, since radiative corrections induce (even for $\alpha=0$) the four-ghost interaction, 
\begin{eqnarray}
   z_{4c} g^2 \epsilon^{ab} \epsilon^{cd} \bar C^a \bar C^b C^c C^d,
\quad z_{4c} = 4 N {g^2 \over (4\pi)^2} \ln {\mu \over \mu_0}  ,
\label{gsi}
\end{eqnarray}
owing to the existence of the vertex
$- i g^2 \epsilon^{ad} \epsilon^{cb} \bar C^a C^b A^\mu{}^c A_\mu^d$, see eq.(2.52) and Appendix B of the paper \cite{KondoI}.
This is because the MA gauge is a nonlinear gauge.  For the renormalizability of the Yang-Mills theory in the MA gauge, therefore, we need the four-ghost interaction from the beginning.  In fact, the renormalizability of the Yang-Mills theory supplemented with the four-ghost interaction was proved to all orders in perturbation theory \cite{MLP85}.

\end{enumerate}
In order to completely fix the gauge degrees of freedom, we add the GF+FP term for the diagonal component $a_\mu^i$ to (\ref{GF20}) or (\ref{GF2}):
\begin{equation}
  S_{GF+FP} = - \int d^4x i \delta_B \left[ 
\bar C^i  \left( \partial^\mu a_\mu^i + {\beta \over 2} B^i \right)
 \right] ,
\label{GFabel}
\end{equation}
where we have adopted the gauge fixing condition of the Lorentz type, $\partial^\mu a_\mu^i=0$.
The choice of the modified MA gauge is essential in deriving the off-diagonal gluon mass.

\par
\subsection{\label{sec:massgeneration}Step 4: Dynamical mass generation for off-diagonal components}

It is shown that the four-ghost self-interaction is indispensable for the renormalizability of Yang-Mills theory in the MA gauge.  Moreover, it has been shown \cite{Schaden99,KS00a} that the attractive four-ghost interaction in the modified MA gauge causes the ghost--anti-ghost condensation and that this condensation provides masses for the off-diagonal gluons and off-diagonal ghosts and anti-ghosts.  The massive off-diagonal fields do not propagate in the long distance.  Therefore, we can neglect off-diagonal components in the low-energy or long-distance region, except for the renormalization of the remaining diagonal fields.   This result strongly supports the infrared Abelian dominance conjectured by Ezawa and Iwazaki \cite{EI82}. 
\par
The dynamical mass generation of the off-diagonal components is understood based on the argument of Coleman-Weinberg type. See the paper\cite{KS00a}.
The off-diagonal gluon propagator is given by
\begin{eqnarray}
  \langle A_\mu^a(x) A_\nu^b(y) \rangle &=& \int {d^4k \over (2\pi)^4} e^{ikx}D_{\mu\nu}^{ab}(k)  ,
\\
 D_{\mu\nu}^{ab}(k) := \delta^{ab} D_{\mu\nu}(k), \quad
D_{\mu\nu}(k) &:=& {1 \over k^2-M_A^2} \left[ g_{\mu\nu} - (1-\alpha){k_\mu k_\nu \over k^2-\alpha M_A^2} \right] .
\end{eqnarray}
The mass $M_A$ of the off-diagonal gluon comes from the ghost--anti-ghost condensation caused by the four-ghost interaction.
Especially, in the SU(2) case, we obtain%
\footnote{Here we have used the minimal subtraction scheme (MS) in the dimensional regularization.
}
\begin{equation}
\mbox{\fboxsep=.1in \framebox{$\displaystyle
 M_A^2 = g^2 \langle i \bar C^a C^a \rangle 
= {g^2 \over 16\pi} 4\pi e^{1-\gamma_E} \mu^2 \exp \left[ {-16\pi^2 \over b_0 g^2(\mu)}  \right] 
= {g^2 \over 16\pi} 4\pi e^{1-\gamma_E} \Lambda_{QCD}^2 .
$}}
\label{offmass}
\end{equation}
The SU(3) case is more complicated, see \cite{KS00a} for details.

\par
\subsection{\label{LEET}Step 5: Low-energy effective theory for diagonal fields}

We are going to calculate the VEV of the {\it Abelian} components in the given non-Abelian gauge theory.  If the non-Abelian gauge theory can be rewritten into the effective {\it Abelian} gauge theory which is expressed exclusively in terms of the Abelian components only, we can calculate the above VEV in the resulting effective Abelian theory.
\par
In the previous work \cite{KondoI}, the author has derived an effective Abelian gauge theory by integrating out the off-diagonal components.  The resulting theory was called the Abelian-projected effective gauge theory (APEGT).  The APEGT is expected to be able to describe the low-energy region of gluodynamics or QCD.  
In order to obtain APEGT, we have introduced an antisymmetric tensor field $B_{\mu\nu}$ which enables us to derive the dual (magnetic) theory which is expected to be more efficient for describing the low-energy region.  The magnetic theory can be obtained by the electro-magnetic duality transformation from the electric theory and vice versa.
We have imposed the following duality in the tree level,
\begin{equation}
\mbox{\fboxsep=.1in \framebox{$\displaystyle
 B_{\mu\nu}^i \leftrightarrow *(\rho f_{\mu\nu}^i + \sigma g f^{iab} A_\mu^a A_\nu^b ) := *Q_{\mu\nu}
$}}
\label{duality}
\end{equation}
where $*$ denotes the Hodge star operation (or duality transformation) \cite{NS83,Nakahara90} defined by
\begin{equation}
  *Q_{\mu\nu}  := {1 \over 2} \epsilon_{\mu\nu\rho\sigma}Q^{\rho\sigma} .
\end{equation}  
\par
First, the Yang-Mills Lagrangian is decomposed as
\begin{equation}
{\cal L}_{\rm YM}
 ={\cal L}_{\rm YM}^{(i)}
  +{\cal L}_{\rm YM}^{(a)},
\quad
{\cal L}_{\rm YM}^{(i)}
  =-\frac14\left({\cal F}_{\mu\nu}^i\right)^2 ,
\quad
{\cal L}_{\rm YM}^{(a)}
  =-\frac14\left({\cal F}_{\mu\nu}^a\right)^2 .
\end{equation}
The first piece is expanded as
\begin{eqnarray}
{\cal L}_{\rm YM}^{(i)}
 &=&-\frac14\left[f_{\mu\nu}^i
                  +gf^{ibc}A_\mu^bA_\nu^c\right]^2
    \nonumber\\
 &=&-\frac14\left(f_{\mu\nu}^i\right)^2
    -\frac g2f_{\mu\nu}^if^{ibc}A^{\mu b}A^{\nu c}
    -\frac{g^2}4\left(f^{ibc}A_\mu^bA_\nu^c\right)^2 .
\end{eqnarray}
The simplest form satisfying the duality requirement (\ref{duality}) is given by
\begin{eqnarray}
{\cal L}_{\rm YM}^{(i)} &=&-\frac{1-\rho^2}4\left(f_{\mu\nu}^i\right)^2
    -\frac{1-\rho\sigma}2gf_{\mu\nu}^if^{ibc}A^{\mu b}A^{\nu c}
    -\frac{1-\sigma^2}4g^2\left(f^{ibc}A_\mu^bA_\nu^c\right)^2
    \nonumber\\
 & &\quad
    -\frac14\left(B_{\mu\nu}^i\right)^2
    +\frac i2B_{\mu\nu}^i *Q^{\mu\nu i} .
\label{L_inv^i}
\end{eqnarray}
On the other hand, by defining the covariant derivative with respect to the Abelian gauge filed,
\begin{equation}
D_\mu{\mit\Phi}^A
  :=\left(\partial_\mu\delta^{AB}
         +gf^{AiB}A_\mu^i\right){\mit\Phi}^B ,
\end{equation}
the second piece is rewritten as
\begin{equation}
{\cal L}_{\rm YM}^{(a)}
  =-\frac14\left[D_\mu A_\nu^a
                 -D_\nu A_\mu^a
                 +gf^{abc}A_\mu^bA_\nu^c\right]^2 .
\label{L_inv^a}
\end{equation}
\par
The tensor field $B_{\mu\nu}$ is introduced in such a way that $B_{\mu\nu}$-integration recovers the original Yang-Mills theory.
$B_{\mu\nu}$ is an auxiliary field, since it doesn't have the corresponding kinetic term.  However, the duality requirement leads to ambiguities for the identification as to what is the dual of $B_{\mu\nu}$. In fact, existence of two parameters $\rho, \sigma$ in (\ref{duality}) reflects this ambiguity.

\par
In particular, when $\rho=\sigma$, $Q_{\mu\nu}^i$ is nothing but the diagonal component of the non-Abelian field strength, 
$Q_{\mu\nu}^i = \rho {\cal F}_{\mu\nu}^i$, and hence,
$B_{\mu\nu}^i   =i\rho *\!{\cal F}_{\mu\nu}^i$.
In view of this, the choice (\ref{duality}) is a generalization of that of the previous paper \cite{KondoI} in which two special cases, 
$\rho=\sigma=1$ and $\rho=0, \sigma=1$ for $SU(2)$ have been discussed as eq.~(2.9) and eq.~(2.12) respectively \cite{KondoI}.
The latter case has been first discussed by Quandt and Reinhardt\cite{QR98}.
In the $SU(2)$ case where $f^{abc}=0$, the choice $\sigma=1$ completely eliminates the quartic self-interaction among off-diagonal gluons. 
\par
The derivation of the APEGT was improved recently \cite{KS00b} so as to obtain the APEGT in the systematic way to the desired order where we have required the renormalizability of the resulting effective Abelian gauge theory as a guiding principle.  
We will discuss how to choose $\rho$ and $\sigma$ in subsection~\ref{sec:parameter}.

\par
The strategy of deriving the APEGT is not unique.
A way for obtaining the APEGT is to separate each field $\Phi^A$ into the high-energy mode $\tilde \Phi^A$ and the low-energy mode $\bar \Phi^A$ (i.e.,
$\Phi^A=\tilde \Phi^A+\bar \Phi^A$) and then to integrate out the {\it high-energy modes},  
$$
\tilde a_\mu^i, \tilde A_\mu^a, \tilde B_{\mu\nu}, \tilde C^i, \tilde {\bar C^i}, \tilde C^a, \tilde {\bar C^a},
$$
of all the fields according to the idea of the Wilsonian renormalization group (RG).
The resulting theory will be written in terms of the {\it low-energy modes} $\bar \Phi^A$.  However, we can neglect the low-energy modes of $A_\mu^a$ and $C^a, \bar C^a$ due to Abelian dominance and the final theory can be written in terms of the low-energy modes, $a_\mu^i, B_{\mu\nu}, C^i, \bar C^i$.
In other words, the off-diagonal components $A_\mu^a$ and $C^a, \bar C^a$ have only the high-energy modes. 
To one-loop level,  the high-energy modes 
$\tilde a_\mu^i, \tilde B_{\mu\nu}, \tilde C^i, \tilde {\bar C^i}$ of the diagonal components don't contribute to the results.  Therefore, we can identify $A_\mu^a$ and $C^a, \bar C^a$ with the high-energy modes to be integrated out for obtaining the LEET.  This strategy was adopted in the paper \cite{KS00b}. We do not adopt this method in this paper.
\par
Another way is to integrate out all the massive fields 
\begin{equation}
   A_\mu^a, C^a, \bar C^a .
\end{equation}
  Then the resultant theory will be written in terms of the massless or light fields $a_\mu^i, B_{\mu\nu}^i, C^i, \bar C^i$.  The effect of the massive fields will appear only through the renormalization of the resultant theory.  This is an example of the decoupling theorem \cite{AC75}.  
The only role of the heavy fields in the low momentum behavior of graphs without external heavy  fields is their contribution to coupling constant and field-strength renormalization.  The heavy fields effectively decouple and the low-momentum behavior of the theory is described by a renormalizable Lagrangian consisting of the massless fields only.  The decoupling theorem applies not only to theories with massless fields but in fact to any renormalizable theory with different mass scales.  At momentum smaller compared to the larger masses, the dynamics is determined by the light sector of the theory.
In this paper we adopt this strategy.
See also section \ref{subsec:higher-order-expansion}.
\par
Consequently, the APEGT which was heuristically obtained in the paper \cite{KondoI} and improved systematically in the paper \cite{KS00b} is further modified by taking into account the mass of off-diagonal field components.  The simplest derivation of the modified APEGT is to replace the massless off-diagonal propagator given in the previous paper \cite{KS00b} with the massive off-diagonal propagator in the QCD vacuum with ghost condensation.%
\footnote{However, the following steps can be performed irrespective of the origin of the off-diagonal gluon mass.}

\unitlength=0.001in
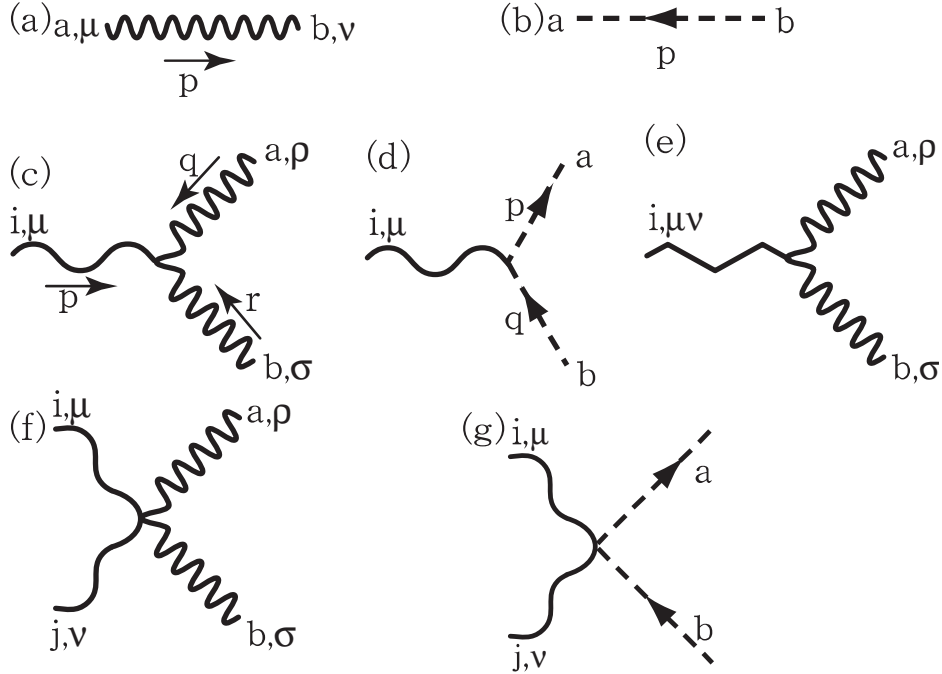
\begin{figure}[t]
\begin{center}
\begin{picture}(5000,3500)
\put(150,3000){%
   \put(0,0){\epsfbox{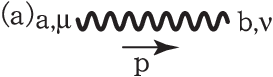}}%
   }%
\put(2700,3150){%
   \put(0,0){\epsfbox{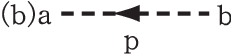}}%
   }%
\put(150,1500){%
   \put(0,0){\epsfbox{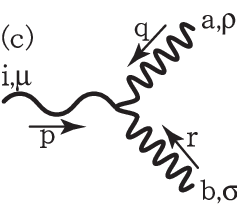}}%
   }%
\put(2000,1500){%
   \put(0,0){\epsfbox{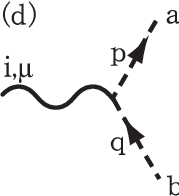}}%
   }%
\put(3450,1500){%
   \put(0,0){\epsfbox{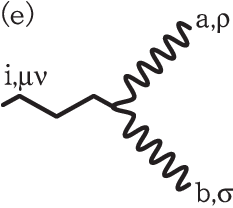}}%
   }%
\put(150,100){%
   \put(0,0){\epsfbox{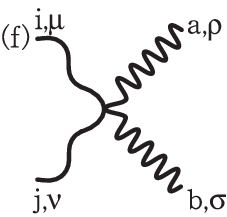}}%
   }%
\put(2500,0){%
   \put(0,0){\epsfbox{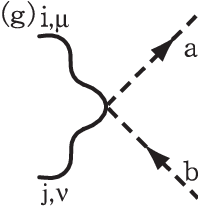}}%
   }%
\end{picture}
\caption[]{Feynman rules. 
           (a),(b): off-diagonal propagators.
           The (rapidly vibrating) wavy line denotes the  
           off-diagonal gluon $A_\mu^a$ and the broken line
           the ghost $C^a$ or anti-ghost
           $\bar C^a$.
           (c),(d),(e): three-point vertices, (f),(g): four-point vertices.
           The (slowly vibrating) wavy line corresponds to the 
           diagonal gluon $a_\mu^i$, while the zig-zag line to the
           (diagonal) anti-symmetric tensor field $*B_{\mu\nu}^i$.
           }
\label{fig:FeynmanRules}
\end{center}
\end{figure}

\par
The Feynman rules are given as follows, see Fig.~\ref{fig:FeynmanRules}.
We enumerate only a part of the rules which are necessary for the renormalization at one-loop level.
The two-loop result will be reported in a subsequent paper\cite{KS00c}.

\subsubsection{Feynman rules} %

{\bf Propagators}: %
\begin{enumerate}
\item[(a)] Off-diagonal gluon propagators: 
\begin{equation}
iD_{\mu\nu}^{ab}
  =-\frac i{p^2-M_A^2}
    \left[g_{\mu\nu}-(1-\alpha)\frac{p_\mu p_\nu}{p^2-\alpha M_A^2}\right]\delta^{ab} .
\end{equation}

\item[(b)] Off-diagonal ghost propagators:%
\footnote{
This is the ghost propagator for $G=SU(2)$.  For $G=SU(3)$, see the paper \cite{KS00a}.  
In this paper, however, we don't need the explicit form of the ghost propagator in the condensed vacuum.  The details will be given in a forthcoming paper \cite{KM00}.
}
\begin{equation}
i\Delta^{ab}
  =-{(k^2+v_d) \delta^{ab}+v_o \epsilon^{ab} \over (-k^2-v_d)^2+v_o^2} .
\end{equation}
\end{enumerate}

{\bf Three-point vertices}: %
\begin{enumerate}
\item[(c)] one diagonal and two off-diagonal gluons:
\begin{eqnarray}
&&i\left< a_\mu^i(p) A_\rho^a(q)
                  A_\sigma^b(r)\right>_{\rm bare}
  \nonumber\\
&&\textstyle
 =gf^{iab}\left[\!\!\left[ (q-r)_\mu
                +\left\{r-(1-\rho\sigma)p+\frac q\alpha\right\}_\rho
                +\left\{(1-\rho\sigma)p-q-\frac r\alpha\right\}_\sigma
                \right]\!\!\right] ,
\end{eqnarray}
where we have introduced the abbreviated notation,
\begin{equation}
[\![A_\mu+B_\rho+C_\sigma]\!]
  =A_\mu g_{\rho\sigma}
   +B_\rho g_{\sigma\mu}
   +C_\sigma g_{\mu\rho} .
\end{equation}

\item[(d)] diagonal gluon, off-diagonal ghost and anti-ghost:
\begin{equation}
i\left< a_\mu^i {\bar C}^a(p) C^b(q)\right>_{\rm bare}
  =i(p+q)_\mu gf^{aib} .
\end{equation}

\item[(e)] diagonal tensor and two off-diagonal gluons:
\begin{equation}
i\left< *B_{\mu\nu}^i A_\rho^a
                  A_\sigma^b\right>_{\rm bare}
  =i 2\sigma g I_{\mu\nu,\rho\sigma} f^{iab} ,
\end{equation}
where
\begin{equation}
 I_{\mu\nu,\alpha\beta} := {1 \over 2}(g_{\mu\alpha}g_{\nu\beta}-g_{\mu\beta}g_{\nu\alpha}) .
\end{equation}

\end{enumerate}

{\bf Four-point vertices}: %
\begin{enumerate}
\item[(f)] two diagonal gluons and two off-diagonal gluons:
\begin{equation}
i\left< a_\mu^i a_\nu^j
        A_\rho^a A_\sigma^b\right>_{\rm bare}
  =ig^2f^{aic}f^{cjb}
    \left[2g_{\mu\nu}g_{\rho\sigma}
          -\left(1-\frac1\alpha\right)
          (g_{\mu\rho}g_{\nu\sigma}+g_{\mu\sigma}g_{\nu\rho})\right] .
\end{equation}

\item[(g)] two diagonal gluons, off-diagonal ghost and anti-ghost:
\begin{equation}
i\left< a_\mu^i a_\nu^j
        {\bar C}^a C^b\right>_{\rm bare}
  =-2g^2f^{aic}f^{cjb}g_{\mu\nu} .
\end{equation}

\end{enumerate}

\unitlength=0.001in
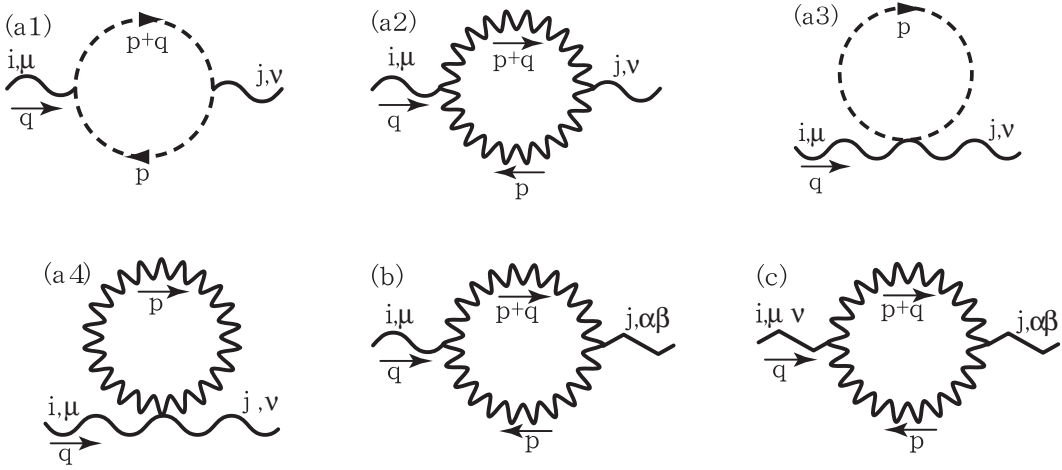
\begin{figure}
\begin{center}
\begin{picture}(5800,2800)
\put(0,1600){%
   \put(0,0){\epsfysize=.9in\epsfbox{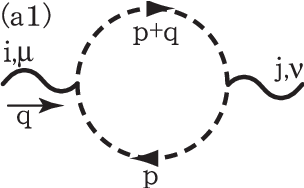}}%
   }%
\put(1900,1520){%
   \put(0,0){\epsfysize=1in\epsfbox{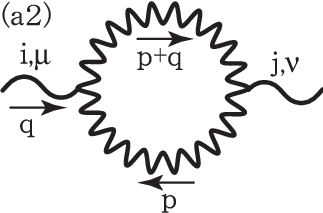}}%
   }%
\put(4100,1550){%
   \put(0,0){\epsfysize=1in\epsfbox{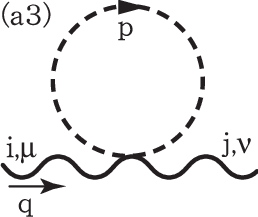}}%
   }%
\put(200,130){%
   \put(0,0){\epsfysize=1.1in\epsfbox{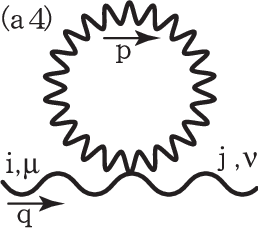}}%
   }%
\put(1900,0){%
   \put(0,200){\epsfysize=1in\epsfbox{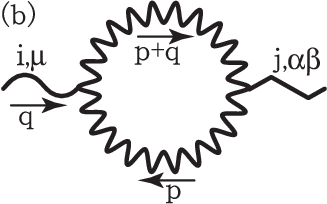}}%
   }%
\put(3900,200){%
   \put(0,0){\epsfysize=1in\epsfbox{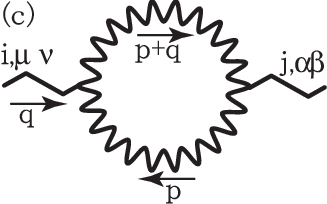}}%
   }%
\end{picture}
\end{center}
\caption[]{Vacuum polarization graphs which are necessary to obtain the APEGT.}
\label{fig:APEGT}
\end{figure}

\subsubsection{APEGT}
\par
Thus the modified APEGT is written as%
\footnote{
In Fig.~\ref{fig:APEGT}, we have omitted Feynman graphs which include external diagonal-ghost and diagonal-anti-ghost lines, and internal off-diagonal-ghosts (-anti-ghost) or internal off-diagonal-gluon lines.  The diagonal ghost and diagonal anti-ghost comes from the GF+FP term (\ref{GFabel}) for the diagonal gluon $a_\mu^i$.  The higher-order terms with more than two external lines are suppressed by $N^{-2}$ in the larege $N$ expansion, as shown in section~\ref{subsec:higher-order-cumulants}.  If we neglect such higher-order terms, the  bilinear terms in the external ghost and anti-ghost decouple from the (\ref{APEGT0}).  Therefore, we don't discuss the contribution from diagonal ghosts and anti-ghosts to (\ref{APEGT0}) in this paper.
}
\begin{equation}
 {\cal L}_{APEGT} = {\cal L}_{APEGT}^0
+ \delta_{(1)} {\cal L}_{APEGT} ,
\label{APEGT0}
\end{equation}
where
${\cal L}_{APEGT}^0$ is the bare diagonal part,
\begin{equation}
 {\cal L}_{APEGT}^0 = - {1-\rho^2 \over 4} (f_{\mu\nu}^i)^2 
- {1 \over 2} \rho  *B_{\mu\nu}^i f^{\mu\nu}{}^i
-{1 \over 4} (B_{\mu\nu}^i)^2 ,
\end{equation}
and the quantum part $\delta_{(1)}{\cal L}_{APEGT}$ is obtained by calculating the vacuum polarization graphs in Fig.~\ref{fig:APEGT}.   In consistent with the renormalizability of the diagonal part, 
\begin{eqnarray}
 \delta_{(1)}{\cal L}_{APEGT} = - {\delta_1 \over 4} (f_{\mu\nu}^i)^2 
- {\delta_2 \over 2}  *B_{\mu\nu}^i f^{\mu\nu}{}^i
-{\delta_3 \over 4} (B_{\mu\nu}^i)^2
+ O \left( {p^2 \over M_A^2} \right) ,
\end{eqnarray}
where $O \left( {p^2 \over M_A^2} \right)$ is the modification term which is obtained in the next section.
We use the dimensional regularization to determine $\delta_1, \delta_2$ and $\delta_3$ which corresponds to Fig.~\ref{fig:APEGT}(a), (b), (c) respectively.
Apart form the finite part, the divergent part of $\delta_1$, $\delta_2$ and $\delta_3$ (proportional to $\epsilon^{-1}$ where $\epsilon :=2-D/2$) is given by
\begin{eqnarray}
\delta_1
  &=&\delta_{{\rm a1}}+\delta_{{\rm a2}},
\nonumber\\
 &&
 \Bigg\{
 \begin{array}{l}
  \displaystyle
  \delta_{{\rm a1}}=
   \left[(2-\rho_{\rm R}\sigma_{\rm R})^2-\frac23
         +\frac{1-\alpha_{\rm R}}2
          (2-\rho_{\rm R}\sigma_{\rm R})\rho_{\rm R}\sigma_{\rm R}\right]
   \frac{(\mu^{-\epsilon}g_{\rm R})^2}{(4\pi)^2}\frac{C_2(G)}{\epsilon} ,
   \cr
  \displaystyle
  \delta_{{\rm a2}}=
    \frac13
    \frac{(\mu^{-\epsilon}g_{\rm R})^2}{(4\pi)^2}\frac{C_2(G)}{\epsilon} ,
 \end{array}
\nonumber\\
\delta_2
  &=&\left[\sigma_{\rm R}(2-\rho_{\rm R}\sigma_{\rm R})
           -\frac{1-\alpha_{\rm R}}2\sigma_{\rm R}
            (1-\rho_{\rm R}\sigma_{\rm R})\right]
     \frac{(\mu^{-\epsilon}g_{\rm R})^2}{(4\pi)^2}
     \frac{C_2(G)}{\epsilon},
\nonumber\\
\delta_3
  &=&-\frac{1+\alpha_{\rm R}}2\sigma_{\rm R}^2
      \frac{(\mu^{-\epsilon}g_{\rm R})^2}{(4\pi)^2}
      \frac{C_2(G)}{\epsilon}
\label{eq:delta2} ,
\end{eqnarray}
where $\delta_{{\rm a1}}$ and $\delta_{{\rm a2}}$ are the contributions from the graphs (a1) and (a2) in Fig.~\ref{fig:APEGT} respectively, and 
$C_2(G)$ is a quadratic Casimir operator defined by 
$C_2(G) \delta^{AB} = f^{ACD}f^{BCD}$ ($C_2(G)=N$ for $G=SU(N)$).%
\footnote{
For a special choice of the parameters, $\rho=\sigma=0$ and $\alpha=0$, only the $\delta_1$ has been calculated by Quandt and Reinhardt\cite{QR98} for $SU(2)$.
}
\par
Note that the Lagrangian ${\cal L}_{APEGT}^0$ is bilinear in the diagonal fields $f_{\mu\nu}$ and $B_{\mu\nu}$.   
So is the term $\delta_{(1)}{\cal L}_{APEGT}$.  Therefore, the divergence can be absorbed if we renormalize the theory.
The mass generation of off-diagonal components also justifies neglecting the higher-derivative terms in the APEGT in the low-energy region.
The additional term of order $O(p^2/M_A^2)$ is discussed in the next step.
In the SU(N) case, it is shown that the higher-order terms which is quartic and more in the fields can be neglected within the framework of the large $N$ expansion, see section~\ref{subsec:higher-order-expansion}.
\par
A non-trivial consequence is that the $\beta$-function in the APEGT has exactly the same form as that in the original Yang-Mills theory and is independent of the gauge parameter $\alpha$ and two parameters $\rho, \sigma$,
\begin{equation}
  \beta(g) := \mu {dg(\mu) \over d\mu} = - b_0 g^3(\mu) + O(g^5) ,
\quad b_0 = {11 \over 3}N > 0 .
\label{beta}
\end{equation}

\subsection{Step 6: Dynamical generation of the kinetic term of $B_{\mu\nu}$}

\par
The antisymmetric tensor field $B_{\mu\nu}$ was introduced as an auxiliary field, since it has no kinetic term.  In this subsection, we show that the  $B_{\mu\nu}$ can have its kinetic term as a consequence of radiative corrections, i.e., {\it the kinetic term of $B_{\mu\nu}$ is dynamically generated}.  It turns out that {\it the dynamical generation of the kinetic terms occurs only when the off-diagonal gluons are massive}.
This fact plays the most important role in deriving the dual superconductivity.
\par
  We proceed to calculate the vacuum polarization for the  $B_{\mu\nu}$ field up to one-loop of the massive off-diagonal gluons, see Fig.\ref{fig:APEGT}(c).
Following the Feynmann rule given in Fig.~\ref{fig:FeynmanRules}, the vacuum polarization of Fig.~\ref{fig:APEGT}(c) is written in momentum space as
\begin{eqnarray}
 \Pi_{\mu\nu,\alpha\beta}^{ij}(k) &:=& {1 \over 2}\int {d^4p \over (2\pi)^4} D_{\sigma_1 \sigma_2}(p) \delta^{d_1d_2}
[-2\sigma gf^{ic_1d_1}I_{\mu\nu,\rho_1 \sigma_1}]
\nonumber\\
&& \times D_{\rho_1 \rho_2}(p+k)\delta^{c_1c_2}
[-2\sigma gf^{jc_2d_2}I_{\alpha\beta,\rho_2 \sigma_2}] ,
\end{eqnarray}
where 
\begin{eqnarray}
D_{\mu\nu}(k) &:=& {1 \over k^2-M_A^2} \left[ g_{\mu\nu} - (1-\alpha){k_\mu k_\nu \over k^2-\alpha M_A^2} \right] 
\\
&=& {1 \over k^2-M^2} \left( g_{\mu\nu} - {k_\mu k_\nu \over M^2} \right) + {k_\mu k_\nu \over M^2}{1 \over k^2-\alpha M^2} .
\end{eqnarray}
Hence the additional contribution coming from Fig.~\ref{fig:APEGT}(c) to the APEGT is given by
\begin{equation}
 \delta_{(1)}^c {\cal L}_{APEGT}  = \int{d^4k \over (2\pi)^4} *B_{\mu\nu}^i(k)\Pi_{\mu\nu,\alpha\beta}^{ij}(k)  
 *B_{\alpha\beta}^j(-k) .
\end{equation}
\par
By making use of the  MS scheme in the dimensional regularization method, we have arrived the following expression after straightforward but tedious calculations: $\epsilon:=2-D/2$
\begin{eqnarray}
 \Pi_{\mu\nu,\alpha\beta}^{ij}(k) &=& {-2C_2 \sigma^2 g^2 \over 16\pi^2}\delta^{ij}
I_{\mu\nu,\alpha\beta} \left[ \epsilon^{-1}{1+\alpha \over 2}\right]
\nonumber\\
&&+ {-2C_2 \sigma^2 g^2 \over 16\pi^2}\delta^{ij}
I_{\mu\nu,\alpha\beta} \int_0^1 dx \Biggr\{  \left[ -x(1-x){k^2 \over M_A^2} \right] \ln \left[{M_A^2 \over \mu^2} - x(1-x){k^2 \over \mu^2} \right]
\nonumber\\
&&- \left[ \alpha+(1-\alpha)x-x(1-x){k^2 \over M_A^2} \right] 
\ln \left[ \{ \alpha+(1-\alpha)x \} {M_A^2 \over \mu^2} - x(1-x){k^2 \over \mu^2} \right] 
\nonumber\\
&&+ (\gamma_E-1)(1-\alpha)(1-x) -\gamma_E + \ln 4\pi \Biggr\} 
\nonumber\\
&&+ {-2C_2 \sigma^2 g^2 \over 16\pi^2}\delta^{ij}
{1 \over 2}{k^2 \over M_A^2}(I-P)_{\mu\nu,\alpha\beta}[...]+ O(\epsilon) ,
\end{eqnarray}
where $C_2\delta^{ij} \equiv f^{iAB}f^{jAB}=f^{iab}f^{jab}$ and $\gamma_E$ is Euler's constant $\gamma_E=0.5772\cdots$.  See Appendix~\ref{sec:vp} for complete expression.
\par
In the neighborhood of $k^2=0$, i.e, in the low-energy region such that $k^2/M^2 \ll 1$, we use the {\it low-energy} (or {\it large mass}) expansion, e.g.,
\begin{eqnarray}
  \ln \left[{M_A^2 \over \mu^2} - x(1-x){k^2 \over \mu^2} \right]
&=& \ln  {M_A^2 \over \mu^2} + \ln \left[1 - x(1-x){k^2 \over M_A^2} \right]
\nonumber\\
&=& \ln  {M_A^2 \over \mu^2} - \sum_{n=1}^{\infty} {1 \over n}  x^n(1-x)^n \left({k^2 \over M_A^2} \right)^n .
\end{eqnarray}
Note that this expansion is possible only when $M_A\not=0$.
This is a reason why the dynamical generation of the kinetic term takes place.
Thus we obtain
\begin{eqnarray}
 && \Pi_{\mu\nu,\alpha\beta}^{ij}(k) 
\nonumber\\
&=&  
- { 2C_2 \sigma^2 g^2 \over 16\pi^2}\delta^{ij}  
\Biggr\{ I_{\mu\nu,\alpha\beta} \Biggr[  \epsilon^{-1} {1+\alpha \over 2} 
+  f_0(\alpha) + f_1(\alpha) {k^2 \over M_A^2} + f_2(\alpha) {k^4
\over M_A^4} \Biggr] 
\nonumber\\
&&+ {1 \over 2}{k^2 \over M_A^2}(I-P)_{\mu\nu,\alpha\beta} \left[ h_0(\alpha)+h_1(\alpha){k^2 \over M_A^2}   \right]
\Biggr\} + O\left(  {k^6 \over M_A^6} \right) ,
\end{eqnarray}
where 
\begin{eqnarray}
 f_0(\alpha) &:=&  - {1+\alpha \over 2} \ln {M_A^2 \over \mu^2} 
+ (\gamma_E-\ln 4\pi -1)(1-\alpha){1 \over 2} - \gamma_E + \ln 4\pi
\nonumber\\&&
+ {1 \over 1-\alpha} \left[ {\alpha^2 \over 2} \ln \alpha +
{1 \over 4}-{1 \over 4}\alpha^2 \right] ,
\\
 f_1(\alpha) &:=& \int_0^1 dx \ x(1-x) \{ 1 + \ln [\alpha+(1-\alpha)x] \}
\\
&=& {1 \over 6} + {1 \over (1-\alpha)^3} \Biggr\{ {\alpha^3 \over 3} \ln \alpha - {\alpha^3 \over 9} + {1 \over 9} - (1+\alpha) \left[ {\alpha^2 \over 2} \ln \alpha - {\alpha^2 \over 4} + {1 \over 4} \right] 
\nonumber\\&&
+ \alpha(\alpha \ln \alpha - \alpha +1) \Biggr\} ,
\\
 f_2(\alpha) &:=& \int_0^1 dx \ \left[ x^2(1-x)^2 - {1 \over 2}{x^2(1-x)^2 \over \alpha+(1-\alpha)x} \right] \\
&=& {1 \over 30} - {1 \over (1-\alpha)^5}\left[ 
 {1 \over 24} -{1 \over 3}\alpha + {1 \over 3}\alpha^3 - {1 \over 24}\alpha^4 
- {1 \over 2} \alpha^2 \ln \alpha
\right] .  
\end{eqnarray}
The function $f=f_1(\alpha)$ is a monotonically increasing function in $\alpha$ defined for $\alpha>0$ and positive $f_1(\alpha)>0$ when $\alpha>0$.  In particular, $f_1(0)=0.0278, f_1(1)=1/6$, and $f_1(11/3)=0.302$.
On the other hand, $f_2(\alpha)$ is a monotonically increasing function defined for $\alpha>0$.  Hence, $0<f_2(\alpha)<1/30$ for $\alpha_0<\alpha<+\infty$ and $-0.00833<f_2(\alpha)<0$ for $0<\alpha<\alpha_0$ with $\alpha_0 \cong 0.12$. Incidentally, $f_2(1)=1/60$, $f_2(2)=0.0220$, $f_2(11/3)=0.0258$.
\par
In the following derivation, we don't need the explicit form of the functions $h_i(\alpha)$.
This is because it can be shown that the term proportional to $(I-P)$ (the so-called boundary term) does not contribute to the area law.
Moreover, if we impose the gauge fixing condition 
$\partial^\mu *B_{\mu\nu}=0$, then such terms give vanishing contributions.
See Appendix~\ref{sec:vp} and a subsequent paper \cite{preparation} for more details.
\par
In the configuration space, therefore, we obtain
\footnote{For an antisymmetric tensor $A_{\mu\nu}$, it is easy to see 
$I_{\mu\nu,\alpha\beta}A^{\alpha\beta}=A_{\mu\nu}$
and 
$A^{\mu\nu}I_{\mu\nu,\alpha\beta}=A_{\alpha\beta}$.
}
\begin{equation}
 \delta_{(1)}^c {\cal L}_{APEGT} = \int d^4x {C_2 \sigma^2 g^2 \over 8\pi^2} *B_{\mu\nu}^i(x)   I^{\mu\nu,\alpha\beta} \left[ f_0(\alpha) 
- f_1(\alpha) {\partial_\lambda \partial^\lambda \over M_A^2} 
+ f_2(\alpha) {(\partial_\lambda \partial^\lambda)^2 \over M_A^4}  
\right] 
 *B_{\alpha\beta}^i(x) .
\end{equation}
The first term in the RHS of $\delta_{(1)}^c {\cal L}_{APEGT}$ gives a piece of the counter terms of the ${\cal L}_{APEGT}^0$ (together with a finite part of $\delta_3$) and the remaining parts give new terms which can not be absorbed into the bare part, the non-renormalizable contribution.   In the limit $M_A \rightarrow \infty$, the non-renormalizable terms disappear.
It is very important to notice that {\it the kinetic term for the auxiliary tensor field $B_{\mu\nu}$ is generated due to radiative corrections} and hence {\it $B_{\mu\nu}$ is regarded as the massive propagating field (at least in the low-energy region less than $M_A$)}.
This is one of the main results of this paper.
The implications of this fact to low-energy QCD will be discussed in what follows.
The other contributions from Fig.~\ref{fig:APEGT}(a),(b) can also be evaluated in the similar manner.  Explicit form will be given in a subsequent paper, since we don't need them in this paper.
\par
Thus, we have obtained the improved version of the APEGT where the VEV of the functional $f[a,h]$ of $a,h$ is given by
\begin{eqnarray}
\mbox{\fboxsep=.1in \framebox{$\displaystyle
 \langle f[a,h] \rangle_{APEGT} 
= Z_{APEGT}^{-1} \int {\cal D}a_\mu^i {\cal D}h_{\mu\nu}^i 
  \exp \left\{i S_{APEGT}[a,h] \right\} f[a,h],
$}}
\end{eqnarray}
using the action
\begin{eqnarray}
\mbox{\fboxsep=.1in \framebox{$\displaystyle
 S_{APEGT}[a,h] = \int d^4x \left[
 {\cal L}_{e}[a]  
+ {1 \over 2} \rho K^{-1/2}  h_{\mu\nu}^i f^{\mu\nu}{}^i  
+ {\cal L}_d[h] \right],
$}}
\\
\mbox{\fboxsep=.1in \framebox{$\displaystyle
 {\cal L}_{e}[a] := - {1-\rho^2 \over 4} (f_{\mu\nu}^i)^2 
+ O\left( {f^4 \over M_A^4} \right),
$}}
\\
\mbox{\fboxsep=.1in \framebox{$\displaystyle
{\cal L}_d[h] := -{1 \over 4} (h_{\mu\nu}^i)^2
- {1 \over 4 \eta^2} h_{\mu\nu}^i \partial_\lambda \partial^\lambda h^{\mu\nu}{}^i 
+ {1 \over 4\gamma^4} h_{\mu\nu}^i (\partial_\lambda \partial^\lambda)^2 h^{\mu\nu}{}^i 
+ O\left( {(\partial^2)^3 \over M_A^6} \right)  ,
$}}
\end{eqnarray}
where we have introduced the rescaling factor $K$ and the rescaled tensor field $h_{\mu\nu}$ by
\begin{eqnarray}
&&
\mbox{\fboxsep=.1in \framebox{$\displaystyle
K := 1+ {C_2 \sigma^2 g^2 \over 2\pi^2}  f_0(\alpha) >0, $}} 
\\&&
\mbox{\fboxsep=.1in \framebox{$\displaystyle
  h_{\mu\nu}^i(x)  :=  K^{1/2}  *B_{\mu\nu}^i(x),
$}}
\end{eqnarray}
and two quantities $\eta$ and $\gamma$ with mass dimension by
\begin{eqnarray}
&& \mbox{\fboxsep=.1in \framebox{$\displaystyle
 \eta^2 := f_1(\alpha)^{-1} {2\pi^2 \over g^2 C_2 \sigma^2}K M_A^2  ,
$}}
\label{eta}
\\&&
\mbox{\fboxsep=.1in \framebox{$\displaystyle
 \gamma^4 := f_2(\alpha)^{-1}  {2\pi^2 \over g^2 C_2 \sigma^2}K M_A^4 .
$}}
\label{gamma}
\end{eqnarray}
The action $S_{APEGT}[a,h]$ has  U(1) invariance of $a_\mu \rightarrow a_\mu + \partial_\mu \theta$, i.e.,
\begin{equation}
 S_{APEGT}[a+d\theta,h]=S_{APEGT}[a,h] .
\end{equation}
Note that the Lagrangian of the modified APEGT is still bilinear in the fields,
$f_{\mu\nu}$ and $B_{\mu\nu}$.
\par
\subsection{Step 7: Dual transformations}

 We use an idea inspired by the (Abelian) electric-magnetic duality to calculate the Abelian field correlators in the cumulant expansion (\ref{cumu}).
Note that the Abelian diagonal field correlation functions 
$\langle f_{\mu\nu}^\xi(x) f_{\rho\sigma}^\xi(y) \rangle_{YM}$
can be calculated in the APEGT, since the APEGT is obtained by integrating out all off-diagonal components.  Hence we have
\begin{eqnarray}
  \langle f_{\mu\nu}^\xi(x) f_{\rho\sigma}^\xi(y) \rangle_{YM}
= \langle f_{\mu\nu}^\xi(x) f_{\rho\sigma}^\xi(y) \rangle_{APEGT} .
\end{eqnarray}
Then the expectation value of the Wilson loop obeys
\begin{eqnarray}
 && \langle W(C) \rangle_{YM} 
\nonumber\\
&=& \int d\mu_C(\xi) \exp \left[ -{J^2g^2 \over 2} 
  \int_{S_C} dS^{\mu\nu}(x) \int_{S_C} dS^{\rho\sigma}(y) \langle f_{\mu\nu}^\xi(x) f_{\rho\sigma}^\xi(y) \rangle_{APEGT} + \cdots \right] 
\end{eqnarray}
We use repeatedly the integration by parts to rewrite the expectation value of the electric quantity into that of the magnetic quantity as follows:%
\footnote{
Here we have used a property of the measure ${\cal D}h$,
$
 \int {\cal D}h {\delta \over \delta h}(\cdots) = 0 .
$
}
\begin{eqnarray}
 && Z_{APEGT} \langle f^{\mu\nu}(x) f^{\rho\sigma}(y) \rangle_{APEGT} 
\nonumber\\
&=& \int {\cal D}a_\mu {\cal D}h_{\mu\nu} \exp \left\{ i \int d^4x \left( 
 {\cal L}_e[a] + {\cal L}_{d}[h] 
 \right) \right\} 
\nonumber\\&& \times
\left({2K^{1/2} \over i\rho}\right)^2 {\delta \over \delta h_{\mu\nu}(x)} 
{\delta \over \delta h_{\rho\sigma}(y)} 
\exp \left\{ i \int d^4x  
 {1 \over 2} \rho K^{-1/2}  h_{\mu\nu} f^{\mu\nu}
  \right\} 
\nonumber\\
&=& \int {\cal D}a_\mu {\cal D}h_{\mu\nu} 
\exp \left\{ i \int d^4x  
 {1 \over 2} \rho K^{-1/2}  h_{\mu\nu} f^{\mu\nu}
  \right\} 
\left({2K^{1/2} \over i\rho}\right)^2
\nonumber\\&& \times
{\delta \over \delta h_{\rho\sigma}(y)} 
{\delta \over \delta h_{\mu\nu}(x)}  \exp \left\{ i \int d^4x \left( 
{\cal L}_e[a] + {\cal L}_{d}[h]  
 \right) \right\} 
\nonumber\\
&=& \int {\cal D}a_\mu {\cal D}h_{\mu\nu} 
\exp \left\{ i \int d^4x  
\left( {1 \over 2} \rho K^{-1/2}  h_{\mu\nu} f^{\mu\nu}
+{\cal L}_e[a] \right)  \right\} 
\left({2K^{1/2} \over i\rho}\right)^2
\nonumber\\&& \times
{\delta \over \delta h_{\rho\sigma}(y)} 
{\delta \over \delta h_{\mu\nu}(x)}  \exp \left\{ i \int d^4x 
  {\cal L}_{d}[h]  
\right\} .
\end{eqnarray}
The functional derivatives are performed as  
\begin{eqnarray}
&& {\delta \over \delta h_{\rho\sigma}(y)} 
{\delta \over \delta h_{\mu\nu}(x)}  \exp \left\{ i \int d^4x  {\cal L}_{d}[h] \right\} 
\nonumber\\
&=& {\delta \over \delta h_{\rho\sigma}(y)} \left(
 {1 \over i}{\cal D}[\partial_x]h_{\mu\nu}(x) \exp \left\{ i \int d^4x  {\cal L}_{d}[h] \right\} \right)
\nonumber\\
&=& \left[ {1 \over i}{\cal D}[\partial_x]\delta(x-y) (\delta_{\mu\rho}\delta_{\nu\sigma}-\delta_{\nu\rho}\delta_{\mu\sigma})
+ ({1 \over i})^2 {\cal D}[\partial_x]h_{\mu\nu}(x) {\cal D}[\partial_y]h_{\rho\sigma}(y) \right]
\nonumber\\&& \times
\exp \left\{ i \int d^4x   {\cal L}_{d}[h]  \right\} ,
\end{eqnarray}
where we have defined an operator,
\begin{equation}
{\cal D}[\partial] := 1 + {\partial^2 \over \eta^2} - {\partial^4 \over \gamma^4} .
\end{equation}
Here ${\cal D}[\partial]$ denotes quantum corrections for the $h$ field.
\par
Therefore, we obtain the relationship reflecting the duality between electric and magnetic sector,
\begin{eqnarray}
 &&  \langle f_{\mu\nu}(x) f_{\rho\sigma}(y) \rangle_{APEGT} 
\nonumber\\
&=& \left({2K^{1/2} \over i\rho}\right)^2 \left( {1 \over i} \right)
{\cal D}[\partial_x]\delta(x-y) (\delta_{\mu\rho}\delta_{\nu\sigma}-\delta_{\nu\rho}\delta_{\mu\sigma})
\nonumber\\
&&+ \left({2K^{1/2} \over i\rho}\right)^2 \left({1 \over i}\right)^2
\langle  {\cal D}[\partial_x]h_{\mu\nu}(x) {\cal D}[\partial_x]h_{\rho\sigma}(y) \rangle_{APEGT} .
\label{tran}
\end{eqnarray}
Thus the VEV of the Wilson loop in the Yang-Mills theory is rewritten in terms of the correlation functions of the magnetic quantity in the APEGT as
\begin{eqnarray}
   \langle W(C) \rangle_{YM} 
&=& \int d\mu_C(\xi) \exp \Biggr[ - 2J^2g^2 \rho^{-2} K 
  \int_{S_C} dS^{\mu\nu}(x) \int_{S_C} dS^{\rho\sigma}(y) 
\nonumber\\ && \quad \quad \quad \quad \quad \quad 
 \times 
\langle {\cal D}[\partial_x]h_{\mu\nu}^\xi(x) {\cal D}[\partial_y]h_{\rho\sigma}^\xi(y) \rangle_{APEGT} + \cdots \Biggr] .
\end{eqnarray}
This is identified as the cumulant expansion of
\begin{equation}
 \langle W(C) \rangle_{YM} = e^{[\cdots]}
\int d\mu_C(\xi) \Biggr\langle \exp \left[ 2i \rho^{-1} K^{1/2} Jg
  \int_{S_C} dS^{\mu\nu}(x) {\cal D}[\partial_x] h_{\mu\nu}^\xi(x)  \right] \Biggr\rangle_{APEGT}  ,
\end{equation}
where $[\cdots]$ denotes the field-independent-constant part (a phase factor) in the first term in the RHS of (\ref{tran}) in which we have no interest.
It should be remarked that the above result is totally independent from the explicit form of  ${\cal L}_e[a]$, even if we include higher order corrections of low-energy expansions.
\par
The above result implies that we can calculate the Wilson loop by making use of the magnetic theory which is written in terms of tensor field $h_{\mu\nu}$ alone.
By performing the integration over $a_\mu$ field, we can obtain such a dual magnetic theory.   
\par
By including the gauge fixing term of $a_\mu^i$ into the action $S_{APEGT}[a,h]$,
\begin{equation}
  {\cal L}_{GF}[a] := - {1 \over 2\beta} (\partial^\mu a_\mu^i)^2 ,
\end{equation}
the action $S_{APEGT}[a,h]$ reads 
\begin{eqnarray}
S_{APEGT}[a,h] &=& \int d^4x \left\{
 {\cal L}_{e}^{tot}[a] 
-   \rho K^{-1/2}  a^\nu{}^i \partial^\mu h_{\mu\nu}^i    
+ {\cal L}_d[h] \right\} ,
\end{eqnarray}
where%
\footnote{
It should be understood that ${\cal L}_{e}^{tot}[a]$ is obtained after integrating out $B^i, C^i, \bar C^i$.
}
\begin{eqnarray}
 {\cal L}_{e}^{tot}[a]  &:=& {\cal L}_{e}[a] + {\cal L}_{GF}[a]
\nonumber\\ 
&=&  {1 \over 2} a_\mu^i \left[ 
(1-\rho^2) g^{\mu\nu} \partial^2
- (1-\rho^2 - \beta^{-1})\partial^\mu \partial^\nu   \right] a_\nu^i 
+ O\left( {f^4 \over M_A^4} \right) .
\end{eqnarray}
For $\rho \not=1$, integrating out the $a_\mu$ field yields
\begin{eqnarray}
 {\cal L}_{d}[h]' &=& {\cal L}_d[h]
+ {\rho^2K^{-1} \over 2(1-\rho^2)}  \partial^\tau h_{\tau\mu}^i
{1 \over \partial^4} \left\{ 
 g^{\mu\nu} \partial^2
- [1-(1-\rho^2)\beta] \partial^\mu \partial^\nu   \right\} 
\partial^\lambda h_{\lambda\nu}^i 
\nonumber\\
&=& {\cal L}_d[h]
+ {\rho^2K^{-1} \over 2(1-\rho^2)}  \partial^\tau h_{\tau\mu}^i
{1 \over \partial^2}  
 g^{\mu\nu}    \partial^\lambda h_{\lambda\nu}^i .
\label{dd}
\end{eqnarray}
Thus the VEV of the Wilson operator is calculated from
\begin{eqnarray}
 \langle W(C) \rangle_{YM} &=&  
\int d\mu_C(\xi) \Biggr\langle \exp \left[ 2i \rho^{-1} K^{1/2} Jg
  \int_{S_C} dS^{\mu\nu}(x) {\cal D}[\partial_x] h_{\mu\nu}^\xi(x)  \right] \Biggr\rangle_{d} 
\\
&=& Z_{d}^{-1} \int {\cal D}h_{\mu\nu}^i  
  \exp \left\{i \int d^4x {\cal L}_d[h]' \right\} {\cal J}[h]  ,
\end{eqnarray}
where
\begin{equation}
 {\cal J}[h] := \exp \left[ 2i \rho^{-1} K^{1/2} Jg
  \int_{S_C} dS^{\mu\nu}(x) {\cal D}[\partial_x] h_{\mu\nu}^\xi(x)  \right].
\end{equation}
When $\rho=1$, ${\cal L}_{e}[a]$ contains only the higher-order terms.
In the limit $\rho \rightarrow 1$, the second term in (\ref{dd}) yields the constraint 
$\delta(\partial^\lambda h_{\lambda\nu}^i)$ in the measure,
\begin{eqnarray}
 \langle W(C) \rangle_{YM} 
&=& Z_{d}^{-1} \int {\cal D}h_{\mu\nu}^i \delta(\partial^\lambda h_{\lambda\nu}^i)
  \exp \left\{i \int d^4x {\cal L}_d[h] \right\} {\cal J}[h] .
\end{eqnarray}
\par
If we integrate out tensor field $h_{\mu\nu}$ in $S_{APEGT}[a,h]$, we will obtain the electric theory which is written in terms of the diagonal gluon field $a_\mu$ alone.  This theory can lead to the area law too, as will be discussed in a subsequent paper \cite{preparation}.

\subsection{Step 8: Recovery of hypergauge symmetry and gauge fixing}
\par
The action $S_{APEGT}[a,h]$ has the U(1) gauge invariance for the gauge transformation of the diagonal gluon field: $a_\mu \rightarrow a_\mu + \partial_\mu \theta$, i.e.,
$S_{APEGT}[a+d\theta,h]=S_{APEGT}[a,h]$.
In this section we consider the symmetry for the tensor field $h$.
\par
Now we introduce the field strength $H$ of an antisymmetric tensor field $h$ (the so-called Kalb-Ramond field), $H:=dh$, i.e, 
\begin{equation}
 H_{\mu\nu\lambda} := \partial_\lambda h_{\mu\nu} + \partial_\mu h_{\nu\lambda}
+ \partial_\nu h_{\lambda\mu}
\end{equation}
Note that the field strength $H$ is invariant, 
$H_{\mu\nu\lambda}  \rightarrow H_{\mu\nu\lambda} $,
under the hypergauge transformation, $h \rightarrow h+d\zeta$, i.e.,
\begin{equation}
 h_{\mu\nu} \rightarrow h_{\mu\nu}^\zeta := h_{\mu\nu} + \partial_\mu \zeta_\nu  - \partial_\nu \zeta_\mu  .
\label{gtr1}
\end{equation}
We require the invariance of the measure ${\cal D}h_{\mu\nu}$ under the hypergauge transformation.  
However,  the Lagrangian ${\cal L}_{APEGT}[a,h]$ or ${\cal L}_d[h]'$ does not have the invariance under the hypergauge transformation due to the existence of the mass term $(h_{\mu\nu})^2$.  
Nevertheless, we can recover the invariance by introducing new degrees of freedom%
\footnote{This field plays the similar role to the St\"uckelberg field in the massive vector theory which recovers the gauge invariance of the vector field.}
 $\Lambda_\mu$ where
 $\Lambda_\mu$ transforms as 
\begin{equation}
 \Lambda_\mu \rightarrow \Lambda_\mu^\zeta := \Lambda_\mu - \zeta_\mu .
\label{gtr2}
\end{equation}
In fact, the combination $h^\Lambda:=h+d\Lambda$, i.e., 
\begin{equation}
 h_{\mu\nu}^\Lambda = h_{\mu\nu} + \partial_\mu \Lambda_\nu  - \partial_\nu \Lambda_\mu ,
\end{equation}
is invariant under the combined transformations, (\ref{gtr1}) and (\ref{gtr2}).
Therefore, the Lagrangian,
\begin{equation}
{\cal L}_m[h,\Lambda] :=  {\cal L}_m[h^\Lambda] = -{1 \over 4} (h_{\mu\nu}^i+ \partial_\mu \Lambda_\nu^i  - \partial_\nu \Lambda_\mu^i)^2
  + {1 \over 12\eta^2} (H_{\mu\nu\lambda}^i)^2 
+ {1 \over 4\gamma^4} (\partial^\lambda H_{\lambda\mu\nu}^i)^2  ,
\label{magtheory}
\end{equation}
is also invariant, i.e.,
${\cal L}_m[h,\Lambda]={\cal L}_m[h^\zeta, \Lambda^\zeta]$.

 As we have recovered the hypergauge invariance, we need to fix the hypergauge invariance in quantizing the dual magnetic theory ${\cal L}_m[h,\Lambda]$.  
From this viewpoint, we adopt the condition,%
\footnote{In section \ref{subsec:comparison} we discuss the relationship between this condition and the setting up of the previous paper \cite{KondoI}.} 
\begin{equation}
\partial^\nu h_{\mu\nu}^i=0  ,
\end{equation}
 as a gauge fixing condition for the antisymmetric tensor field.%
\footnote{In the manifestly covariant quantization of the gauge theory, we need to introduce the ghost as is well known.  However, it is not enough for the antisymmetric tensor gauge theory, since we need to introduce the ghost for ghost in order to completely fix the gauge degrees of freedom.  Such a theory is called a reducible theory. In this subsection we treat the theory in a naive manner.  However, the result is unchanged if we take into account the reducibility of the theory.  See Appendix~\ref{sec:tensorfield}.} 
Under this  condition,
 the derivative terms of $h_{\mu\nu}$ in ${\cal L}_m[h^\Lambda]$ recover the corresponding terms of ${\cal L}_d[h]$,
\begin{eqnarray}
(H_{\mu\nu\lambda}^i )^2 = - 3 h_{\mu\nu}^i \partial^2 h_{\mu\nu}^i
-6 \partial_\mu h_{\mu\nu}^i \partial_\lambda h^{\nu\lambda}{}^i
\rightarrow 
- 3 h_{\mu\nu}^i \partial_\lambda \partial^\lambda h^{\mu\nu}{}^i ,
\\
(\partial^\lambda H_{\lambda\mu\nu}^i )^2  
= (\partial^\lambda \partial_\lambda h_{\mu\nu} + \partial_\mu \partial^\lambda h_{\nu\lambda}+\partial_\nu \partial^\lambda h_{\lambda\mu})^2
\rightarrow 
  h_{\mu\nu}^i (\partial_\lambda \partial^\lambda)^2 h^{\mu\nu}{}^i .
\end{eqnarray}
Here it turns out that all the terms proportional to $(1-P)$ in $\delta_{(1)}^c{\cal L}_{APEGT}$ vanish under this condition.
Therefore, ${\cal L}_m[h,\Lambda]$ under the condition reduces to 
\begin{equation}
{\cal L}_m[h,\Lambda] = -{1 \over 4} (h_{\mu\nu}^i+ \partial_\mu \Lambda_\nu^i  - \partial_\nu \Lambda_\mu^i)^2
  - {1 \over 4\eta^2} h_{\mu\nu}^i \partial_\lambda \partial^\lambda h^{\mu\nu}{}^i 
+ {1 \over 4\gamma^4} h_{\mu\nu}^i (\partial_\lambda \partial^\lambda)^2 h^{\mu\nu}{}^i  .
\end{equation}
It is shown that the theory with ${\cal L}_m[h,\Lambda]$ reduces to the original theory given by ${\cal L}_d[h]$ by integrating out $\Lambda$ field after fixing the gauge freedom of $\Lambda_\mu$, see Appendix~\ref{sec:tensorfield}. 
\par
Thus we obtain an alternative dual description of low-energy Gluodynamics in terms of $h_{\mu\nu}$ and $\Lambda_\mu$. Especially, for $G=SU(2)$, we obtain
\begin{eqnarray}
    \langle W(C) \rangle_{YM} 
= Z_{M}^{-1}   \int  {\cal D}h_{\mu\nu}  
  \delta(\partial^\nu h_{\mu\nu}) \int  {\cal D}\Lambda_\mu
\exp \left\{i S_{M}[h^\Lambda;C]  \right\}  ,
\end{eqnarray}
where
\begin{equation}
S_{M}[h^\Lambda;C] =  \int d^4x {\cal L}_m[h^\Lambda]
+   2 \rho^{-1} K^{1/2}  Jg \int_{S_C} dS^{\mu\nu}(x) {\cal D}[\partial_x] h_{\mu\nu}(x) .
\label{Mth}
\end{equation}
Apart from the third term in ${\cal L}_m[h]$, the above action (\ref{magtheory}) coincides with the action of confining string proposed by Polyakov \cite{Polyakov96} in the weak field limit, see section~\ref{sec:string}.

\subsection{Step 9: Change of variables (path-integral duality transformation)}

 We proceed to rewrite the LEET (\ref{Mth}) into another form which is useful to derive the dual Ginzburg-Landau theory.
\par
First of all, we rewrite the surface integral in $S_{APEGT}[h^\Lambda;C]$ into the volume integral as follows.
Let $\sigma=(\sigma^1,\sigma^2)$ be two-dimensional coordinate on the surface which is bounded by the Wilson loop $C$.  Then
\begin{eqnarray}
 \int_{S_C} dS^{\mu\nu}(x(\sigma)) \tilde h_{\mu\nu}(x(\sigma)) 
&=& {1 \over 2} \int d^2 \sigma \epsilon^{ab}{\partial x^\mu \over \partial \sigma^a}{\partial x^\nu \over \partial \sigma^b} \tilde h_{\mu\nu}(x(\sigma))
\nonumber\\
&=& {1 \over 2} \int d^2 \sigma X^{\mu\nu}(\sigma) \tilde h_{\mu\nu}(x(\sigma))  
\nonumber\\
&=& {1 \over 2} \int d^2 \sigma X^{\mu\nu}(\sigma)  \int d^4x \tilde h_{\mu\nu}(x) \delta^4(x-x(\sigma))
\nonumber\\
&=& \int d^4x \tilde h_{\mu\nu}(x) \Theta_{\mu\nu}(x)  
\\
&=& \int d^4x  h_{\mu\nu}(x) \tilde \Theta_{\mu\nu}(x) ,
\end{eqnarray}
where we have introduced the Jacobian,
\begin{eqnarray}
X^{\mu\nu}(\sigma) := \epsilon^{ab}{\partial x^\mu \over \partial \sigma^a}{\partial x^\nu \over \partial \sigma^b} 
= {\partial(x^\mu,x^\nu) \over \partial(\sigma^1,\sigma^2)} ,
\end{eqnarray}
and an antisymmetric tensor of rank two,
\begin{eqnarray}
 \Theta_{\mu\nu}(x) := {1 \over 2} \int d^2 \sigma X^{\mu\nu}(\sigma) \delta^4(x-x(\sigma)) = - \Theta_{\nu\mu}(x) .
\end{eqnarray}
We call $\Theta_{\mu\nu}(x)$ the vorticity tensor current which has the support on the surface spanned by the Wilson loop $C$.
Note that
\begin{equation}
  \Theta^{\mu\nu}(x) = {1 \over 2} \int_S d^2 S^{\mu\nu}(x(\sigma)) \delta^4(x-x(\sigma)) ,
\quad 
 d^2 S^{\mu\nu} := d\sigma^1 d\sigma^2 {\partial(x^\mu,x^\nu) \over \partial(\sigma^1,\sigma^2)} .
\end{equation}
Hence we can start from the action,
\begin{equation}
S_{M}[h^\Lambda;\Theta] =  \int d^4x \left\{ {\cal L}_m[h^\Lambda] 
+  2Jg \rho^{-1}K^{1/2}  h_{\mu\nu}(x) \tilde \Theta_{\mu\nu}(x) \right\} ,
\label{APEGTh}
\end{equation}
and the expectation value of the Wilson loop is given by
\begin{eqnarray}
    \langle W(C) \rangle_{YM} 
&=& Z_{M}[h^\Lambda;\Theta]/Z_{M}[h^\Lambda;0],
\label{vevh}
\end{eqnarray}
where
\begin{eqnarray}
Z_{M}[h^\Lambda;\Theta] &:=& 
\int  {\cal D}h_{\mu\nu}   \delta(\partial^\nu h_{\mu\nu}) \int {\cal D}\Lambda_\mu 
\exp \left\{i S_{M}[h^\Lambda;\Theta]  \right\}  .
\nonumber\\
&=& 
 \int  {\cal D}h_{\mu\nu}  \delta(\partial^\nu h_{\mu\nu})
\exp \left\{ i \int d^4x \left[ {1 \over 12 \eta^2} (H_{\lambda\mu\nu})^2
+ {1 \over 4\gamma^4} (\partial^\lambda H_{\lambda\mu\nu})^2 \right] \right\}
\nonumber\\
&& \times \exp \left\{ i \int d^4x 2Jg \rho^{-1}K^{1/2}  h_{\mu\nu}(x) \tilde \Theta_{\mu\nu}(x) \right\}
\nonumber\\
&& \times \int {\cal D}\zeta_\mu 
\exp \left\{ i  \int d^4x  {-1 \over 4} (h_{\mu\nu}^\zeta
)^2  \right\} .
\end{eqnarray}
Note that the path integral transformation holds,
\begin{eqnarray}
&&  \int {\cal D}\zeta_\mu 
\exp \left\{ i  \int d^4x  {-1 \over 4} (h_{\mu\nu}^\zeta
)^2  \right\}
\nonumber\\
&=&
  \int {\cal D}\ell_{\mu\nu} \delta(\epsilon^{\mu\nu\rho\sigma}\partial_\rho
(\ell_{\mu\nu}-h_{\mu\nu} 
))
\exp \left\{ i  \int d^4x  {-1 \over 4} (\ell_{\mu\nu})^2  \right\} ,
\label{ell}
\end{eqnarray}
where the constraint, 
\begin{equation}
 \epsilon^{\mu\nu\rho\sigma}\partial_\rho
(\ell_{\mu\nu}-h_{\mu\nu} ) = 0 ,
\end{equation}
is solved by 
\begin{equation}
  \ell_{\mu\nu}-h_{\mu\nu} 
= \partial_\mu \zeta_\nu- \partial_\nu \zeta_\mu ,
\quad i.e, \quad
  \ell_{\mu\nu} = h_{\mu\nu}^{\zeta} .
\end{equation}
\par
Moreover, we introduce the auxiliary (Abelian) vector field $b_\mu$ by
\begin{eqnarray}
  \delta(\epsilon^{\mu\nu\rho\sigma}\partial_\rho
(\ell_{\mu\nu}-h_{\mu\nu} 
))
= \int {\cal D}b_\mu \exp \left\{-  i \int d^4x    *b_{\mu\nu}
 (\ell_{\mu\nu}-h_{\mu\nu} 
)  \right\} ,
\label{iden}
\end{eqnarray}
where $b_{\mu\nu}$ is the (dual) field strength defined by
\begin{equation}
 b_{\mu\nu}:=\partial_\mu b_\nu - \partial_\nu b_\mu .
\end{equation}
By using the identity (\ref{iden}),  the integration over $\ell_{\mu\nu}$ in (\ref{ell}) can be performed as
\begin{eqnarray}
 &&  (\ref{ell})
\nonumber\\
&=& \int {\cal D}b_\mu \exp \left\{ i \int d^4x   *b_{\mu\nu}
 (h_{\mu\nu} 
)  \right\}
\int {\cal D}\ell_{\mu\nu} \exp \left\{ i  \int d^4x  \left[ 
{-1 \over 4} (\ell_{\mu\nu})^2 - *b_{\mu\nu} \ell_{\mu\nu} 
\right] \right\}
\nonumber\\
&=& \int {\cal D}b_\mu \exp \left\{ i \int d^4x  \left[ 
-  (b_{\mu\nu})^2 + *b_{\mu\nu}
 (h_{\mu\nu} 
) \right] \right\} .
\end{eqnarray}
\par
Another way of deriving the equality just derived, i.e,
\begin{eqnarray}
 \int {\cal D}\zeta_\mu 
\exp \left\{ i  \int d^4x  {-1 \over 4} (h_{\mu\nu}^\zeta)^2  \right\}
=  \int {\cal D}b_\mu \exp \left\{ i \int d^4x  \left[ 
-  (b_{\mu\nu})^2 + *b_{\mu\nu}
 (h_{\mu\nu} 
) \right] \right\}
\end{eqnarray}
is as follows.  
The argument of the exponential in the LHS is
\begin{eqnarray}
 (h+d\zeta, h+d\zeta) = (h,h) + (h,d\zeta) + (d\zeta, h) + (d\zeta,d\zeta)
\sim (h,h) + (d\zeta,d\zeta),
\end{eqnarray}
under the condition $\delta h=0$.
The last term decouples after the Gaussian integration of $\zeta$.
On the other hand, the argument of the the exponential in the RHS is cast into
\begin{eqnarray}
 \int d^4x  \left[  (b_{\mu\nu})^2 - *b_{\mu\nu} (h_{\mu\nu} 
) \right] 
= (db,db) - (*db,h) 
= (b,\delta db) - (b,*dh) .
\end{eqnarray}
Suppose the Lorentz type gauge condition $\delta b=0$. We introduce the NL  (zero-form) field $\phi$. Then the Gaussian integration over $b_\mu$ field yields
\begin{eqnarray}
 && (b,\Delta b)-(b,*dh) - (\delta b,\phi)
= (b,\Delta b)-(b,*dh+d\phi)  
\nonumber\\
&&\rightarrow 
(*dh+d\phi,{1 \over \Delta} *dh+d\phi)
= (h, {\delta d \over \Delta}h) + (\phi, {\delta d \over \Delta}\phi)
\sim (h,h) + (\phi,\phi),
\end{eqnarray}
under the condition $\delta h=0$.  In this derivation, we must insert the constraint $\delta(\partial^\mu b_\mu)$ in the measure ${\cal D}b_\mu$.
The identity implies that there are many ways of extracting the transverse modes of $h$.
\par
Thus, the theory is rewritten in terms of $b_\mu$ and $h_{\mu\nu}$ as 
\begin{eqnarray}
 && Z_{M}[b,h;\Theta] 
\nonumber\\
&=&  \int {\cal D}b_\mu
\exp \left\{ i \int d^4x  \left[ -  (b_{\mu\nu})^2  \right] \right\}
\int  {\cal D}h_{\mu\nu}  \delta(\partial^\nu h_{\mu\nu})
\nonumber\\&& \times
\exp \left\{ i \int d^4x \left[ 
  *b_{\mu\nu} h_{\mu\nu}  
+ {1 \over 12 \eta^2} (H_{\lambda\mu\nu})^2 
+ {1 \over 4\gamma^4} (\partial^\lambda H_{\lambda\mu\nu} )^2
\right] \right\}  
\nonumber\\
&& \times \exp \left\{ i \int d^4x 2Jg \rho^{-1}K^{1/2}  h_{\mu\nu}(x) \tilde \Theta_{\mu\nu}(x) \right\} .
\end{eqnarray}
By change of variable 
$b_{\mu\nu} \rightarrow b_{\mu\nu}-2\rho^{-1}K^{1/2} Jg * \tilde \Theta_{\mu\nu}$, we arrive at the expression,
\begin{eqnarray}
 && Z_{M}[b,h;\Theta] 
\nonumber\\
&=&  \int {\cal D}b_\mu
\exp \left\{ i \int d^4x  \left[ -  (b_{\mu\nu}-2\rho^{-1}K^{1/2} Jg * \tilde \Theta_{\mu\nu})^2  \right] \right\}
\nonumber\\&& \times
\int  {\cal D}h_{\mu\nu}  \delta(\partial^\nu h_{\mu\nu})
\exp \left\{ i \int d^4x \left[ 
  *b_{\mu\nu} h_{\mu\nu}  
+ {1 \over 12 \eta^2} (H_{\lambda\mu\nu})^2 
+ {1 \over 4\gamma^4} (\partial^\lambda H_{\lambda\mu\nu} )^2
\right] \right\} .
\nonumber\\
\label{Last}
\end{eqnarray}

\par

\subsection{Step 10: Dual Ginzburg-Landau theory in the London limit}
\par
First of all, we examine a special case $\gamma=\infty$ for simplicity, although our derivation suggest $\gamma<\infty$.  So the last term in (\ref{Last}) is neglected.  The case of a finite $\gamma$ is treated in the next section.
We change the variable $h_{\mu\nu}$ into the new variable $V_\mu$ as follows.\footnote{
From $\delta h^{(2)}=0$, there exists a three-form $Y^{(3)}$ such that 
$h^{(2)}=\delta Y^{(3)}=\delta *W^{(1)}=*dW^{(1)}$.  Then 
$H^{(3)}:=dh^{(2)}=d*dW^{(1)}=*\delta dW^{(1)}=*V^{(1)}$, or
$V^{(1)}=*H^{(3)}$.  Therefore, 
$\delta V^{(1)}=\delta *H^{(3)}=*dH^{(3)}=*ddh^{(2)}=0.$
}
\begin{eqnarray}
 && \int  {\cal D}h_{\mu\nu}  \delta(\partial^\nu h_{\mu\nu})
\exp \left\{ i \int d^4x \left[ {1 \over 12 \eta^2} (H_{\lambda\mu\nu})^2 
+ *b_{\mu\nu} h_{\mu\nu}  \right] \right\}
\nonumber\\
&=& \int  {\cal D}h_{\mu\nu}  \delta(\partial^\nu h_{\mu\nu})
\exp \left\{ i \int d^4x \left[ {1 \over 12 \eta^2} (H_{\lambda\mu\nu})^2 
+  \epsilon^{\mu\nu\rho\sigma} b_{\mu} \partial_\nu h_{\rho\sigma}  \right] \right\}
\\
&=& \int {\cal D}V_\mu \delta(\partial_\mu V^\mu) \exp \left\{ i \int d^4x \left[
{-1 \over 2\eta^2}V_\mu^2 +  2b_\mu V^\mu  \right] \right\} ,
\label{int}
\end{eqnarray}
since the constraint $\partial_\mu V^\mu=0$ can be solved by an antisymmetric tensor field in the form,
\begin{equation}
 V^\mu := {1 \over 2} \epsilon^{\mu\nu\rho\sigma} \partial_\nu h_{\rho\sigma} 
= {1 \over 6} \epsilon^{\mu\nu\rho\sigma} H_{\nu\rho\sigma} 
= \partial_\nu *h^{\mu\nu} .
\label{constr2}
\end{equation}
The massive antisymmetric tensor field $h_{\mu\nu}$ denotes the massive spin-1 field $V_\mu$ whose canonical mass dimensions is three.\footnote{The massless antisymmetric tensor field stands for the massless spin-0 field, see \cite{Townsend79,Kimura80,HKO81}.}
In this step, the number of independent degrees of freedom is conserved, since $V_\mu$ and $h_{\mu\nu}$ have three independent components.
\par
Furthermore, the integration over $V_\mu$ is performed after introducing the new variable $\theta$ to remove the delta function of the constraint $\partial_\mu V^\mu=0$,
\footnote{
In view of the definition (\ref{constr2}), the $V_\mu$ plays the similar role to the magnetic monopole current $k_\mu$ defined in the next section.
 Here the variable $\theta$ is introduced to keep the constraint 
$\partial^\mu k_\mu=0$.  In other words, the introduction of $\theta$ keeps the magnetic $U(1)_m$ symmetry.  So, putting $\theta=0$ breaks the $U(1)_m$ symmetry.
}
\begin{eqnarray}
 (\ref{int})
&=& \int {\cal D}V_\mu \int {\cal D}\theta  \exp \left\{  \int d^4x \left[
{-i \over 2\eta^2}V_\mu^2 + 2iV^\mu b_\mu + i \theta  \partial_\mu V^\mu 
  \right] \right\}
\\
&=& \int {\cal D}V_\mu \int {\cal D}\theta  \exp \left\{ i \int d^4x \left[
{-1 \over 2\eta^2}V_\mu^2 +  V^\mu (2b_\mu - \partial_\mu \theta)  \right] \right\}
\\
&=&  \int {\cal D}\theta \exp \left\{ i \int d^4x \left[
{1 \over 2}\eta^2 (2b_\mu - \partial_\mu \theta)^2  \right] \right\} .
\end{eqnarray}

\par
Finally, we obtain the dual Abelian gauge theory,
\begin{eqnarray}
 && Z_{M}[b,\theta;\Theta] 
\nonumber\\
&=&    \int {\cal D}b_\mu \int {\cal D}\theta 
\exp \left\{ i \int d^4x  \left[ - {1 \over 4} (b_{\mu\nu}+ b_{\mu\nu}^S)^2  
+ {1 \over 2}\eta^2 (b_\mu - \partial_\mu \theta)^2  \right] \right\} ,
\label{dAGT}
\end{eqnarray}
where
\begin{eqnarray}
  b_{\mu\nu}^S(x) &:=& 4\rho^{-1} K^{1/2} Jg   * \tilde \Theta_{\mu\nu}(x), 
\\
 \partial^\nu *b_{\mu\nu}^S(x) = \rho^{-1} K^{1/2} {\cal D}[\partial] J_\mu^S , 
\quad J_\mu^S &:=& 4J g \int_0^1 d\tau {dx_\mu(\tau) \over d\tau} \delta^4(x-x(\tau)) ,
\end{eqnarray}
and $g$ is the Yang-Mills coupling constant of the original Yang-Mills theory.
\par
This model has dual U(1) symmetry, say $U(1)_m$ symmetry, 
\begin{equation}
 b_\mu \rightarrow b_\mu + \partial_\mu \vartheta, \quad
\theta \rightarrow \theta + \vartheta .
\end{equation}
\par
This model is identified with the London limit $\lambda \rightarrow \infty$ of the dual Abelian Higgs model or the dual Ginzburg-Landau theory with the Lagrangian,
\begin{equation}
 {\cal L}_{DGL}[b,\phi] = {-1 \over 4}(b_{\mu\nu}+b_{\mu\nu}^S)^2 
+ |(\partial_\mu-ig_m b_\mu) \phi|^2
-  \lambda (|\phi|^2-v^2)^2 
\end{equation}
where $g_m$ is the magnetic charge subject to the Dirac quantization condition,
\begin{equation}
  g_m g = 4\pi .
\end{equation}
The London limit is equivalent to putting 
$|\phi(x)|=v=const.$, i.e, $\phi(x)=v \exp[i\theta(x)]$. 
The dual U(1) symmetry is broken in the London limit,
\begin{equation}
\mbox{\fboxsep=.1in \framebox{$\displaystyle
 {\cal L}_{DGL}[b] = {-1 \over 4}(b_{\mu\nu}+b_{\mu\nu}^S)^2 
+ {1 \over 2}m_b^2 b_\mu b^\mu . 
$}}
\end{equation}
This corresponds to the infinitesimaly thin flux tube connecting the quark and anti-quark.  In the London limit, the Higgs mass 
$m_\phi=2\sqrt{\lambda}v$ diverges, i.e., $m_\phi=\infty$ or $m_\phi^{-1}=0$.  This is the extreme case of the type II superconductor where $m_\phi>m_b$.  
It turns out that the mass $m_b$ of the dual gauge field $b_\mu$ is given by $\eta$,
\begin{equation}
\mbox{\fboxsep=.1in \framebox{$\displaystyle
 \eta = m_b = \sqrt{2} g_m v \equiv {\sqrt{2}4\pi \over g}v .
$}}
\end{equation}
The monopole condensation is shown to occur in section~\ref{sec:monopole}.

\section{\label{sec:finalstep}Final step: Dual Ginzburg-Landau theory of the general type}
\setcounter{equation}{0}

In this section, we discuss how the LEET given by (\ref{Last}) is related to the dual Ginzburg-Landau theory of type II.  
In section~\ref{sec:monopole}, we give another evidence of equivalence between the LEET (\ref{Last}) and the dual Ginzburg-Landau theory on the border between type II and type I.

\subsection{Dual gauge theory}
We return to eq.(\ref{Last}):
\begin{eqnarray}
 && Z_{M}[b,h;\Theta] 
\nonumber\\
&=& \int {\cal D}b_\mu
\exp \left\{ i \int d^4x  \left[ -  (b_{\mu\nu}-2\rho^{-1}K^{1/2} Jg  *\Theta_{\mu\nu})^2  \right] \right\}
\nonumber\\&& \times
\int  {\cal D}h_{\mu\nu}  \delta(\partial^\nu h_{\mu\nu})
\exp \left\{ i \int d^4x \left[ 
  *b_{\mu\nu} h_{\mu\nu}  
+ {1 \over 12 \eta^2} (H_{\lambda\mu\nu})^2 
+ {1 \over 4 \gamma^4} (\partial^\lambda H_{\lambda\mu\nu} )^2
\right] \right\} .
\nonumber
\end{eqnarray}
By making use of the change of variable (\ref{constr2}), we change the variable $h_{\mu\nu}$ into the new variable $V_\mu$,
\begin{eqnarray}
 && \int  {\cal D}h_{\mu\nu}  \delta(\partial^\nu h_{\mu\nu})
\exp \left\{ i \int d^4x \left[ 
  *b_{\mu\nu} h_{\mu\nu}  
+ {1 \over 12 \eta^2} (H_{\lambda\mu\nu})^2 
+ {1 \over 4 \gamma^4} (\partial^\lambda H_{\lambda\mu\nu} )^2
\right] \right\}
\nonumber\\
&=& \int  {\cal D}h_{\mu\nu}  \delta(\partial^\nu h_{\mu\nu})
\exp \left\{ i \int d^4x \left[ 
 \epsilon^{\mu\nu\rho\sigma} b_{\mu} \partial_\nu h_{\rho\sigma}
+ {1 \over 12 \eta^2} (H_{\lambda\mu\nu})^2 
+ {1 \over 4 \gamma^4} (\partial^\lambda H_{\lambda\mu\nu} )^2
  \right] \right\}
\nonumber\\
&=& \int {\cal D}V_\mu \delta(\partial_\mu V^\mu) \exp \left\{ i \int d^4x \left[
2b_\mu V^\mu - {1 \over 2\eta^2}V_\mu^2 
- {1 \over 4 \gamma^4}(\partial_\mu V_\nu - \partial_\nu V_\mu)^2    \right] \right\} .
\label{inter}
\end{eqnarray}
After introducing the Lagrange multiplier field $\theta$ for the constraint $\partial_\mu V^\mu=0$, the $V_\mu$ integration is performed as
\begin{eqnarray}
 && (\ref{inter})
\nonumber\\
&=& \int {\cal D}V_\mu \int {\cal D}\theta \exp \left\{ i \int d^4x \left[
(2b_\mu -\partial_\mu \theta) V^\mu - {1 \over 2\eta^2}V_\mu^2 
- {1 \over 4 \gamma^4}(\partial_\mu V_\nu - \partial_\nu V_\mu)^2    \right] \right\}
\nonumber\\
&=& \int {\cal D}\theta \exp \left\{ i \int d^4x \left[ 
{1 \over 2}(2b_\mu -\partial_\mu \theta)
{\gamma^4  \over \Delta -\gamma^4/\eta^2} \left( g^{\mu\nu} - {\eta^2 \over \gamma^4} \partial^\mu \partial^\nu \right) 
(2b_\nu -\partial_\nu \theta)
\right] \right\} .
\label{intt}
\nonumber\\
\end{eqnarray}
Then we obtain
\begin{eqnarray}
 && Z_{M}[b,\theta;\Theta] 
\nonumber\\
&=& \int {\cal D}b_\mu \delta(\partial^\mu b_\mu) \int  {\cal D}\theta
\exp \Biggr\{ i \int d^4x  \Biggr[ 
- {1 \over 4} (b_{\mu\nu}+b_{\mu\nu}^S)^2 
\nonumber\\
&& + {1 \over 2}
(b_\mu -\partial_\mu \theta)
{\gamma^4  \over \Delta -\gamma^4/\eta^2} \left( g^{\mu\nu} - {\eta^2 \over \gamma^4} \partial^\mu \partial^\nu \right) 
(b_\nu -\partial_\nu \theta)
\Biggr] \Biggr\} ,
\label{dGL2}
\end{eqnarray}
where we have inserted the delta function $\delta(\partial^\mu b_\mu)$ for fixing the gauge for $b_\mu$.  Note that (\ref{dGL2}) reproduces the London limit when $\gamma \rightarrow \infty$.  In the case of finite $\gamma$, 
\begin{eqnarray}
 && {1 \over 2}
 (b_\mu - \partial_\mu \theta) \eta^2
\left( 1- {\eta^2 \over \gamma^4}\Delta \right)^{-1} \left( g^{\mu\nu} - {\eta^2 \over \gamma^4} \partial^\mu \partial^\nu \right) 
 (b_\nu - \partial_\nu \theta)
\\
&\sim& {1 \over 2}
  b_\mu  \eta^2
\left( 1+ {\eta^2 \over \gamma^4}\Delta  + \cdots \right)
 b^\mu  
 - {1 \over 2} \eta^2 \theta \Delta \theta + O(\Delta^2)  
\\
&=& {1 \over 2} \eta^2 b_\mu b^\mu
 + {1 \over 2} {\eta^4 \over \gamma^4} b_\mu \Delta b^\mu  
 - {1 \over 2} \eta^2 \theta \Delta \theta + O(\Delta^2) ,
\label{dGL3}
\end{eqnarray}
where we have used $\partial^\mu b_\mu=0$. 
Thus we arrive at a LEET of the Yang-Mills theory,
\begin{eqnarray}
  Z_{APEGT}[b,\theta;0] 
&=& \int {\cal D}b_\mu \delta(\partial^\mu b_\mu)  
\exp \Biggr\{ i \int d^4x {\cal L}_{K}[b,\theta]  \Biggr\} ,
\\
 {\cal L}_{K}[b,\theta] &:=& - {1 \over 4} \left( 1+ {\eta^4 \over \gamma^4} \right) (b_{\mu\nu})^2 
+ {1 \over 2}\eta^2 b_\mu b^\mu 
- {1 \over 2} \eta^2 \theta \Delta \theta .
\label{dGL4}
\end{eqnarray}
The LEET just obtained is of the same form as (\ref{dAGT}), except for the renormalization of the kinetic term of the dual gauge field.
The dual gauge field becomes massive, whereas the $\theta$ field remains massless.  This is reasonable, since the field $\theta$ corresponds to the Nambu-Goldstone (NG) mode associated with the spontaneous breakdown of the magnetic U(1) symmetry.

\subsection{Low-energy effective theory of dual Abelian Higgs model}

In the following we discuss how this theory (\ref{dGL4}) is related to the dual Ginzburg-Landau theory of type II.
We remember that the London limit $m_H \rightarrow \infty$ corresponds to the limit $\gamma=\infty$.
Therefore, we expect that the LEET with the Lagrangian (\ref{dGL4}) can be reproduced from the dual Abelian Higgs model, in the region $0 = m_\theta \ll m_b \ll m_H$.
\par
We show that the dual Abelian Higgs model with the Lagrangian,
\begin{equation}
  {\cal L}_{DGL}[b,\phi] = {-1 \over 4}(b_{\mu\nu}+b_{\mu\nu}^S)^2 
+ |(\partial_\mu-ig_m b_\mu) \phi|^2
-  \lambda (|\phi|^2-v^2)^2 ,
\end{equation}
reduces to the LEET with the Lagrangian (\ref{dGL4}) in the low-energy region.
We adopt the renormalizable gauge \cite{tHooft71,FLS72},
\begin{equation}
  \partial^\mu b_\mu + \xi m_b \varphi_2 = 0 ,
\end{equation}
where the scalar field $\phi$ is parameterized as
\begin{equation}
  \phi(x) = {1 \over \sqrt{2}}[ v + \varphi_1(x) + i \varphi_2(x)] .
\end{equation}
The GF+FP term of the renormalizable gauge is given by
\begin{equation}
 {\cal L}_{GF+FP} = -{1 \over 2\xi}(\partial^\mu b_\mu + \xi m_b \varphi_2)^2
+ i \bar c (\partial^2 + \xi m_b^2) c 
+ i g_m \xi m_b \bar c c \varphi_1 .
\end{equation}
Even in th Abelian gauge theory,  the renormalizable gauge require the FP ghost and anti-ghost which have a non-trivial interaction with the Higgs scalar $\varphi_1$.
A merit of the renormalizable gauge is that the mixing term $m_b b_\mu \partial^\mu \varphi_2$ between $b_\mu$ and $\varphi_2$ in ${\cal L}_{GF+FP}$ cancels the same term in the original Lagrangian,
\begin{eqnarray}
  |D_\mu[b]\phi|^2 &=& |(\partial_\mu -ig_m b_\mu)\phi|^2
\\
&=& {1 \over 2}(\partial_\mu \varphi_1+g_m b_\mu \varphi_2)^2 
+ {1 \over 2}(\partial_\mu \varphi_2-g_m b_\mu \varphi_1)^2
\nonumber\\
&&- g_m v b^\mu(\partial_\mu \varphi_1 + gb_\mu \varphi_1)
+ {g^2v^2 \over 2}b_\mu b^\mu .
\end{eqnarray}
Note that $\varphi_2$ is the would-be Nambu-Goldstone (NG) boson which corresponds to the phase factor $\theta$ in the polar coordinate,
\begin{equation}
  \phi(x) = {1 \over \sqrt{2}}[v+\rho(x)]e^{i\theta(x)} . 
\end{equation}
However, we don't use this parameterization in this section, since it is not suitable for performing the loop calculation. 
Then the total Lagrangian of the DGL theory
\begin{equation}
 {\cal L}_{DGL}[b,\phi,c,\bar c] = {-1 \over 4}(b_{\mu\nu})^2 
+ |(\partial_\mu-ig_m b_\mu) \phi|^2
-  \lambda (|\phi|^2-v^2)^2 
+  {\cal L}_{GF+FP} ,
\end{equation}
is decomposed into the free part and the interaction part as
\begin{eqnarray}
  {\cal L}_0 &=& {1 \over 2}[(\partial_\mu \varphi_1)^2 - m_{\varphi_1}^2 \varphi_1^2]
+ {1 \over 2}[(\partial_\mu \varphi_2)^2 - m_{\varphi_2} \varphi_2^2]
\nonumber\\&&
-{1 \over 4}(\partial_\mu b_\nu - \partial_\nu b_\mu)^2 
+ {1 \over 2} m_b^2 b_\mu b^\mu - {1 \over 2\xi}(\partial^\mu b_\mu)^2
+ i \bar c (\partial^2 + \xi m_b^2) c ,
\\
  {\cal L}_1 &=& g_m b_\mu (\partial^\mu \varphi_1 \varphi_2 - \partial^\mu \varphi_2 \varphi_1) 
+ g^2 v b_\mu b^\mu \varphi_1 
+ {1 \over 2}g_m^2 b_\mu b^\mu (\varphi_1^2+\varphi_2^2)
\nonumber\\&&
- \lambda v \varphi_1 (\varphi_1^2+\varphi_2^2) 
- {\lambda \over 4}(\varphi_1^2+\varphi_2^2)^2   
+ i g_m \xi m_b \bar c c \varphi_1 ,
\end{eqnarray}
where the dual gauge boson mass $m_b$, the Higgs scalar mass $m_{\varphi_1}$ and the would-be Nambu-Goldstone mass $m_{\varphi_2}$ are given by
\begin{eqnarray}
  m_b := g_m v, \quad m_{\varphi_1}^2=2\mu^2=\lambda v^2 ,
\quad m_{\varphi_2} := \xi m_b^2 .
\end{eqnarray}
The Feynman rules is given in Fig.~\ref{fig:dGL}.
The propagators are given as follows.
\par
\noindent
{\bf Propagators}:
\begin{enumerate}
\item[(a)] Higgs scalar $\varphi_1$ propagator:
\begin{equation}
 iD_1(k) = {i \over k^2-m_{\varphi_1}^2+i\epsilon}, \quad 
 (m_{\varphi_1}^2=2\mu^2 = \lambda v^2) .
\end{equation}

\item[(b)] Would-be Nambu-Goldstone boson $\varphi_2$ propagator:
\begin{equation}
 iD_2(k) = {i \over k^2-m_{\varphi_2}^2+i\epsilon}, 
\quad (m_{\varphi_2}^2= \xi m_b^2 = \xi g_m^2 v^2) .
\end{equation}

\item[(c)] Gauge boson $b_\mu$ propagator:
\begin{equation}
 iD_{\mu\nu}(k) = {-i \over k^2-m_b^2+i\epsilon}
\left[ g_{\mu\nu}-(1-\xi){k_\mu k_\nu \over k^2-\xi m_b^2+i\epsilon} \right] .
\end{equation}
 
\item[(d)] Ghost (anti-ghost) $c$ ($\bar c$) propagator:
\begin{equation}
 iD_c(k) = {i \over k^2-\xi m_b^2+i\epsilon} .
\end{equation}

\end{enumerate}
The relevant vertices in the Feynmann rules are given in (e) to (j) of Fig.\ref{fig:dGL}. 

\unitlength=0.001in
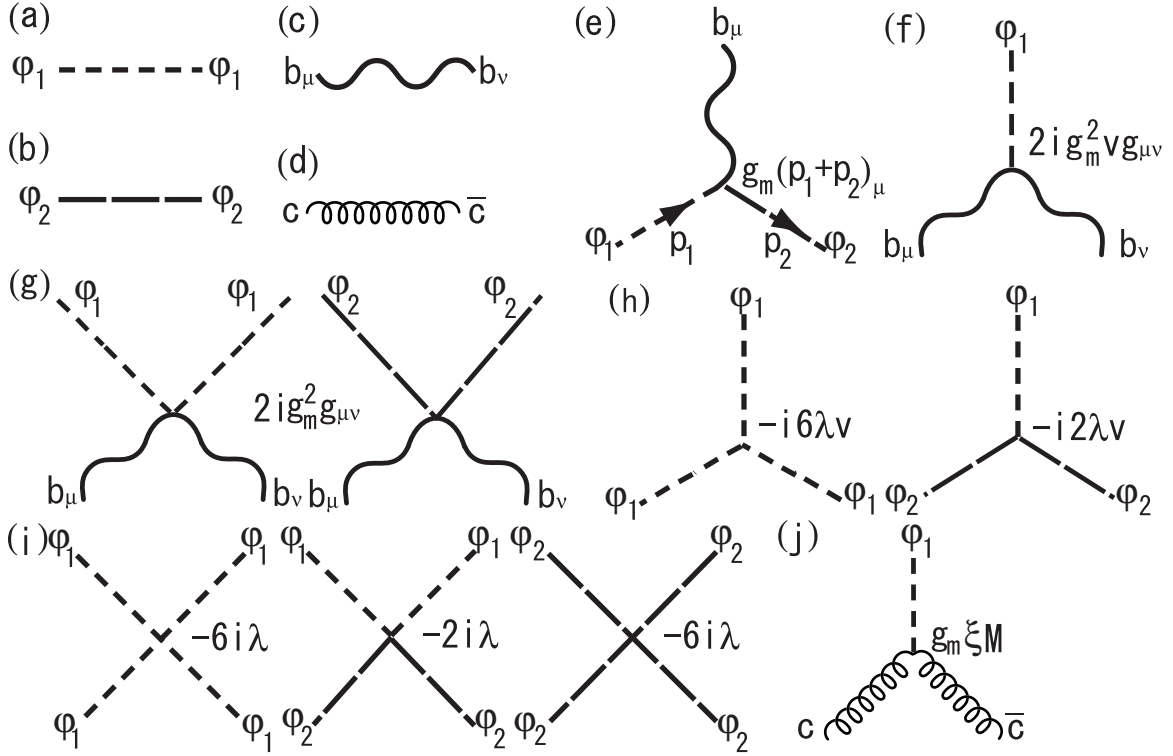
\begin{figure}[t]
\begin{center}
\begin{picture}(5000,3500)
\put(-450,3300){%
   \put(0,0){\epsfbox{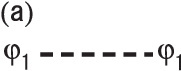}}%
   }%
\put(-450,2600){%
   \put(0,0){\epsfbox{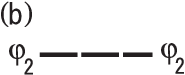}}%
   }%
\put(950,3300){%
   \put(0,0){\epsfbox{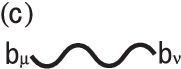}}%
   }%
\put(950,2600){%
   \put(0,0){\epsfbox{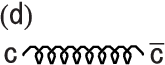}}%
   }%
\put(2500,2300){%
   \put(0,0){\epsfbox{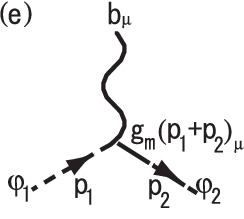}}%
   }%
\put(4100,2400){%
   \put(0,0){\epsfbox{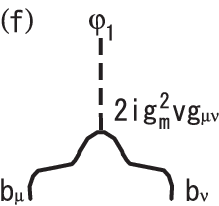}}%
   }%
\put(-450,1120){%
   \put(0,0){\epsfbox{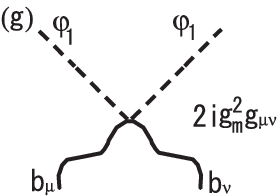}}%
   }%
\put(1050,1100){%
   \put(0,0){\epsfbox{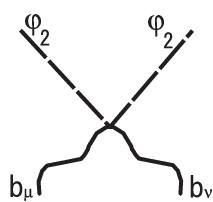}}%
   }%
\put(2650,1000){%
   \put(0,0){\epsfbox{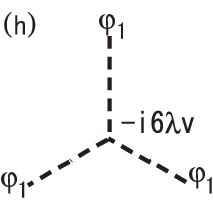}}%
   }%
\put(4100,1080){%
   \put(0,0){\epsfbox{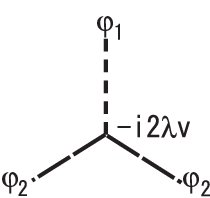}}%
   }%
\put(-450,-150){%
   \put(0,0){\epsfbox{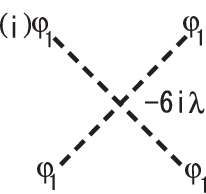}}%
   }%
\put(950,-130){%
   \put(0,0){\epsfbox{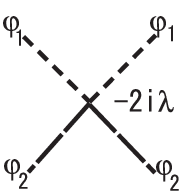}}%
   }%
\put(2150,-150){%
   \put(0,0){\epsfbox{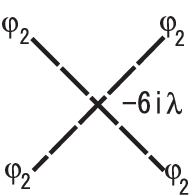}}%
   }%
\put(3550,-150){%
   \put(0,0){\epsfbox{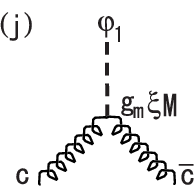}}%
   }%
\end{picture}
 \caption[]{Feynman rules for the dual Ginzburg-Landau model:
Propagators:
(a) Higgs scalar $\varphi_1$ propagator,
(b) Would-be Nambu-Goldstone boson $\varphi_2$ propagator:
(c) Gauge boson $b_\mu$ propagator,
(d) ghost $C$ propagator;
Vertices:
(e) $\varphi_1-\varphi_2-b$,
(f) $\varphi_1-b-b$,
(g) $\varphi_1-\varphi_1-b-b$, $\varphi_2-\varphi_2-b-b$,
(h) $\varphi_1-\varphi_1-\varphi_1$, $\varphi_1-\varphi_2-\varphi_2$,
(i) $\varphi_1-\varphi_1-\varphi_1-\varphi_1$,
$\varphi_1-\varphi_1-\varphi_2-\varphi_2$,
$\varphi_2-\varphi_2-\varphi_2-\varphi_2$,
(j) $\varphi_1-c-\bar c$ .

}
 \label{fig:dGL}
\end{center}
\end{figure}

\par
The basic strategy of showing the equivalence is to integrate out the massive scalar (dual Higgs) field $\varphi_1$ with mass squared $m_{\varphi_1}^2=2\mu^2=\lambda v^2$, since the theory (\ref{dGL2}) is written in terms of $b_\mu$ and $\theta$ ($\theta \sim \varphi_2/\eta$).
The renormalized mass $m_H=m_{\varphi_2}$ of the heavy field is made large while all other parameters are held finite. The $m_b$ and $m_{\varphi_1}$ are the masses of the light fields.
The decoupling theorem \cite{AC75} asserts that phenomena on energy scales much less than the Higgs mass $m_H=m_{\varphi_2}$ are described by a low-energy effective theory with the Lagrangian
\begin{eqnarray}
 {\cal L}_b
&=& -{1 \over 4}(\partial_\mu b_\nu - \partial_\nu b_\mu)^2 
+ {1 \over 2} m_b^2 b_\mu b^\mu
\nonumber\\
&=& - {1 \over 4}Z_b (\partial_\mu b_\nu^R - \partial_\nu b_\mu^R)^2
+ {1 \over 2} m_b^2 Z_b b_\mu^R b^\mu{}^R ,
\label{rL1}
\end{eqnarray}
where we have substituted the renormalization relation,
$b_\mu:=Z_b^{1/2}b_\mu^R$.
On the other hand, the renormalized Lagrangian with the counter term is given by
\begin{equation}
 {\cal L}_b 
= - {1 \over 4} (\partial_\mu b_\nu^R - \partial_\nu b_\mu^R)^2
+ {1 \over 2} (m_b^R)^2  b_\mu^R b^\mu{}^R 
\\
- {1 \over 4} \delta_b (\partial_\mu b_\nu^R - \partial_\nu b_\mu^R)^2
+ {1 \over 2} \delta_m  b_\mu^R b^\mu{}^R .
\label{rL2}
\end{equation}
Equating (\ref{rL1}) and (\ref{rL2}), we obtain the relationship,
\begin{equation}
 \delta_b := Z_b - 1, \quad \delta_m := Z_b m_b^2 - (m_b^R)^2 .
\end{equation}
We use the renormalized perturbation theory in the (dual) coupling constant
\begin{equation}
  g_m :=  {4\pi \over g} .
\end{equation}
\par
In order to see  coincidence of the resultant theory with Lagrangian (\ref{dGL4}), it is useful to choose the Landau gauge $\xi=0$ in which the would-be NG boson $\varphi_2$ is massless, since $m_{\varphi_2}^2=\xi m_b^2$.  In Appendix~\ref{sec:dGL}, we evaluate one-loop graphs for the self-energy of the would-be NG boson $\varphi_2$ and the vacuum polarization of the dual gauge boson $b_\mu$ as shown in Fig.~\ref{fig:NG} and Fig.\ref{fig:dualvector} respectively, based on the Feynman rules given in Fig.~\ref{fig:dGL}.
The radiative correction causes the mass renormalization and wavefunction renormalization of $b_\mu$ field, while the would-be NG boson field $\varphi_2$ field remains massless for $\xi=0$.
\par
Now the renormalized mass is written as
\begin{equation}
  (m_b^R)^2 = Z_b m_b^2 - \delta_m
= m_b^2 + m_b^2 \delta_b - \delta_m ,
\label{rmass}
\end{equation}
where $(m_b^R)^2$ is of order $g_m^2 m_b^2$, whereas $\delta_m$ is of order $g_m^2 m_b^2$ or $g_m^2 m_H^2$.
Actual calculations in Appendix~\ref{sec:dGL} give
\begin{eqnarray}
 \delta_b &=& 
 -{g_m^2 \over (4\pi)^{2}}  \left( {N_\epsilon \over 3}+{13 \over 36} + 6 \ln {m^2 \over \mu^2} \right) + O(g_m^4) ,
\nonumber\\
\delta_m  &=& 
{4g_m^2 \over (4\pi)^{2}} m_b^2 \left[ {3 \over 4}(N_\epsilon + 1)
- {m_H^2 \ln {m_H^2 \over \mu^2} - m_b^2 \ln {m_b^2 \over \mu^2} \over m_H^2-m_b^2}  
 - {1 \over 8} {m_H^2 \over m_b^2} \left(-1+2 \ln {m_H^2 \over \mu^2} \right) \right] 
\nonumber\\&& 
-{g_m^2 \over (4\pi)^{2}}  m_H^2  \left( -N_\epsilon+{1 \over 2}+\ln {m_H^2 \over \mu^2} \right) + O(g_m^4) ,
\label{delta}
\end{eqnarray}
where
\begin{equation}
 N_\epsilon := {1 \over \epsilon} + \ln 4\pi - \gamma_E .
\end{equation}
Substituting (\ref{delta}) into (\ref{rmass}), we obtain
\begin{eqnarray}
 (m_b^R)^2  &=& m_b^2   
  - {g_m^2 \over (4\pi)^{2}} m_H^2  \left( {4 \over 3}N_\epsilon +{49 \over 36} + 5 \ln {m_H^2 \over \mu^2} - \ln {m_b^2 \over \mu^2} \right)  
\nonumber\\&&
-{4g_m^2 \over (4\pi)^{2}} m_b^2 \left[ {3 \over 4}(N_\epsilon + 1)
- {m_H^2 \ln {m_H^2 \over \mu^2} - m_b^2 \ln {m_b^2 \over \mu^2} \over m_H^2-m_b^2}   \right] 
+ O(g_m^4) .
\end{eqnarray}
In order to keep the physical mass $m_b^R$ of $b_\mu$ finite even for the large dual Higgs mass, $m_H \rightarrow \infty$, we must let $m_b^2$ have a term proportional to $g_m^2 m_H^2$,
\begin{eqnarray}
 m_b^2  &=&  \mbox{finite term}
+  {g_m^2 \over (4\pi)^{2}} m_H^2  \left( {4 \over 3}N_\epsilon +{49 \over 36} + 5 \ln {m_H^2 \over \mu^2} - \ln {m_b^2 \over \mu^2} \right)  .
\end{eqnarray}
Thus we arrive at the conclusion,
\begin{eqnarray}
 (m_b^R)^2  &=&  \mbox{finite term in $m_b$}
\nonumber\\&&
-{4g_m^2 \over (4\pi)^{2}} m_b^2 \left[ {3 \over 4}(N_\epsilon + 1)
- {m_H^2 \ln {m_H^2 \over \mu^2} - m_b^2 \ln {m_b^2 \over \mu^2} \over m_H^2-m_b^2}  
 \right] 
+ O(g_m^4) .
\end{eqnarray}
The comparison with (\ref{dGL4}) implies that 
\begin{eqnarray}
 Z_b = 1 + {\eta^4 \over \gamma^4}, \quad (m_b)^2 Z_b = \eta^2 .
\end{eqnarray}

\subsection{\label{subsec:comparison}Comparison with the previous work}

\par
In the previous paper \cite{KondoI}, we have used the Hodge decomposition for the two-form $B$,
\begin{equation}
 B = db + *d\chi ,
\end{equation}
with two one-forms $b$ and $\chi$.
Then the definition of $h$ yields
\begin{equation}
 h := *B = *db + **d\chi = \delta *b - d\chi  ,
\end{equation}
which leads to
\begin{eqnarray}
 \delta h &=& - \delta d \chi,
\\
 H := dh &=& d\delta *b = *\delta db ,
\end{eqnarray}
where we have used $dd=0=\delta \delta$.
Moreover,
\begin{equation}
 \delta H = \delta dh = \delta * \delta db = * d\delta db
= *d(\delta d+d \delta) b = *d \Delta b .
\end{equation}
Hence, if we require $\delta h=0$, then $\Delta \chi=0$  under the gauge fixing condition $\delta \chi=0$.  It is known \cite{Nakahara90} that a $p$-form $\omega$ is harmonic ($\Delta \omega=0$) if and only if $\omega$ is closed ($d\omega=0$) and co-closed ($\delta \omega=0$).  The harmonic form $\omega$ does not exist on the topologically trivial manifold, since the dimension of the set of exact $p$-forms is equal to the Betti number, i.e.,
$\mbox{dim Harm}^p(M)=b^p$.
In this case, $\chi$ is divergenceless and rotation free vector field on the four-dimensional Minkowski spacetime.
Hence we can eliminate the variable $\chi$ and hence $\chi$ does not appear in the result.  This corresponds to the situation discussed in the paper \cite{KondoI}.  

\par
It is instructive to compare the above result with the previous one.
Replacing $V^\mu=\partial_\nu *h^{\mu\nu}$ with the magnetic monopole current $k_\mu$ (this definition is suggested from $h \sim *B$ since $k \sim \delta B$), we obtain the theory,
\begin{eqnarray}
   Z_{APEGT}[b,k;\Theta] 
&=&  \int {\cal D}b_\mu \int {\cal D}k_\mu  
\exp \Biggr\{ i \int d^4x  \Biggr[ - {1 \over 4} (b_{\mu\nu}+b_{\mu\nu}^S)^2  
\nonumber\\ &&
 +    b_\mu k^\mu  + {1 \over 2\eta^2}k_\mu^2
+ {1 \over 2\gamma^4} (k, \Delta k) \Biggr] \Biggr\} .
\end{eqnarray}
After integrating out the dual gauge field $b_\mu$, this action leads to the theory of magnetic monopole written in terms only of the monopole current $k_\mu$.  See section \ref{sec:monopole}.

\subsection{\label{sec:parameter}How to determine the parameters $\rho,\sigma,\alpha$}

It has been shown \cite{QR98,KondoI,KS00b} that the renormalization of the Yang-Mills coupling constant $g$ does not depend on $\rho, \sigma$ and the gauge parameter $\alpha$.  Therefore, the $\beta$-function is also independent of the choice of $\rho, \sigma$ and of the gauge parameter $\alpha$.  The resulting $\beta$-function exactly coincides with the one-loop $\beta$-function of the original $SU(N)$ Yang-Mills theory, 
\begin{equation}
 \beta(g_{\rm R}) := \mu {\partial g_{\rm R} \over \partial \mu} 
 = - g_{\rm R} \mu {\partial \over \partial \mu} \ln Z_g = - {b_0 \over (4\pi)^2} g_{\rm R}^3 + O(g_{\rm R}^5) ,
\quad b_0 = {11 \over 3} C_2(G) ,
\end{equation}
exhibiting the asymptotic freedom.
Moreover, some of the anomalous dimensions have been calculated.%
\footnote{The  anomalous dimensions of the fields $a_\mu^i$, $B_{\mu\nu}^i$ and
 the parameters $\rho, \beta$  are calculated in the previous paper \cite{KS00b}.
To obtain the anomalous dimension for $\sigma, \alpha$, we need to calculate more Feynman graphs than those in \cite{KS00b}, see \cite{preparation}.
For $G=SU(2)$, $Z_\alpha$ and $Z_\zeta$ were calculated by Hata and Niigata \cite{HN93} and Schaden \cite{Schaden99},  $Z_A$ by Schaden \cite{Schaden99}
and $Z_{C^a}$ by Quandt and Reinhardt \cite{QR98}.
}
The anomalous dimensions of the fields in the $SU(N)$ Yang-Mills theory are evaluated as
\begin{eqnarray}
 \gamma_{a^i}(g) &:=& {1 \over 2} \mu {\partial \over \partial\mu} \ln Z_a 
= \frac{11}3 C_2(G) \frac{g_{\rm R}^2}{(4\pi)^2} \quad [SU(N)] ,
\nonumber\\
 \gamma_{B^i}(g) &:=& {1 \over 2} \mu {\partial \over \partial\mu} \ln Z_B 
= - \frac{1+\alpha_{\rm R}}2\sigma_{\rm R}^2
    C_2(G) \frac{g_{\rm R}^2}{(4\pi)^2} \quad [SU(N)],
\nonumber\\
 \gamma_{A^a}(g) &:=& {1 \over 2}  \mu {\partial \over \partial\mu} \ln Z_A 
= \left( {22 \over 3} - {9 \over 2} - {\alpha_R \over 2} - \beta_R \right) C_2(G) \frac{g_{\rm R}^2}{(4\pi)^2}  \quad [SU(2)].
\nonumber\\
 \gamma_{C^a}(g) &:=& {1 \over 2}  \mu {\partial \over \partial\mu} \ln Z_{C^a} 
=   (\beta_R - 3) \frac{g_{\rm R}^2}{(4\pi)^2}  \quad [SU(2)],
\nonumber\\
 \gamma_{C^i}(g) &:=& {1 \over 2}  \mu {\partial \over \partial\mu} \ln Z_{C^i} 
= ? ,
\end{eqnarray}
where $?$ denotes that the result is not yet unavailable. 
For the parameters $\rho, \sigma, \alpha, \beta$ and $\zeta$,
\begin{eqnarray}
 \gamma_\rho(g) &:=&   \mu {\partial \rho_R \over \partial\mu} 
= - \rho_R \mu {\partial \over \partial\mu} \ln Z_\rho 
\nonumber\\
&=& - \rho_R \left[-\frac{11}6-\frac{\sigma_{\rm R}^2}2
        +2\frac{\sigma_{\rm R}}{\rho_{\rm R}}
        -\frac{1-\alpha_{\rm R}}2
        \left(\frac{\sigma_{\rm R}}{\rho_{\rm R}}
              -\frac{\sigma_{\rm R}^2}2\right)\right]
  C_2(G)\frac{g_{\rm R}^2}{(4\pi)^2} \quad [SU(N)],
\label{ad-rho}
\nonumber\\
 \gamma_\sigma(g) &:=&   \mu {\partial \sigma_R \over \partial\mu}  
=  - \sigma_R \mu {\partial \over \partial \mu} \ln Z_\sigma 
= ?,
\nonumber\\
 \gamma_\alpha(g) &:=&   \mu {\partial \alpha_R \over \partial\mu}  
= - \alpha_R \mu {\partial \over \partial\mu} \ln Z_\alpha 
=  - \alpha_R \left( {3 \over \alpha_R} + 6 - {22 \over 3} + \alpha_R \right) {g_R^2 \over 8\pi^2} \quad [SU(2)],
\label{ad-alpha}
\nonumber\\
 \gamma_\beta(g) &:=&  \mu {\partial \beta_R \over \partial\mu}  
= -\beta_{\rm R} \mu {\partial \over \partial\mu} \ln Z_{a^i} 
= -2 \gamma_{a^i}(g) \beta_{\rm R} \quad [SU(N)],
\nonumber\\
 \gamma_\zeta(g) &:=&   \mu {\partial \zeta_R \over \partial\mu}  
=  - \zeta_R \mu {\partial \over \partial \mu} \ln Z_\zeta 
= \alpha_R (1+\zeta_R) \left( \zeta_R + {3 \over \alpha_R^2} \right)  \quad [SU(2)] .
\end{eqnarray}
Note that the parameters $\rho, \sigma, \alpha, \beta$ do run and changes according to the scale $\mu$.  
We would like to obtain the renormalization scale $\mu$ independent result.  The simplest way is to search for the fixed point for these parameters 
on which the parameters are kept fixed irrespective of the renormalization scale $\mu$.
The fixed point is determined by solving the simultaneous equations,
$\gamma_\rho(g)=0, \gamma_\sigma(g)=0, \gamma_\alpha(g)=0$ and
$\gamma_\zeta(g)=0$. 
\par
Finally, we give a simple argument how to determine the parameters $\rho, \sigma, \alpha$ in our theory.  
The parameter $\rho$ obeys the differential equation,
\begin{eqnarray}
 \gamma_\rho(g) :=  \mu {\partial \rho_R \over \partial\mu} 
=  {1 \over 2} \left[ \left(  {11 \over 3} + {1+\alpha \over 2}\sigma^2 \right) \rho_R - (3+\alpha) \sigma \right]
  C_2(G)\frac{g_{\rm R}^2}{(4\pi)^2} .
\end{eqnarray}
This implies that the point $\rho=\rho^*$,
\begin{equation}
 \rho=\rho^* := {3+\alpha \over {11 \over 3} + {1+\alpha \over 2}\sigma^2 },
\end{equation}
is the infrared fixed point, 
as far as 
$
 {11 \over 3} + {1+\alpha \over 2}\sigma^2 > 0. 
$
At least in the present stage of investigations, the simplest choice of the parameters is
\begin{equation}
  \alpha_R  = 1, \quad \sigma_R  = 1.
\end{equation}
The choice $\alpha=1$ greatly simplifies the evaluation and the expression of the result for the vacuum polarization of $B_{\mu\nu}$.  The choice $\sigma=1$ completely eliminates the quartic gluon interaction, see (\ref{L_inv^i}).  Usually, the auxiliary field $B_{\mu\nu}$ is introduced so as to achieve this situation.
  In this case, the value 
\begin{equation}
  \rho_R = {6 \over 7}
\label{rho}
\end{equation}
 is the infrared fixed point.  
However,  (\ref{ad-alpha}) implies that $\alpha_R(\mu)$ monotonically decreasing in $\mu$ and that there is no fixed point for $\alpha$ in the $SU(2)$ case, although the $SU(N)$ case can be different from the $SU(2)$ case.  
\par
The complete list of the anomalous dimensions in the $SU(N)$ case  and the details of the RG properties of the APEGT will be given in a subsequent paper \cite{preparation}.

\section{\label{sec:estimate}Estimation of neglected higher-order terms}
\setcounter{equation}{0}

We proceed to discuss the issue whether the neglected terms in the above derivation do not invalidate the above result in the range of parameters and the energy scale in question.  

\unitlength=0.001in
\begin{figure}
\begin{center}
\begin{picture}(5800,2800)
\put(3900,0){%
   \put(-3000,0){\epsfysize=3in\epsfbox{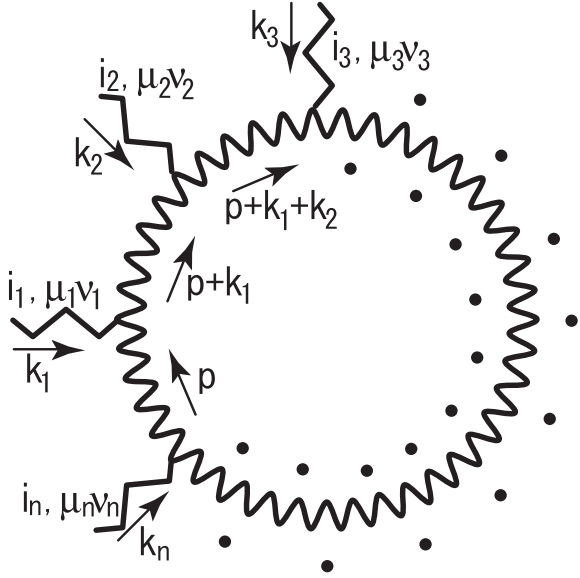}}%
   }%
\end{picture}
\end{center}
\caption[]{Higher order terms for obtaining corrections of APEGT.}
\label{fig:Bvp}
\end{figure}

\subsection{\label{subsec:higher-order-cumulants}Higher-order cumulants and large N suppression}

In this paper we have obtained the APEGT which is bilinear with respect to the diagonal fields, $a_{\mu}$ and $B_{\mu\nu}$.  Hence the bilocal approximation in section~\ref{sec:cumulant} is exact within this framework.  
In the framework of large $N$ expansion, moreover, neglecting higher-order cumulants can be justified as follows.%
\footnote{
The author would like to thank Toru Shinohara for helpful discussion on this point.
}  
In the large N expansion, 
$\lambda :=g^2 N$ is kept fixed.
\par
For example, we consider the diagram in Fig.\ref{fig:Bvp} with $n$ external $B_{\mu\nu}$ lines. 
According to the Feynman rules in Fig.~\ref{fig:FeynmanRules}, it corresponds to 
\begin{eqnarray}
 && \Pi_{\mu_1\nu_1,\cdots,\mu_n \nu_n}^{i_1 \cdots i_n}(k_1,\cdots,k_n)
\nonumber\\
&=& {1 \over n} \int {d^4p \over (2\pi)^4} 
 D_{\sigma_n \rho_1}^{d_n c_1}(p)
[-2g\sigma f^{i_1 c_1 d_1} I_{\mu_1 \nu_1,\rho_1 \sigma_1}]
\nonumber\\&& \times
 D_{\sigma_1 \rho_2}^{d_1 c_2}(p+k_1)
[-2g\sigma f^{i_2 c_2 d_2} I_{\mu_2 \nu_2,\rho_2 \sigma_2}]
\nonumber\\&&
 \cdots
\nonumber\\&& \times
 D_{\sigma_{n-1} \rho_n}^{d_{n-1} c_n}(p+k_1+\cdots+k_{n-1})
[-2g\sigma f^{i_n c_n d_n} I_{\mu_n \nu_n,\rho_n \sigma_n}] .
\end{eqnarray}
This quantity is proportional to the factor,
\begin{equation}
 g^n \delta^{d_n c_1} f^{i_1 c_1 d_1}  
 \delta^{d_1 c_2} f^{i_2 c_2 d_2}  
 \cdots
 \delta^{d_{n-1} c_n} f^{i_n c_n d_n} .
\end{equation}
For $n=2$, the factor reads
\begin{equation}
 g^2 \delta^{d_2 c_1} f^{i_1 c_1 d_1}  
 \delta^{d_1 c_2} f^{i_2 c_2 d_2}  
=  g^2 f^{i_1 c_1 c_2}   f^{i_2 c_2 c_1} = - g^2 C_2 \delta^{i_1 i_2} 
= - g^2 N  \delta^{i_1 i_2} 
\sim O(N^0) .
\end{equation}
For $n=3$, it is identically zero,
\begin{equation}
 g^3 \delta^{d_3 c_1} f^{i_1 c_1 d_1}  
 \delta^{d_1 c_2} f^{i_2 c_2 d_2}  
 \delta^{d_{2} c_3} f^{i_3 c_3 d_3} 
=    g^3 f^{i_1 c_1 c_2} f^{i_2 c_2 c_3} f^{i_3 c_3 c_1} 
= 0 .
\end{equation}
For $n=4$, it is of order $O(N^{-2})$ for large $N$, since 
\begin{eqnarray}
&& g^4 \delta^{d_4 c_1} f^{i_1 c_1 d_1}  
 \delta^{d_1 c_2} f^{i_2 c_2 d_2}  
 \delta^{d_2 c_3} f^{i_3 c_3 d_3} 
 \delta^{d_3 c_4} f^{i_4 c_4 d_4} 
\nonumber\\
&=&  g^4
   f^{i_1 c_1 c_2}  
   f^{i_2 c_2 c_3}  
   f^{i_3 c_3 c_4} 
   f^{i_4 c_4 c_1} 
\nonumber\\
&=& g^4 {N \over N+1} 
(\delta^{i_1 i_2}\delta^{i_3 i_4}
+\delta^{i_1 i_3}\delta^{i_2 i_4}
+\delta^{i_1 i_4}\delta^{i_2 i_3})
\sim O(N^{-2}) .
\end{eqnarray}
The last equality is obtained as follows.  From the symmetry under the exchange of the indices $i_1,i_2,i_3,i_4$, we can put
\begin{eqnarray}  
   f^{i_1 c_1 c_2}  
   f^{i_2 c_2 c_3}  
   f^{i_3 c_3 c_4} 
   f^{i_4 c_4 c_1} 
= A 
(\delta^{i_1 i_2}\delta^{i_3 i_4}
+\delta^{i_1 i_3}\delta^{i_2 i_4}
+\delta^{i_1 i_4}\delta^{i_2 i_3}) .
\end{eqnarray}
By contracting $i_3$ and $i_4$, we obtain
\begin{eqnarray}  
LHS &=&   f^{i_1 c_1 c_2}  
   f^{i_2 c_2 c_3}  
   f^{i_3 c_3 c_4} 
   f^{i_3 c_4 c_1} 
=
   - f^{i_1 c_1 c_2}  
   f^{i_2 c_2 c_3}  
   \delta^{c_1 c_3} 
=  N \delta^{i_1 i_2} ,
\\
RHS 
&=&  A [\delta^{i_1 i_2}(N-1)+ 2 \delta^{i_1 i_2} ]
= A(N+1) \delta^{i_1 i_2} ,
\end{eqnarray}
where we have used the formula (\ref{f3}).  Hence we obtain $A=N/(N+1)$.
This argument can be extended to arbitrary $n$. 
Thus it turns out that the higher-order terms with $n$ external lines of the tensor fields $B$ are suppressed by $1/N^2$ in the large $N$ expansion for $n \ge 4$.  
\par
Thus the contribution from the diagrams with $n$ external lines of diagonal fields is suppressed by a factor $1/N^2$ for $n \ge 4$ where $n$ is the total number of external diagonal gluon fields $a_\mu^i$ and external tensor gauge fields $B_{\mu\nu}^i$ in the off-diagonal one-gluon-loop or one-ghost-loop diagram.  This is because every three-point vertex (c),(d),(e) in Feynman rules has a common factor $gf^{iab}$.
In the leading order of the large N expansion, we have only to take into account the diagrams with two external lines and hence the resulting APEGT is bilinear in the diagonal fields, $a_\mu^i$ (or $f_{\mu\nu}^i$) and $B_{\mu\nu}^i$.  
In this limit, therefore, the bilocal approximation of neglecting the higher-order cumulants is exact within our approach.
In this sense, the bilocal approximation is consistent within the framework of the APEGT.  This situation is in sharp contrast to the bilocal approximation in the analytical treatment of the stochastic vacuum model, although the validity is confirmed by the numerical calculations on a lattice\cite{DDM97}. 

\par

\subsection{\label{subsec:higher-order-expansion}Higher-order terms of low-energy or large mass expansion
and the decoupling theorem}

We have neglected higher-order terms of the large mass or the derivative expansion in powers of $p^2/M_A^2$.  This approximation will be valid in the low-energy region below $M_A$.  This is considered as an example of the Appelquist-Carazzone decoupling theorem\cite{AC75}.  
\par
First, we recall the case of QED.
A typical example of applying the decoupling theorem is a derivative expansion of the photon effective action (known as the Euler-Heisenberg Lagrangian) obtained by integrating out the electron field in QED as
\begin{eqnarray}
  && \Gamma_{eff}[a] 
\nonumber\\
&=& -i \ln \int [d\psi][d\bar \phi] \exp \left\{ iS_{QED} \right\} 
\\
&=& {-1 \over 4} \int d^4x f_{\mu\nu}f^{\mu\mu} 
- {e^2 \over 3(4\pi)^2} z(\mu)  \int d^4x f_{\mu\nu}f^{\mu\mu}
- {e^2 \over 15(4\pi)^2M^2}   \int d^4x f_{\mu\nu}\partial^2 f^{\mu\mu} 
\nonumber\\
&+& {e^4 \over 90(4\pi)^2M^4} \int d^4x \left[ (f_{\mu\nu} f^{\mu\nu})^2
+ {7 \over 4}  (f_{\mu\nu} *f^{\mu\nu})^2 \right]
+ \left( {p^2 \over M^2} \right)^3 ,
\end{eqnarray}
where $M$ is the electron mass and
\begin{equation}
 z(\mu) := {2 \over \epsilon} + \ln 4\pi - \gamma_E - \ln {M^2 \over \mu^2} .
\end{equation}
It turns out that the terms that do not decouple in the $M \rightarrow \infty$ limit have the same form as those appearing in the original Lagrangian and therefore they can be absorbed in the wavefunction renormalization.  The new structures appear as non-renormalizable terms and they vanish in the limit $M \rightarrow \infty$. This is an example of the decoupling theorem \cite{AC75}.  The theorem states that, under some given conditions, the effects of the heavy particle only appear in the light particle physics through corrections proportional to a negative power of $M$ or through renormalization.
The validity of this approach is limited to energies much lower than the  mass $M$.  

\par
In the low-energy effective action of the Yang-Mills theory, all the terms including more than two diagonal fields can be suppressed in the large N limit, as we discussed in the above.  Of course, this argument does not hold for the relatively small $N$.
Even in the case of not-so-large $N$ (e.g., $N=2$), however, such terms are suppressed by the power of the inverse (off-diagonal-gluon ) mass $1/M_A$.  In the limit $M_A \rightarrow \infty$,  the off-diagonal gluons affect the physics described by the diagonal gluons only through renormalization.
In the case of small $N$, the validity of our approach is limited to low energies lower than the off-diagonal gluon mass $M_A$.  
More quantitative estimate of the higher order terms will be given in a subsequent paper \cite{preparation}.

\section{\label{sec:monopole}Magnetic monopole condensation and area law}
\setcounter{equation}{0}
\par
  In this section, we show that the LEET given by (\ref{Last}) reproduces the same results as those obtained by the supposed dual Ginzburg-Landau theory of type II.  Here the dual Ginzburg-Landau theory can include the type II on the border of type I.  Therefore, two extreme limits, London limit and Bogomol'nyi  limit, are special cases.

\subsection{Monopole action and monopole condensation}
\par
In order to obtain the monopole action, we return to the action (\ref{APEGTh}),
\begin{equation}
S_{M}[h;\Theta] =  \int d^4x \left\{ {\cal L}_m[h]
+  2Jg  \rho^{-1}K^{1/2}   h_{\mu\nu}(x) \tilde \Theta_{\mu\nu}(x)  \right\} ,
\end{equation}
where the expectation value of the Wilson loop is evaluated as
\begin{eqnarray}
    \langle W(C) \rangle_{YM} 
&=& Z_{M}[h;\Theta]/Z_{M}[h;0],
\\
Z_{M}[h;\Theta] &:=& 
\int  {\cal D}h_{\mu\nu}   \delta(\partial^\nu h_{\mu\nu})
\exp \left\{i S_{M}[h;\Theta]  \right\}  .
\end{eqnarray}
We define the monopole current $k_\mu$ by
\begin{equation}
  k_\mu := g_m \partial^\nu *h_{\mu\nu} =  g_m K^{1/2} \partial^\nu B_{\mu\nu} ,
\end{equation}
which satisfies the (topological) conservation law, $\partial^\mu k_\mu = 0$.
Conversely, the tensor field $h$ is written in terms of the monopole current $k$ as
\begin{equation}
 h_{\mu\nu} = g_m^{-1} \epsilon_{\mu\nu\rho\sigma}{1 \over \partial^2} \partial^\rho k^\sigma 
=  g_m^{-1} {1 \over 2}\epsilon_{\mu\nu\rho\sigma}{1 \over \partial^2} (\partial^\rho k^\sigma - \partial^\sigma k^\rho) ,
\end{equation}
which is subject to the constraint, $\partial^\mu h_{\mu\nu}=0$.
It is easy to show that
\begin{eqnarray}
  \int d^4x {g_m^2 \over 2}(h_{\mu\nu})^2 
&=& (k,\Delta^{-1}k) + (\delta h,\Delta^{-1} \delta h) 
\sim (k,\Delta^{-1}k) ,
\\
  - \int d^4x {g_m^2 \over 6}(H_{\lambda\mu\nu})^2 &=& (k, k) , 
\\
  \int d^4x {g_m^2 \over 2}(\partial^\lambda H_{\lambda\mu\nu})^2 
&=& (k,\Delta k) ,
\end{eqnarray}
where we have defined $\Delta := d\delta + \delta d$ and used $\delta h=0$.
Thus, we obtain the magnetic monopole theory as a LEET of Yang-Mills theory,
\begin{equation}
\mbox{\fboxsep=.1in \framebox{$\displaystyle
S_{MP}[k] =  \int d^4x {1 \over g_m^2} \left[ {-1 \over 2} (k,\Delta^{-1}k)
-  {1 \over 2 \eta^2} (k_\mu)^2  +{1 \over 2\gamma^4} (k,\Delta k) 
+ O\left( {(k,\Delta^2 k) \over M_A^6} \right) \right] ,
$}}
\label{mpaction}
\end{equation}
where $\eta$ and $\gamma$ are given by (\ref{eta}) and (\ref{gamma}) respectively.
Hence the expectation value of the Wilson loop is obtained from
\begin{eqnarray}
\mbox{\fboxsep=.1in \framebox{$\displaystyle
    \langle W(C) \rangle_{YM} 
= Z_{MP}[k;\Xi]/Z_{MP}[k;0],
$}}
\end{eqnarray}
and
\begin{eqnarray}
\mbox{\fboxsep=.1in \framebox{$\displaystyle
Z_{MP}[k;\Xi] := 
\int  {\cal D}k_{\mu}    
\exp \left\{i S_{MP}[k] + i 2Jg \rho^{-1}K^{1/2}g_m^{-1} \int d^4x k_\mu {\cal D}[\partial] \Xi^\mu  \right\} \label{Wmp} ,
$}}
\end{eqnarray}
where $\Xi_\mu$ denotes the four-dimensional solid angle under which the surface%
\footnote{An apparent $S$ dependence should drop out after summing over branches of the multivalued potential.}
$S$ with the two-dimensional coordinate $(\sigma^1,\sigma^2)$ is seen by an observer at the point $x$,
\begin{eqnarray}
 \Xi_\mu(x) &:=& {1 \over 2} \epsilon_{\mu\nu\rho\sigma} \partial^\nu_x \int_S dS^{\rho\sigma}(x(\sigma)) \Delta^{-1}(x-x(\sigma)) 
\\
&=& {1 \over 8\pi^2} \epsilon_{\mu\nu\rho\sigma} \partial^\nu_x \int_S dS^{\rho\sigma}(x(\sigma)) {1 \over (x-x(\sigma))^2} .
\end{eqnarray}
Here note that the total number of independent degrees of freedom is unchanged under the change of variables from $h_{\mu\nu}$ to $k_\mu$ 
and that the associated Jacobian in the integration measure ${\cal D}k_\mu$ is field independent and hence omitted.\par

From the monopole action, we can demonstrate that the monopole condensation  does really occur in the sense
$\langle k^\mu k_\mu \rangle  \not= 0$.
In the London limit $\gamma^{-1}=0$, the propagator of the monopole current is given by
\begin{equation}
 \langle k_\mu(x) k_\nu(y) \rangle = g_{\mu\nu} \int {d^4p \over (2\pi)^4i}
 \left( {1 \over p^2}- {1 \over \eta^2} \right)^{-1}e^{ip(x-y)}
= \eta^2  g_{\mu\nu}  \int {d^4p \over (2\pi)^4i}
  {p^2 \over \eta^2-p^2} e^{ip(x-y)} .
\end{equation}
Therefore, we obtain non-vanishing monopole condensation for non-vanishing off-diagonal gluon mass,
\begin{equation}
 \langle k_\mu(x) k_\mu(x) \rangle 
=  4 \int {d^4p \over (2\pi)^4}
  {\eta^2 p^2 \over \eta^2-p^2} 
=  {1 \over 4\pi^2} \eta^6 \left( \ln 4\pi - \gamma_E + 1 - \ln \eta^2 \right)  ,
\end{equation}
where we have used the MS scheme of the dimensional regularization.
This should be compared with the mass $m_b$ of the dual gauge field $b_\mu$. 
An close relationship between the monopole condensation and the mass of the dual gauge field was conjectured in the previous work \cite{KondoI}.  In fact, the above propagator for the monopole current leads to
\begin{equation}
 \langle k_\mu(x) k_\nu(y) \rangle 
= - \eta^2 \delta_{\mu\nu}  \int {d^4p \over (2\pi)^4}
  \left( 1-{\eta^2 \over p^2} \right)^{-1} e^{ip(x-y)}
= - \eta^2 \delta_{\mu\nu}  \delta(x-y) + O(\eta^4) ,
\end{equation}
which is nothing but eq.(4.22) predicted in the previous paper \cite{KondoI}.
In this paper we have shown that {\it the origin of monopole condensation is the existence of off-diagonal gluon mass $M_A$ which provides also the mass $m_b$ of the dual gauge field $b_\mu$},
\begin{equation}
\mbox{\fboxsep=.1in \framebox{$\displaystyle
  \langle k_\mu k_\mu \rangle \not= 0 
\leftrightarrow M_A \not= 0 \leftrightarrow m_b \not= 0
 .
$}}
\end{equation}
\par
The monopole action (\ref{mpaction}) is written in the form, 
\begin{equation}
 S_{MP}[k] = - {1 \over 2g_m^2}(\tilde k, D_m(p^2) \tilde k) ,
\label{mpaction2}
\end{equation}
where the inverse of $D_m(p)$ is the propagator given by
\begin{eqnarray}
 D_m^{-1}(p^2) :=
 \left( {1 \over p^2} - {1 \over \eta^2} +{p^2 \over \gamma^4} \right)^{-1}
= \chi
\left( {p^2 \over p^2-m_1^2} - {p^2 \over p^2-m_2^2} \right) ,
\end{eqnarray}
where
\begin{eqnarray}
&&
\mbox{\fboxsep=.1in \framebox{$\displaystyle
 m_{1,2}^2 := {\gamma^4 \over 2\eta^2} \left( 1 \pm \sqrt{1-4\eta^4/\gamma^4} \right) \quad (m_1 \ge m_2) ,
$}}
\\
&&
\mbox{\fboxsep=.1in \framebox{$\displaystyle
 \chi := {m_1^2 m_2^2 \over m_1^2-m_2^2} 
= {\eta^2 \over \sqrt{1-4\eta^4/\gamma^4}} .
$}}
\end{eqnarray}
 Finally, the monopole condensation is calculated according to
\begin{eqnarray}
 \langle k_\mu k_\mu \rangle 
= 4 \int {d^Dp \over (4\pi)^D} D_m^{-1}(p) ,
\end{eqnarray}
which yields
\begin{eqnarray}
\mbox{\fboxsep=.1in \framebox{$\displaystyle
\langle k_\mu k_\mu \rangle 
= {\chi \over 4\pi^2} \left[ ( \ln 4\pi-\gamma_E +1) (m_1^4-m_2^4) - m_1^4 \ln m_1^2 + m_2^4 \ln m_2^2 \right] .
$}}
\label{mpcond}
\end{eqnarray} 
Thus we obtain non-zero monopole condensate.

\subsection{Area law of the Wilson loop}
\par
Thanks to the NAST, the Wilson loop operator has an alternative form expressed in terms of the monopole current.  The expectation value is expressed by (\ref{Wmp}).
By making use of the monopole action (\ref{mpaction2}), the VEV of the large Wilson loop is calculated by performing the Gaussian integration over $k_\mu$ as
\begin{equation}
 \langle W(C) \rangle_{YM} =
 \exp \left\{ -{1 \over 2}(2Jg\rho^{-1}K^{1/2})^2  (\tilde \Xi_\mu, D_m^{-1} \tilde \Xi^\mu) 
\right\} .
\label{Wc1}
\end{equation}
It is not difficult to show 
(see Appendix~\ref{sec:WLcalc}) 
that (\ref{Wc1}) leads to the area law, 
\begin{equation}
\mbox{\fboxsep=.1in \framebox{$\displaystyle
 \langle W(C) \rangle_{YM} \cong \exp [- \sigma_{st} A(C)] ,
$}}
\end{equation}
for the large Wilson loop.
The string tension is obtained as
\begin{eqnarray}
\mbox{\fboxsep=.1in \framebox{$\displaystyle
 \sigma_{st} = {J^2g^2  \over 2\pi} \rho^{-2}K \chi \ln {m_1^2 \over m_2^2} .
$}}
\label{tension}
\end{eqnarray}
This is one of main results of this paper.
The static potential is defined for the rectangular loop with side lengths $T,R$  as
\begin{equation}
  V(R) := - \lim_{T \rightarrow \infty} {1 \over T} \ln \langle W(C) \rangle_{YM} .
\end{equation}
Then we obtain the linear static potential, 
\begin{equation}
\mbox{\fboxsep=.1in \framebox{$\displaystyle
  V(R)=\sigma_{st} R ,
$}}
\end{equation}
for large separation $R$.
This result is consistent with the claim that the QCD vacuum is the dual superconductor (of type II).  
This result indicates that quark in any representation is confined in the SU(2) case.  For SU(3), our result can be applied only to quark in the fundamental representation due to a restriction coming from NAST.  
\par
\par
The factor $\rho^{-2}K$ is estimated as follows.  In particular, when $\alpha=1$, $f_0$ is given by
\begin{eqnarray}
 f_0 = - \ln {M_A^2 \over \mu^2} - \gamma_E + \ln 4\pi 
= - \ln {g^2 \over 16\pi^2} -1 + {16\pi^2 \over b_0 g^2} ,
\end{eqnarray}
where we have used the expression of off-diagonal gluon mass (\ref{offmass}).
Then  $K$ reads
\begin{eqnarray}
  K := 1 + {N g^2 \over 2\pi^2}\sigma^2 f_0
= 1  + {24 \over 11} \sigma^2  
-  {N g^2 \over 2\pi^2}\sigma^2  \left( \ln {g^2 \over 16\pi^2} +1 \right) := K(g) ,
\end{eqnarray}
since $C_2=N$ and $b_0={11 \over 3}N$ for $G=SU(N)$.
For the choice $\sigma=1$ and $\rho=6/7$ in (\ref{rho}),  the factor reads
\begin{eqnarray}
  \rho^{-2}K 
= {49 \over 36} \left[ {35 \over 11} - {N g^2 \over 2\pi^2}  \left( \ln {g^2 \over 16\pi^2} +1 \right) \right] ,
\end{eqnarray}
which is positive for $0 \le g < 9.6$ and monotonically increases from $[4.3, 6.5]$ for $g \in [0,4.6]$ with a peak at $g \cong 4.6$ and monotonically decreases with $[6.5,0]$ for $g \in [4.6,9.6]$.
Therefore, we can put $\rho^{-2}K \cong 5$ for $g \cong 2$ in the SU(2) case.
The numerical estimation of $\chi, m_1, m_2$ will be given in section~\ref{sec:numerical}.
\par
From the viewpoint of dual Ginzburg-Landau theory, two constants $m_1$ and $m_2$ may be regarded as the coherence length $m_\phi$ and the penetration depth $m_b$ (apart from a factor $\sqrt{2}$).  Therefore, we can identify
\begin{equation}
 m_1 \rightarrow m_\phi = 2\sqrt{\lambda} v , \quad m_2 \rightarrow m_b = \sqrt{2} g_m v ,
\end{equation}
which leads to the ratio given  by
\begin{equation}
 {m_1^2 \over m_2^2} = {\lambda g^2 \over 8\pi^2} = 2{\lambda \over g_m^2} .
\end{equation}
In the Bogomolny limit, the coupling $\lambda$ is solely given by the Yang-Mills coupling constant $g$ as 
$\lambda=8\pi^2/g^2$.  The value of $\lambda$ in the Bogomolny limit is smaller than that in the London limit, but it is still large $\lambda \sim 20$ even for $g \sim 2$, see section~\ref{sec:numerical}.

\par
In the Bogomolny limit $m_1=m_2$, the mass satisfies
\begin{equation}
 m_1^2=m_2^2=m^2=\gamma^2=2\eta^2 .
\end{equation}
The string tension in the Bogomolny limit is given by
\begin{eqnarray}
\mbox{\fboxsep=.1in \framebox{$\displaystyle
 \sigma_{st} \cong  {J^2g^2  \over 2\pi} \rho^{-2}K \eta^2
=  {J^2g^2 \over 4\pi}  \rho^{-2}K m^2@ ,
$}}
\end{eqnarray}
since 
$
 \ln (m_1^2/m_2^2) = \ln (1+\sqrt{1-4\eta^4/\gamma^4}) - \ln (1-\sqrt{1-4\eta^4/\gamma^4}) \cong 2 \sqrt{1-4\eta^4/\gamma^4}.
$
In this limit, the monopole condensate reads
\begin{eqnarray}
\mbox{\fboxsep=.1in \framebox{$\displaystyle
\langle k_\mu k_\mu \rangle 
= {1 \over 4\pi^2} \left[ 2( \ln 4\pi-\gamma_E)+1
-2 \ln m^2 \right] m^6 .
$}}
\end{eqnarray} 

\par
In the London limit, $m_2$ reduces to the mass $m_b$ of the dual gauge field and $m_1$ to the diverging mass $m_\phi$ of the Higgs field, 
\begin{equation}
 m_2^2 \rightarrow \eta^2 (=m_b^2), 
\quad m_1^2 \rightarrow {\gamma^4 \over \eta^2}(=m_\phi^2 \rightarrow \infty) .
\end{equation}
Actually, in the near London case of type II, we have
\begin{eqnarray}
\mbox{\fboxsep=.1in \framebox{$\displaystyle
 \sigma_{st} \cong  {J^2g^2  \over 2\pi} \rho^{-2}K \eta^2 \ln {\gamma^2 \over \eta^2} ,
$}}
\end{eqnarray}
which diverges in the naive London limit $\gamma \rightarrow \infty$.
\par

\subsection{Type of dual superconductivity}
\par
In the LEET, the ratio $\eta^4/\gamma^4$ determines the type of the QCD vacuum as the dual superconductor,
\begin{eqnarray}
  {\eta^4 \over \gamma^4} = {2\pi^2 \over g^2C_2 \sigma^2}{f_2(\alpha) \over f_1(\alpha)^2} K .
\end{eqnarray}
If the ratio $\eta^4/\gamma^4$ is in the range $[0,1/4)$, the QCD vacuum is the type II dual superconductor.  Then $\sigma=0$ seems to be excluded.
If this ratio is zero, the QCD vacuum is the dual superconductor in the London limit.
On the other hand, if the ratio $\eta^4/\gamma^4$ approaches $1/4$, the dual superconductor is on the border between type II and type I.
The Bogomolny limit $m_1=m_2$ is achieved if the ratio is 
$\eta^2/\gamma^2=1/2$.  
The function $f_2(\alpha)/f_1(\alpha)^2$ is positive for $\alpha> \alpha_0$ and it has a peak $0.21$ at $\alpha \cong 2.65$, although it approaches zero slowly as $\alpha$ increases.
The negative factor can come from $\ln {M_A^2 \over \mu^2}$ in $K$.
We can reproduce both types of the dual superconductor by choosing $\mu$. 
depending on the ratio $M_A/\mu$.

\par
Finally, we discuss how the above results change if we consider the effect of ${\cal D}[\partial]$.
\begin{eqnarray}
  D_m(p^2)  {\cal D}[p]^2 &=&
 \left( {1 \over p^2} - {1 \over \eta^2} +{p^2 \over \gamma^4} \right)
\left( 1- {p^2 \over \eta^2} - {p^4 \over \gamma^4} \right)^2
\nonumber\\
&=& {1 \over p^2} - {3 \over \eta^2} + p^2 \left( {3 \over \eta^4} - {1 \over \gamma^4} \right) ,
\end{eqnarray}
Therefore, the effect is equivalent to the following replacement of the parameters,
\begin{equation}
 \eta^2 \rightarrow {1 \over 3}\eta^2, \quad
 {1 \over \gamma^4} \rightarrow {3 \over \eta^4} - {1 \over \gamma^4} .
\end{equation}
Consequently, $m_1, m_2, \chi$ are modified as
\begin{eqnarray}
 m_{1,2}^2 &=& {\eta^2 \over 2 \left(1 - {1 \over 3}{\eta^4 \over \gamma^4}\right)} \left( 1 \pm {2 \over 3}
\sqrt{{\eta^4 \over \gamma^4}-{3 \over 4}} \right) ,
\\
\chi &=& {1 \over 2}{\eta^2 \over \sqrt{{\eta^4 \over \gamma^4}-{3 \over 4}}}.
\end{eqnarray}
For $m_{1,2}^2$ to be real and positive, $\eta^4/\gamma^4$ must be in the range $[3/4,3)$.  The type II theory lies in between the London limit $\eta/\gamma \rightarrow 0$ and the Bogomolny limit $\eta^4/\gamma^4 \rightarrow 3/4$.
Therefore, the dual superconductivity of type II  is excluded.  This result suggests that the dual superconductivity of QCD is of type I or on the border between type I and type II.
Therefore, the correction is important to determine the type of dual supercoductivity. 
The above result will be compared with the lattice results in section~\ref{sec:numerical}.
\par
A simple estimation is as follows. When $\alpha=1, \sigma=1$, $\eta$ and $\gamma$ are 
\begin{eqnarray}
  \eta^2   &=& 6 {2\pi^2 \over Ng^2}K M_A^2  = 10 F(g) M_A^2 ,
\nonumber\\
  \gamma^2 &=& {2\sqrt{30} \over \sqrt{Ng^2}} K^{1/2} M_A^2 
= 10 F(g)^{1/2} M_A^2 ,
\end{eqnarray}
and the ratio is
\begin{eqnarray}
  {\eta^4 \over \gamma^4} = {2\pi^2 \over Ng^2}{3 \over 5} K 
=  {2\pi^2 \over Ng^2}{3 \over 5} \left[  1  + {24 \over 11} \sigma^2  
-  {N g^2 \over 2\pi^2}\sigma^2  \left( \ln {g^2 \over 16\pi^2} +1 \right) \right] := F(g) .
\end{eqnarray}
The function $F=F(g)$ is monotonically decreasing in $g$ and goes into the negative region for $g > 9.6$ which is beyond the reach of the analyses of this paper.
Finally, the string tension in SU(2) Yang-Mills theory is expressed as a function of the Yang-Mills coupling constant $g$ as
\begin{eqnarray}
  {\sigma_{st} \over M_A^2} = {J^2g^2 \over 2\pi}   5 ({7 \over 6})^2 K(g){F(g) \over \sqrt{F(g)-{3 \over 4}}} \ln {1+{2 \over 3}\sqrt{F(g)-{3 \over 4}} \over 1-{2 \over 3}\sqrt{F(g)-{3 \over 4}}} ,
\end{eqnarray}
where $F(g):={2\pi^2 \over Ng^2}{3 \over 5}K(g)$.  More details will be given in a subsequent paper.

\section{\label{sec:string}Confining string theory and string tension}
\setcounter{equation}{0}

Finally, we derive the confining string theory as a LEET of Gluodynamics.  The final expression of the string action indicates the area law of the Wilson loop or the linear static interquark potential where the string tension is represented as a proportional constant.
It is an old idea that the confining phase of gauge theories can be formulated as a string theory, see a review\cite{Polchinski92}.  Especially, the large-$N$ QCD might be exactly reformulated as a string theory\cite{tHooft74a}.  
For earlier approaches of QCD (gluon) string in the last century, see the references \cite{oldstring,Polyakov86,Kleinert86,David89,PWZ93,Lee93,Orland94,SY95,Polyakov96}.
\par
We can decompose the phase variable $\theta$ as $\theta:=\theta^r+\theta^s$ where $\theta^r$ is the regular piece and $\theta^s$ the singular piece.  Then it is shown \cite{ACPZ96} that the integration measure ${\cal D}\theta$ over the field $\theta$ factorizes into the product of measures, i.e.,
${\cal D}\theta={\cal D}\theta^r {\cal D}\theta^s$. 
The singularity of the phase of the scalar field $\phi$ just takes place at the string world sheet.
The location of the world sheet, $x=x(\sigma)$, is determined by the condition
$
 \phi(x(\sigma))=0
$
which implies $|\phi(x)|=0$ on the world sheet where the angle $\theta(x)$ is not determined uniquely.
The singular piece $\theta^s(x)$ describes a configuration of the vortex string whose world-sheet obeys the equation,

\begin{equation}
  {1 \over 4\pi} \epsilon^{\mu\nu\rho\sigma} (\partial_\rho \partial_\sigma - \partial_\sigma \partial_\rho )  \theta^s(x) 
= \Theta_{\mu\nu}(x) .
\label{rel}
\end{equation}
On the other hand, the regular piece $\theta^r$ describes a single-valued fluctuation around the string configuration just mentioned above.
From the correspondence (\ref{rel}), the integration over $\theta^s$ corresponds to the integration over the world-sheet $x_\mu(\sigma)$ of the string, so that the integration measure is transformed as
\begin{equation}
 {\cal D}\theta^s(x) \rightarrow {\cal D}x_\mu(\sigma) J[x] ,
\end{equation}
where $J[x]$ is the Jacobian for the change of the integration variables.%
\footnote{The Jacobian $J[x]$ has been evaluated in \cite{ACPZ96} for the world-sheet of the closed string.  The Jacobian exactly cancels the conformal anomaly.  Such a possibility was already suggested in the paper \cite{PS91}.
Therefore, the vortex string might be self-consistent in $D=4$ space-time dimensions at least in the long-distance limit.  See also reference \cite{Alvarez81,Arvis83,Olesen85,BS00}.}
\par
We return to the expression of the VEV of the Wilson loop (\ref{vevh}) for  the action (\ref{APEGTh}), i.e.,
\begin{equation}
  S_{APEGT} = (h, \Delta D_m h)
 + 2Jg\rho^{-1}K^{1/2} (h, \tilde \Theta) .
\end{equation}
Performing the Gaussian integration over the Kalb-Ramond field $h_{\mu\nu}$,
we obtain the an action written in terms of the vorticity tensor,, 
\begin{eqnarray}
 \mbox{\fboxsep=.1in \framebox{$\displaystyle
  S_{cs} = \int d^4x \int d^4y (2Jg\rho^{-1}K^{1/2})^2  
\tilde \Theta_{\mu\nu}(x) \left( {\chi \over \Delta-m_2^2} - {\chi  \over \Delta-m_1^2} \right)(x,y) \tilde \Theta^{\mu\nu}(y)  ,
$}}
\label{csaction}
\end{eqnarray}
for the expectation value of the Wilson loop,
\begin{eqnarray}
 \mbox{\fboxsep=.1in \framebox{$\displaystyle
    \langle W(C) \rangle_{YM} 
= Z_{cs}[C]/Z_{cs}[0],
\quad 
 Z_{cs}[C] = \int {\cal D}x_\mu(\sigma) J[x] \exp \{i S_{cs}[x] \} .
$}}
\end{eqnarray}
Here $m_2,m_2,\chi$ are the same as those defined in section~\ref{sec:monopole},
\begin{eqnarray}
&&
\mbox{\fboxsep=.1in \framebox{$\displaystyle
 m_{1,2}^2 := {\gamma^4 \over 2\eta^2} \left( 1 \pm \sqrt{1-4\eta^4/\gamma^4} \right) \quad (m_1 \ge m_2) ,
$}}
\\
&&
\mbox{\fboxsep=.1in \framebox{$\displaystyle
 \chi := {m_1^2 m_2^2 \over m_1^2-m_2^2} 
= {\eta^2 \over \sqrt{1-4\eta^4/\gamma^4}} .
$}}
\end{eqnarray}

\par
\par
Let $\sigma=(\sigma^1,\sigma^2)$ be a two-dimensional coordinate on the world sheet $x_\mu=x_\mu(\sigma)$.  Then the infinitesimal surface element,
\begin{equation}
  dS_{\mu\nu}(x(\sigma))=\sqrt{g(\sigma)}t_{\mu\nu}(\sigma) d^2\sigma ,
\end{equation}
is expressed by the determinant $g(\sigma)=\det||g_{ab}(\sigma)||$ calculated from the induced metric tensor of the surface defined by
\begin{equation}
 g_{ab}(\sigma)=\partial_a x_\mu(\sigma) \partial_b x_\mu(\sigma)
\end{equation}
with the derivative,
\begin{equation}
 \partial_a = {\partial \over \partial \sigma^a} \quad (a=1,2) ,
\end{equation} 
and the so-called extrinsic curvature tensor of the surface,
\begin{equation}
  t_{\mu\nu}(\sigma)={\epsilon^{ab} \over \sqrt{g(\sigma)}} \partial_a x_\mu(\sigma) \partial_b x_\nu(\sigma) .
\end{equation}
\par
Given the theory with the action  of the form,
\begin{equation}
  S =    (\Theta, {\kappa \over \Delta^2-m^2}\Theta) ,
\end{equation}
the low-energy limit is obtained by performing a derivative expansion of this action.  The derivative expansion is equivalent to an expansion in powers of $1/m$.
This procedure is well know in the literatures, see e.g. \cite{KC96,DQT96,DT97,AES96,AE98,AE99,Antonov99} and
Appendix~\ref{sec:dexp}.  The result is the Nambu-Goto action with a rigidity term,
\begin{eqnarray}
  S_{cs} = \sigma_{st}^0 \int_S d^2 \sigma \sqrt{g} + {1 \over \alpha_0} \int d^2 \sigma \sqrt{g} g^{ab}\partial_a t_{\mu\nu} \partial_b t^{\mu\nu} 
+ \kappa_t \int d^2\sigma \sqrt{g} R + \cdots .
\label{naivecs}
\end{eqnarray}
The string tension of the Nambu-Goto term is given by
\begin{equation}
  \sigma_{st}^0  =   {\kappa \over 4\pi} K_0 \left( {m \over \Lambda} \right) ,
\label{naivest}
\end{equation}
where $K_0(x)$ is the modified Bessel function and $\Lambda$ is the ultraviolet cut-off.%
\footnote{In the setting up of this paper, the role of the ultraviolet cutoff $\Lambda$ is played by the Higgs mass $m_H/\sqrt{2}=m_1$, as shown below.
}
Moreover, it has been shown that the coefficient of the extrinsic curvature term is  a negative constant,
\begin{equation}
 {1 \over \alpha_0} = - {1 \over 128\pi} < 0 ,
\end{equation}
which is independent of $\kappa, m, \Lambda$.

\par
In the naive confining string theory based on the action (\ref{naivecs}), the string tension diverges $\sigma_{st} \rightarrow \infty$ as $\Lambda \rightarrow \infty$, since $K_0(0)=+\infty$.  This pathology can be automatically avoided in the confining string theory derived in this paper.  It is easy to see that the rigidity term cancels in the confining string theory (\ref{csaction}).
Therefore, the action (\ref{csaction}) is cast into the confining string action,  
\begin{eqnarray}
 \mbox{\fboxsep=.1in \framebox{$\displaystyle
  S_{cs} = \sigma_{st} \int_S d^2 \sigma \sqrt{g} 
+ {1 \over \alpha_0} \int d^2 \sigma \sqrt{g} g^{ab}\partial_a t_{\mu\nu} \partial_b t^{\mu\nu} 
+ \kappa_t \int d^2\sigma \sqrt{g} R + \cdots  ,
$}}
\end{eqnarray}
with the string tension,
\begin{equation}
  \sigma_{st} = (2Jg\rho^{-1}K^{1/2})^2 {\chi \over 4\pi} \left[ K_0 \left( {m_2 \over \Lambda} \right) - K_0 \left( {m_1 \over \Lambda} \right) \right] .
\end{equation}
Note that the asymptotic behavior of the modified Bessel function $K_0(z)$ for $z \ll 1$ is given by
\begin{equation}
 K_0(z) \cong - (\gamma_E + \ln {z \over 2}) = \ln {2e^{-\gamma_E} \over z} ,
\end{equation}
with $\gamma_E$ being Euler's constant $\gamma_E=0.5772\cdots$.
Thus, for sufficiently large $\Lambda$, we obtain the $\Lambda$-independent {\it finite} result,%
\footnote{The string tension is a free energy per unit length of the string.  It is well know that the free energy has a logarithmic dependence in the Ginzburg-Landau theory.
The London limit corresponds to $m_1 \rightarrow \infty$.  Hence, the string tension reduces to the expression,
$
\sigma_{st} = (2Jg\rho^{-1}K^{1/2})^2 {\chi \over 4\pi} K_0 \left( {m_2 \over \Lambda} \right) ,
$
since
$K_0(m_1/\Lambda) \rightarrow 0$.
}
\begin{equation}
 \mbox{\fboxsep=.1in \framebox{$\displaystyle
 \sigma_{st}  
\cong {(2Jg)^2  \over 4\pi} \rho^{-2}K \chi \ln \left( {m_1 \over m_2} \right) ,
$}}
\end{equation}
where $J=1/2$ for the fundamental and $J=1$ for the adjoint quark (sources) in the case of $SU(2)$, and $J=1/3$ for the fundamental quark in the case of $SU(3)$.
The string tension just obtained agrees with that obtained from the monopole action (\ref{mpaction2}).
The mass $m_1$ can be viewed as a dual Higgs mass $m_H$ and $1/m_1$ corresponds to a finite thickness of the string.  The limit $m_1 \rightarrow \infty$ corresponds to the London limit, the thin string. 
This is consistent with a fact that the naive confining string theory with the string tension (\ref{naivest}) is  obtained from the DAH model in the London limit. 
\par
When the theory is near the London limit $m_1 \gg m_2$, 
the expression of the string tension reduces to
\begin{equation}
  \sigma_{st}  
\cong 
 {(2Jg)^2  \over 8\pi} \rho^{-2}K  m_b^2 \ln \left( {m_H \over m_b} \right)^2 ,
\end{equation}
which is similar to the result obtained by Suganuma, Sasaki and Toki \cite{SST95}.
In this case, it is again cast into the form,
\begin{equation}
  \sigma_{st}  
\cong {(2Jg)^2  \over 4\pi} \rho^{-2}K  m_b^2 K_0 \left(\sqrt{2} {m_b \over m_H} \right) ,
\end{equation}
which agrees with the result of Suzuki \cite{Suzuki88} up to a numerical factor.  Thus our results agree with those predicted based on the hypothetical DGL theory.
This fact also supports that the DGL theory is one of the LEET's of Gluodynamics.
\par
The coefficient of the rigidity term is a negative constant,
\begin{eqnarray}
 \alpha_0^{-1}  = - J^2g^2 \rho^{-2}K {1 \over \pi}  <0  .
\end{eqnarray}
Finally, the $\kappa$ is a positive constant,
\begin{eqnarray}
 \kappa = {2 \over 3} J^2g^2 \rho^{-2}K {1 \over \pi}  
= - {2 \over 3}\alpha_0^{-1} > 0  .
\end{eqnarray}
\par
String theory was originally developed as dual resonance models to explain hadronic physics.  It was almost abandoned after the invention of QCD and the discovery of asymptotic freedom.  Nevertheless, string theory might be useful by offering an alternative but tractable method of solving the strong coupling problem at long distance or low energies to which QCD has not yet given the definite answer.
Indeed, it is known that ordinary strings have the notorious troubles such as tachyons and conformal anomaly (critical dimensions). However, they disappear asymptotically at large distance as shown e.g. by Olesen\cite{Olesen85}.  
In other words, a string theory becomes a consistent model at large distances, although it is in a strict sense inconsistent in four dimensions.  
Therefore, the large-distance QCD can be described by a string model.  Of course, such a description breaks down at some distance, since QCD does not contain tachyons and is Lorentz invariant in spacetime dimensions $D$ less than or equal to four.   
In view of this, it should be worthwhile to mention that the static potential for the Nambu-Goto model (i.e., bosonic string model with Nambu-Goto action) has been computed to leading order in the $1/D$ expansion by Alvarez \cite{Alvarez81} and exactly in the whole range of $R$ by Arvis \cite{Arvis83} in arbitrary dimension $D$,
\begin{eqnarray}
 V(R) = \sigma_{st} \left( R^2 - R_c^2 \right)^{1/2} ,
\quad 
R_c = \sqrt{{\pi(D-2) \over 12\sigma_{st}}} .
\end{eqnarray}
In the long-distance region $R>R_c$, $V(R)$ is expanded into
\begin{eqnarray}
 V(R) = \sigma_{st} R - {\pi(D-2) \over 24R} + (1/R^{3}) .
\end{eqnarray}
For the large $R$, therefore, the static interquark potential is given by the linear potential where the string tension $\sigma_{st}$ is the proportional coefficient.
Moreover, the static potential has an additional long-distance Coulomb term which agrees with the earlier observation of L\"uscher, Symanzik and Weisz \cite{LSW80}.
The long-distance Coulomb term is the universal term depending only on the dimensionality of spacetime due to the transverse displacement $x_T$ of the string with fixed end points, under the assumption that $x_T$ is the only relevant dynamical variable at large distances (This is not the case for the superstring \cite{Olesen85}). 
The second term should not be confused with the short-distance Coulomb potential which is consistent with the asymptotic freedom of QCD.  
We see that the expression of V(R) loses the meaning at short distance $R<R_c$.  
\par
In this paper we have shown that the Nambo-Goto action can be a piece of the effective string action as a LEET of QCD.  Therefore, QCD should possess a long-distance Coulomb potential.
If the string theory which is capable of describing the whole energy range of QCD exists, the string theory must reduce in the long-distance limit to the above string theory derived in this paper.
\par
The large $N$ expansion can give another link between QCD and string theory, as pointed out by 't Hooft\cite{tHooft74a}.
The leading order of this expansion is some kind of free string theory that has yet to be identified.  In a free string theory, the surfaces in the sum should be dominated by smooth surfaces with no surface tension and without self-intersections.  
The Nambu-Goto model describes fundamental string without a transverse extension.  The string describing color electric flux tubes in QCD must be thick strings with a fundamental  transverse length scale $m_H^{-1}$ (The London limit $m_H \rightarrow \infty$ corresponds to a thin string) \cite{BS00}. If so, the string action should be responsible to the bending rigidity due to the finite width of the string.  
In order to take into account these features, the string model with the extrinsic curvature (the so-called rigid string) has been introduced by Polyakov \cite{Polyakov86} and Kleinert \cite{Kleinert86}.  The extrinsic curvature stiffness was expected to suppress the crumpled surface with a large number of self-intersectoins. 
However, it turns out that the new term is infrared irrelevant, see \cite{DT97} and references therein for more details.
\par
Recently, new string theories (so-called the confining string theory) of describing the confining phase in gauge theories were proposed by Polyakov \cite{Polyakov96} and by Kleinert and Chervyakov \cite{KC96}.  The confining string theory has a non-local interaction between world-sheet elements and a negative stiffness.  The confining string theories are very promising, since they seem to solve all the problems of rigid strings.  Especially, a negative stiffness is crucial in order to match the correct high-temperature behavior of large $N$ QCD \cite{PY92,KC96}.
The confining string theory can be explicitly derived for Abelian gauge theories, compact U(1) gauge theory \cite{DQT96}, Abelian Higgs model\cite{AE98} and so on, see a review \cite{Antonov99}.
In this paper we have derived the confining string with a negative stiffness directly from QCD at least in the low-energy regime.  This is performed by integrating out the antisymmetric tensor field in the improved version of the APEGT (originally proposed by the author in the paper\cite{KondoI}) which was derived directly from QCD.
The action realizes explicitly the necessary zig-zag invariance of confining string \cite{Polyakov97,AGO98}.
 
\section{\label{sec:numerical}Parameter fitting for numerical estimation}
\setcounter{equation}{0}

For the numerical estimation of physical quantities, we can use the following values which seem to be mutually consistent.

\subsection{Dual Ginzburg-Landau theory}

We use the values suggested in \cite{Suzuki88} 
For the dimensionful quantities,
\begin{eqnarray}
  m_b = \sqrt{2}g_m v = \sqrt{2}{4\pi \over g}v  \sim 0.88 \mbox{GeV}, 
\quad m_H = 2\sqrt{\lambda} v \sim 18 \mbox{GeV},  \quad v \sim 0.1  \mbox{GeV},  
\end{eqnarray}
and for the dimensionless coupling constants,
\begin{eqnarray}
  \alpha_s := {g^2 \over 4\pi} \sim 0.24 \ (g \sim 1.7) .  
\quad  \lambda \sim 8 \times 10^3 .
\end{eqnarray}
These values are consistent with the string tension,
\begin{eqnarray}
\sigma_{st} \sim (0.42  \mbox{GeV})^2 \sim 0.18 ( \mbox{GeV})^2 .
\end{eqnarray}
The off-diagonal gluon mass obtained by Monte Carlo simulation \cite{AS99} for $SU(2)$ is
\begin{eqnarray}
 M_A = 1.2  \mbox{GeV} . 
\end{eqnarray}

\subsection{Monopole action}

In the paper \cite{maKanazawa} the following lattice monopole action was adopted,
\begin{equation}
 S[k] = \sum_{s,s',\mu} k_\mu(s) D_0(s-s') k_\mu(s') ,
\end{equation}
where $D_0$ is parameterized by three parameters as
\begin{equation}
 D_0(s-s') = \bar \alpha \delta_{s,s'} + \bar \beta \Delta_L^{-1}(s-s') + \bar \gamma \Delta_L(s-s') ,
\end{equation}
with the lattice Laplacian 
$
 \Delta_L(s-s') := - \partial \partial'.
$
This leads to 
$
 D_0(p) = \bar \alpha + \bar \beta/p^2 + \bar \gamma p^2 .
$
Its inverse is 
\begin{equation}
 D_0^{-1}(p) = \kappa \left( {m_1^2 \over p^2+m_1^2} - {m_2^2 \over p^2+m_2^2} \right)  ,
\end{equation}
where
\begin{equation}
 m_1^2+m_2^2 = \bar \alpha/\bar \gamma, \quad
m_1^2 m_2^2 = \bar \beta/\bar \gamma, \quad
\kappa := \bar \gamma^{-1}/(m_1^2-m_2^2) .
\end{equation}
The string tension is obtained as $\sigma_{tot} = \sigma_{cl} + \sigma_{q}$ with
\begin{equation}
 \sigma_{cl} = {\pi \over 2} \kappa  \ln {m_1 \over m_2} ,
\end{equation}
and $\sigma_{q}$ being negligible.
The results of \cite{maKanazawa} are
\begin{equation}
 m_1 \cong 1.0 \times 10^4, \quad m_2 \cong 12, \quad \kappa = 4.83 ,
\end{equation}
and the parameters of the monopole action are obtained as
\begin{equation}
 \bar \alpha = 0.207 (0.435), \quad
\bar \beta = 2.49, \quad
\bar \gamma = 2.07 (9.15) \times 10^{-5} ,
\end{equation}
and
\begin{equation}
 \sigma_{phys} \cong (0.44 \mbox{GeV})^2, \quad \sigma_{phys}/\sigma_{cl} \sim (1.4)^{-2} \sim 0.51 .
\end{equation}
Note that our parameterization is
\begin{equation}
m_1^2+m_2^2=\gamma^4/\eta^2, \quad m_1^2m_2^2=\gamma^4, \quad
 m_1^2-m_2^2=(\gamma^4/\eta^2)\sqrt{1-4\eta^4/\gamma^4} .
\end{equation}
Hence the correspondence of the parameters in our theory to the lattice result \cite{maKanazawa} are given by
\begin{equation}
\eta^2 \rightarrow \bar \beta/\bar \alpha = 10, \quad
\gamma^4 \rightarrow \bar \beta/\bar \gamma = 10^5 .
\end{equation}
Hence we obtain
\
\begin{equation}
 m_1 \cong 10^2, \quad m_2 \cong 1, \quad \chi = 10 .
\end{equation}

\subsection{Confining string}

The confining string action has the following parameters (see section 2.1 of the paper \cite{Antonov99}),
\begin{equation}
 \sigma \cong 0.2 ( \mbox{GeV})^2, \quad \kappa_t \cong 0.003, \quad {1 \over \alpha_0} \cong - 0.005 .
\end{equation}
They are obtained from the so-called correlation length of the vacuum and the gluon condensate,
\begin{equation}
 T_g \cong 0.65 ( \mbox{GeV})^{-1} (T_g^{-1} \cong 1.5  \mbox{GeV}) ,
\quad
 \alpha_s \langle ({\cal F}_{\mu\nu}^A)^2 \rangle \sim 0.038 ( \mbox{GeV})^2 .
\end{equation}

\section{\label{sec:conclusion}Conclusion and discussion}
\setcounter{equation}{0}

In this paper, by improving the strategy suggested in the previous paper \cite{KondoI},  we have derived three equivalent LEET's of Gluodynamics, i.e., dual Ginzburg-Landau theory, magnetic monopole theory and confining string theory.  
Although each of them has already been proposed and analyzed as a LEET by various authors, we have given a first-principle derivation of these theories directly from Gluodynamics, i.e., Yang-Mills theory.  
In other words, we have shown their equivalence in this paper as is obvious from the derivation. 
Especially, we have shown that the monopole condensation occurs due to non-zero mass of off-diagonal gluons.
\par
The very origin of our nontrivial results is reduced to quantum correction to the diagonal fields $a_\mu$ and $B_{\mu\nu}$ arising from the massive off-diagonal gluons and off-diagonal ghosts.  This is a novel viewpoint examined in this paper in analyzing the low-energy Gluodynamics.  In the conventional analytical approaches, the off-diagonal components are completely neglected from the beginning by virtue of the infrared Abelian dominance.  However, we notice that it is the off-diagonal components that convey the characteristic properties of the original non-Abelian gauge theory to the LEET written in terms of the diagonal field alone. 
The reproduction of the $\beta$ function of the original Yang-Mills theory in the LEET is a good example of this fact.
\par
As a mechanism for mass generation of the off-diagonal component, we have used the ghost--anti-ghost condensation caused by the quartic ghost interaction in the modified MA gauge (section~\ref{sec:massgeneration}).
To author's knowledge, any other analytical derivation of off-diagonal gluon mass is not known for the Yang-Mills theory in the MA gauge.
However, the following steps after section~\ref{sec:massgeneration} can be performed irrespective of the origin of the off-diagonal gluon mass, once we regard the massive off-diagonal gluons.  Therefore, even if the off-diagonal gluons become massive due to other mechanism, we obtain the same LEET's as given in this paper.

\par
The LEET's of the SU(N) Yang-Mills theory have been obtained by way of the APEGT.
The APEGT is bilinear in the diagonal fields, $a_\mu$ and $B_{\mu\nu}$.  This result is regarded as the leading order result of the large $N$ expansion.  The (higher-order)  correction terms are suppressed by a factor $N^{-2}$.  Since we have obtained the confining string theory by rewriting the APEGT,
this is consistent with the claim that the Yang-Mills theory in the large $N$ limit is equivalent to a certain string theory \cite{tHooft74a}.
Therefore, our result may shed more light on the relationship between the gauge theory and the string theory.  However, we have used the derivative (or low-energy) expansion or weak field approximation \cite{Polyakov96} to derive the confining string theory.  Therefore, it is not clear where the multi-valuedness of the confining string action comes from.  Multi-valuedness is considered as a reflection of the compactness of the residual Abelian gauge group which is embedded in the original compact non-Abelian gauge group.  This problem will be discussed elsewhere.

\par
By way of LEET's, we have succeeded to calculate the string tension of QCD (gluon) string.  The non-zero value of the string tension implies area law of the Wilson loop, i.e., quark confinement.
However, the string tension obtained in this way depends on the parameters $\rho, \sigma$ which were introduced to rewrite the Yang-Mills theory into the dual Abelian gauge theory.  It is possible for the string tension to be independent of the renormalization scale $\mu$.
In a subsequent paper \cite{preparation}, we will discuss in detail the issue how those parameters  are determined within the same framework as that proposed in this paper. 
 We will give a simple model (toy model for the gauge theory) in which the role of parameters introduced in the auxiliary field formalism is clarified in more tractable way.
We will also give more quantitative argument so that the LEET's derived in this paper can reproduce the experimental values of physical quantities and predict unobserved quantities, e.g., the glueball mass.  We hope to discuss the spontaneous chiral symmetry breaking which is expected to occur simultaneously with quark confinement.

\section*{Acknowledgments}
The author would like to thank Atsushi Nakamura for helpful discussion on Appendix~\ref{sec:tensorfield} and bringing his attention to the reference \cite{BS96}, and Seikou Kato for sending the reference \cite{maKanazawa} prior to the publication.  He is grateful to Takahito Imai for drawing Figures 3, 4, 5 and 6.  He also thanks Toru Shinohara, Takeharu Murakami and T.-S. Lee for helpful discussions.
This work is supported in part by
the Grant-in-Aid for Scientific Research from the Ministry of
Education, Science and Culture (10640249).

\appendix
\section{\label{sec:formula}Useful formulae}
\setcounter{equation}{0}

\subsection{Structure constants}
Note that
\begin{equation}
 f^{ACD}f^{BCD} = C_2 \delta^{AB} ,
\label{f1}
\end{equation}
where $C_2$ is called the quadratic Casimir (operator).
This implies that
\begin{equation}
 f^{icd}f^{jcd} = C_2 \delta^{ij} ,
\label{f2}
\end{equation}
since the structure constant with two and three diagonal indices are zero, $f^{ijc}=0=f^{ijk}$ according to 
$f^{ABC}=-2\sqrt{-1}{\rm tr}\{[T^A,T^B]T^C \}$.
Combining the identity,
\begin{equation}
 f^{i a c} f^{i b c} = \delta^{ab} ,
\label{f3}
\end{equation}
with (\ref{f1}) leads to
\begin{equation}
 f^{acd}f^{bcd} = (C_2-2) \delta^{ab} .
\label{f4}
\end{equation}

\subsection{Differential forms}

For the p-form, 
\begin{equation}
 \omega := {1 \over p!} \omega_{\mu_1 \cdots \mu_p} dx^{\mu_1} \wedge \cdots \wedge dx^{\mu_p} ,
\end{equation}
the dual form $*\omega$ in four dimensional Minkowski space is defined by
\begin{eqnarray}
 *\omega :&=& {1 \over (4-p)!} *\omega_{\mu_1 \cdots \mu_{4-p}} dx^{\mu_1} \wedge \cdots \wedge dx^{\mu_{4-p}} ,
\\
 *\omega_{\mu_1 \cdots \mu_{4-p}} &:=& {1 \over p!} \epsilon_{\mu_1 \cdots \mu_{4-p}\nu_1 \cdots \nu_p} \omega^{\nu_1 \cdots \nu_p} .
\end{eqnarray}
The identity,
\begin{eqnarray}
 \epsilon_{\mu_1 \cdots \mu_{p}\alpha_1 \cdots \alpha_{4-p}}
\epsilon^{\mu_1 \cdots \mu_{p}\beta_1 \cdots \beta_{4-p}}
= - g^{-1} p!(4-p)! \delta^{\alpha_1}_{[\beta_1} \cdots \delta^{\alpha_{4-p}}_{\beta_{4-p}]} ,
\end{eqnarray}
leads to
\begin{equation}
 **\omega = g^{-1} (-1)^{p} \omega ,
\end{equation}
since  
\begin{eqnarray}
\epsilon^{\mu_1 \cdots \mu_{p}\beta_1 \cdots \beta_{4-p}}
  = g^{-1} \epsilon_{\mu_1 \cdots \mu_{p}\beta_1 \cdots \beta_{4-p}},
\quad g := \det(g_{\mu\nu}) .
\end{eqnarray}
Note that $g=-1$ for Minkowski spacetime with a Lorentz metric, while
$g=1$ for Euclidean space.

\subsection{Integration formula by dimensional regularization}

Define
\begin{equation}
  \epsilon := 2-{D \over 2} 
\end{equation}
In Minkowski space, 
\begin{equation}
  \int {d^Dk \over i(2\pi)^D} \ln (m^2+2p \cdot k-k^2)
= - {\Gamma(\epsilon-2) \over (4\pi)^{2-\epsilon}} (m^2+p^2)^{2-\epsilon} ,
\end{equation}
\begin{equation}
 \int {d^Dk \over i(2\pi)^D} {1 \over (m^2+2p \cdot k-k^2)^a}
=  {\Gamma(\epsilon+a-2) \over (4\pi)^{2-\epsilon}\Gamma(a)} 
(m^2+p^2)^{2-a-\epsilon} ,
\end{equation}
\begin{equation}
 \int {d^Dk \over i(2\pi)^D} {k_\mu \over (m^2+2p \cdot k-k^2)^a}
=  {\Gamma(\epsilon+a-2) \over (4\pi)^{2-\epsilon}\Gamma(a)} 
p_\mu (m^2+p^2)^{2-a-\epsilon} ,
\end{equation}
\begin{eqnarray}
 && \int {d^Dk \over i(2\pi)^D} {k_\mu k_\nu \over (m^2+2p \cdot k-k^2)^a}
\nonumber\\
&=&  {1 \over (4\pi)^{2-\epsilon}\Gamma(a)} \left[
\Gamma(\epsilon+a-2) p_\mu p_\nu (m^2+p^2)^{2-a-\epsilon}  
- \Gamma(\epsilon+a-3){1 \over 2}g_{\mu\nu}(m^2+p^2)^{3-a-\epsilon} \right],
\nonumber\\
\end{eqnarray}
where $m^2$ is understood as $m^2-i\delta, \delta>0$, i.e., $Im (m^2) <0$.

\subsection{Gamma function}

The Laurent expansion of the Gamma function is as follows.
\begin{eqnarray}
 \Gamma(\epsilon) &=& {1 \over \epsilon} - \gamma_E + O(\epsilon) ,
\\
 \Gamma(\epsilon-1) &=& - {1 \over \epsilon} + \gamma_E -1  + O(\epsilon) ,\\
 \Gamma(\epsilon-2) &=&   {1 \over \epsilon} - \gamma_E + {3 \over 2}  + O(\epsilon) ,
\\
 \Gamma(\epsilon-n) &=& {(-1)^n \over n!}\left[ {1 \over \epsilon} - \gamma_E + \sum_{k=1}^{n}{1 \over k} \right]  + O(\epsilon) ,
\end{eqnarray}
where $\gamma_E$ is Euler's constant $\gamma_E=0.5772\cdots$.

\section{\label{sec:NAST}Derivation of a version of non-Abelian Stokes theorem}
\setcounter{equation}{0}

In an expression of the non-Abelian Stokes theorem,
\begin{equation}
 W(C) = \int d\mu_C(\xi) \exp \left[ig \int_{S_C} dS^{\mu\nu} f_{\mu\nu}^\xi(x) \right] ,
\end{equation}
the argument of the exponential is rewritten as follows. The target space coordinate of the surface spanned by the Wilson loop $C$ is denoted by $x(\sigma)$ where $\sigma$ is the world sheet coordinate.  The surface integral is rewritten in terms of the vorticity tensor as 
\begin{eqnarray}
 {1 \over 2}  \int_{S_C} dS^{\mu\nu}(x(\sigma)) f_{\mu\nu}(x(\sigma))  
&=& \int d^4x \Theta^{\mu\nu}(x) f_{\mu\nu}(x)
\\
&:=& (\Theta,f) = (*\Theta,*f) 
\nonumber\\
&=& (*\Theta,\Delta^{-1}(d\delta+\delta d)*f)
\nonumber\\
&=& (*\Theta,\Delta^{-1}d\delta *f) + (*\Theta,\Delta^{-1}\delta d*f)
\nonumber\\
&=& (\delta \Delta^{-1} *\Theta, \delta *f) + (\Theta,*\Delta^{-1}\delta * \delta f) \nonumber\\
&=& (\delta \Delta^{-1} *\Theta \delta *f) + (\Theta,\Delta^{-1}d  \delta f)  
\nonumber\\
&=& (\delta \Delta^{-1} *\Theta, k) + (\Delta^{-1}\delta \Theta,  j)  ,
\end{eqnarray}
where $k:=\delta *f$ is the magnetic monopole current and 
$j:=\delta f$ is the electric current.
Assuming the absence of the electric source $j=0$, we obtain
\begin{eqnarray}
 W(C) 
= \int d\mu_C(\xi) \exp \left[ ig (\Xi, k^{\xi}) + i g(N,j^\xi) \right] ,
\end{eqnarray}
where $N$ is one-form defined by
\begin{eqnarray}
 \Xi := * d\Theta \Delta^{-1} = \delta *\Theta \Delta^{-1} , \quad
 N := \delta \Theta \Delta^{-1} .
\end{eqnarray}
with the components,
\begin{eqnarray}
 \Xi^\mu(x) &=& {1 \over 2} \epsilon^{\mu\nu\rho\sigma} \partial_\nu^x \int d^4y \Theta_{\rho\sigma}(y) \Delta^{-1}(y-x)
\\
&=& {1 \over 2} \epsilon^{\mu\nu\rho\sigma} \partial_\nu^x \int_S d^2 S_{\rho\sigma}(x(\sigma)) \Delta^{-1}(x(\sigma)-x) ,
\\
 N^\mu(x) &=& \partial_\nu^x \int d^4y \Theta^{\mu\nu}(y) \Delta^{-1}(y-x)
\\
&=& {1 \over 2} \partial_\nu^x \int_S d^2 S^{\mu\nu}(x(\sigma)) \Delta^{-1}(x(\sigma)-x) .
\end{eqnarray}

\section{\label{sec:vp}Calculation of the vacuum polarization for tensor fields}
\setcounter{equation}{0}

We shall evaluate the vacuum polarization of the tensor field $B$.  The contribution from Fig.~\ref{fig:APEGT} is calculated from
\begin{eqnarray}
 \Pi_{\mu\nu,\alpha\beta}^{ij}(k) &:=& {1 \over 2}\int {d^4p \over (2\pi)^4} D_{\sigma_1 \sigma_2}(p) \delta^{d_1d_2}
[-2\sigma gf^{ic_1d_1}I_{\mu\nu,\rho_1 \sigma_1}]
\nonumber\\
&& \times D_{\rho_1 \rho_2}(p+k)\delta^{c_1c_2}
[-2\sigma gf^{jc_2d_2}I_{\alpha\beta,\rho_2 \sigma_2}] ,
\end{eqnarray}
where 
\begin{equation}
 I_{\mu\nu,\alpha\beta} := {1 \over 2}(g_{\mu\alpha}g_{\nu\beta}-g_{\mu\beta}g_{\nu\alpha}) ,
\end{equation}
and the the off-diagonal massive gluon propagator  
$D_{\mu\nu}^{ab}(k)=\delta^{ab}D_{\mu\nu}(k)$ is given by
\begin{eqnarray}
D_{\mu\nu}(k) &:=& {1 \over k^2-M_A^2} \left[ g_{\mu\nu} - (1-\alpha){k_\mu k_\nu \over k^2-\alpha M_A^2} \right] 
\\
&=& {1 \over k^2-M^2} \left( g_{\mu\nu} - {k_\mu k_\nu \over M^2} \right) + {k_\mu k_\nu \over M^2}{1 \over k^2-\alpha M^2} 
\\
&=& {1 \over M^2} \left[ {M^2 g_{\mu\nu} - k_\mu k_\nu \over k^2-M^2}   +  {k_\mu k_\nu \over k^2-\alpha M^2} \right] .
\end{eqnarray}
Hence the additional term to the APEGT is given by
\begin{equation}
 \delta_{(1)}^c {\cal L}_{APEGT} = \int{d^4k \over (2\pi)^4} *B_{\mu\nu}^i(k)\Pi_{\mu\nu,\alpha\beta}^{ij}(k) *B_{\alpha\beta}^j(-k) .
\end{equation}
Using the identity,
\begin{equation}
  \delta^{d_1d_2}f^{ic_1d_1}  \delta^{c_1c_2}f^{jc_2d_2} 
=  f^{ic_1d_1} f^{jc_1d_1} = C_2 \delta^{ij} ,
\end{equation}
we obtain
\begin{eqnarray}
 \Pi_{\mu\nu,\alpha\beta}^{ij}(k) &:=& {[-2\sigma g ]^2 \over 2} 
C_2 \delta^{ij}
I_{\mu\nu,\rho_1 \sigma_1} I_{\alpha\beta,\rho_2 \sigma_2}
\int {d^4p \over (2\pi)^4} D_{\sigma_1 \sigma_2}(p)  D_{\rho_1 \rho_2}(p+k) 
\\
&=&  {2\sigma^2 g^2 C_2 \delta^{ij} \over M^4} 
I_{\mu\nu,\rho_1 \sigma_1} I_{\alpha\beta,\rho_2 \sigma_2}
\int {d^4p \over (2\pi)^4} 
\Biggr[ 
\nonumber\\&&
 {M^2 g_{\sigma_1\sigma_2} - p_{\sigma_1} p_{\sigma_2} \over p^2-M^2}
 {M^2 g_{\rho_1\rho_2} - (p+k)_{\rho_1} (p+k)_{\rho_2} \over (p+k)^2-M^2}
\label{p1}
\\&&
+  {M^2 g_{\sigma_1\sigma_2} - p_{\sigma_1} p_{\sigma_2} \over p^2-M^2}
{(p+k)_{\rho_1} (p+k)_{\rho_2} \over (p+k)^2-\alpha M^2} 
\\&&
+ {p_{\sigma_1} p_{\sigma_2} \over p^2-\alpha M^2} 
{M^2 g_{\rho_1\rho_2} - (p+k)_{\rho_1} (p+k)_{\rho_2} \over (p+k)^2-M^2}
\\&&
+ {p_{\sigma_1} p_{\sigma_2} \over p^2-\alpha M^2} 
{(p+k)_{\rho_1} (p+k)_{\rho_2} \over (p+k)^2-\alpha M^2} \Biggr] .
\end{eqnarray}
Using the Feynman parameter formulas, two denominators are combined into one denominator, e.g,
\begin{equation}
 {1 \over p^2-\alpha M^2}{1 \over (p+k)^2-\beta M^2}
 = \int_0^1 dx {1 \over [p^2+2xk \cdot p + \{ xk^2-(x\beta-x\alpha+\alpha)M^2 \} ]^2} ,
\end{equation}
we have
\begin{eqnarray}
 \Pi_{\mu\nu,\alpha\beta}^{ij}(k) 
&=&  {2\sigma^2 g^2 C_2 \delta^{ij} \over M^4} 
I_{\mu\nu,\rho_1 \sigma_1} I_{\alpha\beta,\rho_2 \sigma_2} \int_0^1 dx
\Biggr[ 
\nonumber\\&&
\int {d^4p \over (2\pi)^4}  {[M^2 g_{\sigma_1\sigma_2} - p_{\sigma_1} p_{\sigma_2}]
 [M^2 g_{\rho_1\rho_2} - (p+k)_{\rho_1} (p+k)_{\rho_2}] \over 
[p^2+2xk \cdot p + \{ xk^2-M^2 \} ]^2}
\label{int1}
\\&&
+ \int {d^4p \over (2\pi)^4}  {[M^2 g_{\sigma_1\sigma_2} - p_{\sigma_1} p_{\sigma_2}]  
[(p+k)_{\rho_1} (p+k)_{\rho_2}] \over 
[p^2+2xk \cdot p + \{ xk^2-(x\alpha-x+1)M^2 \} ]^2} 
\label{int2}
\\&&
+ \int {d^4p \over (2\pi)^4} {[p_{\sigma_1} p_{\sigma_2}]  
[M^2 g_{\rho_1\rho_2} - (p+k)_{\rho_1} (p+k)_{\rho_2}] \over
[p^2+2xk \cdot p + \{ xk^2-(x-x\alpha+\alpha)M^2 \} ]^2}
\label{int3}
\\&&
+ \int {d^4p \over (2\pi)^4} {[p_{\sigma_1} p_{\sigma_2}]  
[(p+k)_{\rho_1} (p+k)_{\rho_2}] \over
[p^2+2xk \cdot p + \{ xk^2-\alpha M^2 \} ]^2} \Biggr] .
\label{int4}
\end{eqnarray}
The momentum integration can be performed by making use of the formula in Appendix~\ref{sec:formula} where $\epsilon:=2-D/2$.
After straightforward but tedious calculations, we are lead to
\begin{eqnarray}
 && \Pi_{\mu\nu,\alpha\beta}^{ij}(k) 
\nonumber\\
&=&  
 - 2C_2 \sigma^2 g^2 \delta^{ij}  i I_{\mu\nu,\alpha\beta} \int_0^1 dx 
{(4\pi)^\epsilon \over (4\pi)^2 \Gamma(2)} 
\nonumber\\&& \times
\Biggr\{   \Gamma(\epsilon)[M^2-x(1-x)k^2]^{-\epsilon}
+ M^{-2}\Gamma(-1+\epsilon) [M^2-x(1-x)k^2]^{1-\epsilon}
\nonumber\\&&
- {1 \over 2}M^{-2}\Gamma(-1+\epsilon)
  [\{\alpha+(1-\alpha)x\}M^2-x(1-x)k^2]^{1-\epsilon}
\nonumber\\&&
- {1 \over 2}M^{-2}\Gamma(-1+\epsilon)
  [\{\alpha+(1-\alpha)x\}M^2-x(1-x)k^2]^{1-\epsilon} 
 \Biggr\} 
\nonumber\\
&&  
 - 2C_2 \sigma^2 g^2 \delta^{ij}  i 
{1 \over 2}k^2 (I-P)_{\mu\nu,\alpha\beta} 
\int_0^1 dx 
{(4\pi)^\epsilon \over (4\pi)^2 \Gamma(2)} 
\nonumber\\&& \times
\Biggr\{   -M^{-2}\Gamma(\epsilon)[M^2-x(1-x)k^2]^{-\epsilon}(2x^2+2x+1)
\nonumber\\&&
- {1 \over 2}M^{-4}\Gamma(-1+\epsilon) [M^2-x(1-x)k^2]^{1-\epsilon}
\\&&
+M^{-2}\Gamma(\epsilon)
  [\{\alpha+(1-\alpha)x\}M^2-x(1-x)k^2]^{-\epsilon}(x^2+2x+1)
\nonumber\\&&
+{1 \over 2}M^{-4}\Gamma(-1+\epsilon) 
  [\{\alpha+(1-\alpha)x\}M^2-x(1-x)k^2]^{1-\epsilon}
\\&&
+M^{-2}\Gamma(\epsilon)
  [\{\alpha+(1-\alpha)x\}M^2-x(1-x)k^2]^{-\epsilon}x^2
\nonumber\\&&
+{1 \over 2}M^{-4}\Gamma(-1+\epsilon)
  [\{\alpha+(1-\alpha)x\}M^2-x(1-x)k^2]^{1-\epsilon}
\\&&
-{1 \over 2}M^{-4}\Gamma(-1+\epsilon)
  [\alpha M^2-x(1-x)k^2]^{1-\epsilon}
 \Biggr\} ,
\end{eqnarray}
where the fourth term does not give the term proportional to $I_{\mu\nu,\alpha\beta}$.
Here we have used the following properties of the projection operator,
\begin{eqnarray}
I_{\mu\nu,\alpha\beta} &=& - I_{\nu\mu,\alpha\beta} = I_{\alpha\beta,\mu\nu}
= - I_{\mu\nu,\beta\alpha} ,
\\
I_{\mu\nu,\rho\sigma}I_{\rho\sigma,\alpha\beta} &=& I_{\mu\nu,\alpha\beta}, 
\\
k^{\rho_1} I_{\mu\nu,\rho_1\sigma} k^{\rho_2} I_{\alpha\beta,\rho_2\sigma}
&=& {1 \over 4}(k_\mu k_\alpha g_{\nu\beta}-k_\mu k_\beta g_{\nu\alpha}
-k_\nu k_\alpha g_{\mu\beta}+k_\nu k_\beta g_{\mu\alpha})
\nonumber\\
&=& {1 \over 2}k^2 (I-P)_{\mu\nu,\alpha\beta} ,
\end{eqnarray}
where we have introduced
\begin{equation}
 P_{\mu\nu,\alpha\beta} := {1 \over 2}(T_{\mu\alpha}T_{\nu\beta}-T_{\mu\beta}T_{\nu\alpha}), \quad 
T_{\mu\nu} := g_{\mu\nu} - {k_\mu k_\nu \over k^2} .
\end{equation}

Using the Laurent expansion of the Gamma function $\Gamma(x)$, 
we obtain
\begin{eqnarray}
&& \Pi_{\mu\nu,\alpha\beta}^{ij}(k) 
\\
&=& {-2C_2 \sigma^2 g^2 \over 16\pi^2}\delta^{ij}
I_{\mu\nu,\alpha\beta}  \epsilon^{-1} \int_0^1 dx [\alpha+(1-\alpha)x]  
\nonumber\\
&&+ {-2C_2 \sigma^2 g^2 \over 16\pi^2}\delta^{ij}
I_{\mu\nu,\alpha\beta} \int_0^1 dx \Biggr\{  \left[ -x(1-x){k^2 \over M^2} \right] \ln \left[  {M^2 \over \mu^2} - x(1-x){k^2 \over \mu^2} \right]
\nonumber\\
&&- \left[ \alpha+(1-\alpha)x-x(1-x){k^2 \over M^2} \right] 
\ln \left[ \{ \alpha+(1-\alpha)x \} {M^2 \over \mu^2} - x(1-x){k^2 \over \mu^2} \right] 
\nonumber\\
&&+ (\gamma_E-\ln 4\pi-1)(1-\alpha)(1-x) -\gamma_E + \ln 4\pi \Biggr\} 
\nonumber\\
&&+ {-2C_2 \sigma^2 g^2 \over 16\pi^2}\delta^{ij}
{1 \over 2} k^2 (I-P)_{\mu\nu,\alpha\beta}
\nonumber\\&& \times 
\Biggr[
  M^{-2} \int_0^1 dx (2x^2+2x+1) \Biggr\{ 
 \ln \left[ {M^2 \over \mu^2}-x(1-x){k^2 \over \mu^2} \right] 
\nonumber\\&& \quad\quad\quad\quad
    - \ln \left[ \{\alpha+(1-\alpha)x\}{M^2 \over \mu^2}-x(1-x){k^2 \over \mu^2} \right] \Biggr\}
\nonumber\\&&
- {1 \over 2}M^{-2} \int_0^1 dx \left[ 1 - x(1-x){k^2 \over M^2} \right] \ln \left[ {M^2 \over \mu^2} - x(1-x){k^2 \over \mu^2} \right] 
\nonumber\\&&
- {1 \over 2}M^{-2} \int_0^1 dx \left[ \alpha - x(1-x){k^2 \over M^2} \right] \ln \left[ \alpha{M^2 \over \mu^2} - x(1-x){k^2 \over \mu^2} \right] 
\nonumber\\&&
+ M^{-2} \int_0^1 dx \left[ \alpha+(1-\alpha)x - x(1-x){k^2 \over M^2} \right] 
\nonumber\\&& \times
\ln \left[ \{ \alpha+(1-\alpha)x \} {M^2 \over \mu^2} - x(1-x){k^2 \over \mu^2} \right] 
\Biggr]
+ O(\epsilon) .
\end{eqnarray}
Here note that the divergent term $\epsilon^{-1}k^2(1-P)$ does not exist.
In the neighborhood of $k^2=0$, i.e, in the low-energy region such that $k^2/M^2 \ll 1$,
\begin{eqnarray}
 && \Pi_{\mu\nu,\alpha\beta}^{ij}(k) 
\nonumber\\
&=&  
-{ 2C_2 \sigma^2 g^2 \over 16\pi^2}\delta^{ij}  
\Biggr\{ I_{\mu\nu,\alpha\beta} \Biggr[   \epsilon^{-1} {1+\alpha \over 2} 
+  f_0(\alpha) + f_1(\alpha) {k^2 \over M^2} + f_2(\alpha) {k^4
\over M^4} \Biggr]  
\nonumber\\
&&+ {1 \over 2}{k^2 \over M^2}(I-P)_{\mu\nu,\alpha\beta} \left[ h_0(\alpha)+h_1(\alpha){k^2 \over M^2}  \right]
\Biggr\} + O\left(  {k^6 \over M^6} \right) ,
\end{eqnarray}
where%
\footnote{
The piece proportional to $1-P$ leads to the perimeter law. 
Therefore, it is neglected when we obtain the string tension.
}
\begin{eqnarray}
 f_0(\alpha) &:=&  - {1+\alpha \over 2} \ln {M^2 \over \mu^2} 
+ (\gamma_E-\ln 4\pi -1)(1-\alpha){1 \over 2} - \gamma_E + \ln 4\pi
\nonumber\\&&
 - \int_0^1 dx \ [\alpha+(1-\alpha)x]  \ln [\alpha+(1-\alpha)x] 
\\
&=& - {1+\alpha \over 2} \ln {M^2 \over \mu^2} 
+ (\gamma_E-\ln 4\pi -1)(1-\alpha){1 \over 2} - \gamma_E + \ln 4\pi
\nonumber\\&&
+ {1 \over 1-\alpha} \left[ {\alpha^2 \over 2} \ln \alpha +
{1 \over 4}-{1 \over 4}\alpha^2 \right] ,
\\
 f_1(\alpha) &:=& \int_0^1 dx \ x(1-x) \{ 1 + \ln [\alpha+(1-\alpha)x] \}
\\
&=& {1 \over 6} + {1 \over (1-\alpha)^3} \Biggr\{ {\alpha^3 \over 3} \ln \alpha - {\alpha^3 \over 9} + {1 \over 9} - (1+\alpha) \left[ {\alpha^2 \over 2} \ln \alpha - {\alpha^2 \over 4} + {1 \over 4} \right] 
\nonumber\\&&
+ \alpha(\alpha \ln \alpha - \alpha +1) \Biggr\} ,
\\
 f_2(\alpha) &:=& \int_0^1 dx \ \left[ x^2(1-x)^2 - {1 \over 2}{x^2(1-x)^2 \over \alpha+(1-\alpha)x} \right] \\
&=& {1 \over 30} - {1 \over (1-\alpha)^5}\left[ 
 {1 \over 24} -{1 \over 3}\alpha + {1 \over 3}\alpha^3 - {1 \over 24}\alpha^4 
- {1 \over 2} \alpha^2 \ln \alpha
\right] ,  
\end{eqnarray}
and
\begin{eqnarray}
 h_0(\alpha) &:=&   \int_0^1 dx \Biggr\{ -{1 \over 2}\alpha \ln \alpha 
+ [ \alpha +(1-\alpha)x ] \ln [ \alpha +(1-\alpha)x] 
\nonumber\\&&
 - (2x^2+2x+1)  \ln [\alpha+(1-\alpha)x] \Biggr\} ,
\\
 h_1(\alpha) &:=& \int_0^1 dx \Biggr\{
 (2x^2+2x+1)  {(1-\alpha)x(1-x)^2 \over \alpha+(1-\alpha)x}  
\nonumber\\&&
+ {1 \over 2} x(1-x) \ln \alpha - x(1-x) \ln [\alpha+(1-\alpha)x] \Biggr\} .
\end{eqnarray}
For $\alpha=1$, the results are greatly simplified; $h_i=0$ and 
$f_0=-  \ln {M^2 \over \mu^2} - \gamma_E + \ln 4\pi$$, f_1=1/6$, $f_2=1/60$.

\section{\label{sec:tensorfield}Manifest covariant quantization of the second rank antisymmetric tensor gauge field}
\setcounter{equation}{0}

\subsection{Second-rank antisymmetric tensor gauge theory}

We discuss the gauge fixing of a second-rank antisymmetric tensor gauge field $A_{\mu\nu}$ whose Lagrangian is given by 
\begin{eqnarray}
  {\cal L}_0 = - {1 \over 8}(\epsilon_{\mu\nu\rho\sigma}\partial^\nu A^{\rho\sigma})^2 .
\end{eqnarray}
This Lagrangian is invariant under the hypergauge transformation,
\begin{equation}
  \delta A_{\mu\nu}(x) = 
\partial_\mu \xi_\nu(x) - \partial_\nu \xi_\mu(x) .
\end{equation}
In order to fix the gauge, we adopt the gauge fixing condition for $A_{\mu\nu}$,
\begin{equation}
 \partial^\nu A_{\mu\nu} = 0 .
\end{equation}
Then the gauge fixing (GF) and Faddeev-Popov (FP) ghost term is obtained based on the prescription of Kugo and Uehara \cite{KU82} as
\begin{equation}
 {\cal L}_1 := - i\delta_B\left[ \bar C^\nu( \partial^\mu A_{\mu\nu} + {\alpha_1 \over 2}B_\nu) \right] ,
\end{equation}
where we have defined the nilpotent BRST transformation,
\begin{eqnarray}
  \delta_B A_{\mu\nu}(x) &=& 
\partial_\mu C_\nu(x) - \partial_\nu C_\mu(x) ,
\nonumber\\
  \delta_B C_\mu(x) &=& i \partial_\mu d(x),
\nonumber\\
  \delta_B d(x) &=& 0 ,
\nonumber\\
  \delta_B \bar C_\mu(x) &=& i B_\mu(x) ,
\nonumber\\
  \delta_B B_\mu(x) &=& 0 .
\end{eqnarray}
Hence, the explicit form of ${\cal L}_1$ reads
\begin{equation}
 {\cal L}_1 =  B^\nu \partial^\mu A_{\mu\nu} 
+ i \bar C^\nu \partial^\mu [\partial_\mu C_\nu - \partial_\nu C_\mu]
+ {\alpha_1 \over 2}(B_\mu)^2 .
\end{equation}
\par
However, the Lagrangian ${\cal L}_1$ and hence ${\cal L}_0+{\cal L}_1$ is still invariant under the transformation of the vector ghosts $C_\mu$ and $\bar C_\mu$, i.e., 
$\delta C_\mu(x) = i \partial_\mu \theta(x),
\delta \bar C_\mu(x) = i \partial_\mu \varphi(x)$.  Note that $C_\mu$ and $\bar C_\mu$ are independent fields and that
$C^\dagger=C$, $\bar C^\dagger=\bar C$. 
 We consider the gauge fixing conditions,
$\partial^\mu \bar C_\mu= 0$ and $\partial^\mu C_\mu=0$.
Therefore, we must add an additional GF+FP term,
\begin{equation}
 {\cal L}_2 := - i\delta_B\left[ \bar d \left( \partial^\mu C_{\mu}
 + \alpha_2 P \right) \right]
- i \delta_B \left[ N \left( \partial^\mu \bar C_\mu +  \alpha_3 B^{(1)} \right) \right] .
\end{equation}
where the nilpotent BRST transformation of the additional fields is supplemented as
\begin{eqnarray}
  \delta_B N(x) &=&   P(x) ,
\nonumber\\
  \delta_B P(x) &=& 0 ,
\nonumber\\
  \delta_B \bar d(x) &=&  B^{(1)}(x) ,
\nonumber\\
  \delta_B B^{(1)}(x) &=& 0 ,
\end{eqnarray}
The explicit form of ${\cal L}_2$ reads 
\begin{equation}
 {\cal L}_2 = -i B^{(1)}\partial^\mu C_{\mu} -i  \alpha_4 B^{(1)}P
+ \bar d \partial^\mu \partial_\mu d
-i P \partial^\mu \bar C_\mu   + N \partial^\mu B_\mu  ,
\end{equation}
where we have defined $\alpha_4 :=\alpha_2-\alpha_3$.  Note that $P$ and $B^{(1)}$ anti-commute.
For the assignment of the ghost number of each field, see Table.1.
Two vector fields $C_\mu, \bar C_\mu$ are two primary ghosts, and three scalar fields $d, \bar d, N$ are three secondary ghosts. Three fields $B_\mu, P, B^{(1)}$ are the Lagrange multiplier fields for the condition,
$\partial^\nu A_{\mu\nu}=0, \partial^\mu \bar C_\mu= 0, \partial^\mu C_\mu=0$,
respectively.  Thus we obtain the GF+FP term for the Lagrangian ${\cal L}_0$, i.e.,
${\cal L}_{GF+FP}={\cal L}_1 + {\cal L}_2$.
\par
Now all the gauge freedom is fixed.  Thus the full Lagrangian density is given by
\begin{eqnarray}
 {\cal L}_{tot} &=& {\cal L}_0 + {\cal L}_{GF+FP} ,
\nonumber\\ 
  &=&  {\cal L}_0  + B^\nu \partial^\mu A_{\mu\nu} 
+ i \bar C^\nu \partial^\mu [\partial_\mu C_\nu - \partial_\nu C_\mu]+ {\alpha_1 \over 2}(B_\mu)^2 
\nonumber\\&&
-i B^{(1)}\partial^\mu C_{\mu} -i  \alpha_4 B^{(1)}P
+ \bar d \partial^\mu \partial_\mu d
-i P \partial^\mu \bar C_\mu   + N \partial^\mu B_\mu  .
\end{eqnarray}
The massless antisymmetric tensor field stands for the massless spin-0 field as a physical mode.
It is possible to show that all the unphysical modes decouple leaving correctly one physical mode \cite{Kimura80}.

The above result is the summary of the results obtained by Townsend \cite{Townsend79}, Hata, Kugo and Ohta \cite{HKO81} and Kimura \cite{Kimura80}.
The same result can be obtained within the framework of the extended theory for the constrained system based on the canonical Hamiltonian formalism on the extended phase space, the so-called the Batalin-Fradkin-Vilkovisky (BFV)formalism, see e.g. the original papers and a review \cite{BFV}.

\begin{table}
\begin{center}
 \begin{tabular}{ccc}
field & rank  & ghost number \\

$A$      & 2 & 0 \\
$C$      & 1 & 1 \\
$d$      & 0 & 2 \\
$\bar C$ & 1 & -1 \\
$B$      & 1 & 0 \\
$N$      & 0 & 0 \\
$P$      & 0 & 1 \\
$\bar d$ & 0 & -2 \\
$B^{(1)}$ & 0 & -1 \\
$\Lambda$ & 1 & 0 \\
$C'$      & 0 & 1 \\
$\bar C'$ & 0 & -1 \\
$B'$      & 0 & 0 
 \end{tabular}
\end{center}
\caption{The ghost number of the field with the indicated rank.}
\label{table1}
\end{table}

\subsection{Inclusion of mass term}

Next, we consider the theory of an antisymmetric tensor field with the mass term,%
\begin{eqnarray}
  {\cal L}_0^m[A] 
= - {1 \over 8}(\epsilon_{\mu\nu\rho\sigma}\partial^\nu A^{\rho\sigma})^2 
 - {1 \over 4} m^2 (A_{\mu\nu})^2 .
\end{eqnarray}
This Lagrangian with the mass term is no longer invariant under the hypergauge transformation of $A_{\mu\nu}$.  However, the invariance is recovered by introducing an additional vector field $\Lambda_\mu$ in such a way that
\begin{eqnarray}
  {\cal L}_0^m{}'[A,\Lambda] 
= - {1 \over 8}(\epsilon_{\mu\nu\rho\sigma}\partial^\nu A^{\rho\sigma})^2 
 - {1 \over 4} (mA_{\mu\nu}+\partial_\mu \Lambda_\nu - \partial_\nu \Lambda_\mu)^2 .
\end{eqnarray}
Actually, this Lagrangian is invariant under the combined transformation,
\begin{equation}
  \delta A_{\mu\nu}(x) = 
\partial_\mu \xi_\nu(x) - \partial_\nu \xi_\mu(x) ,
\quad \delta \Lambda_\mu(x) = - m \xi_\mu(x) .
\end{equation}
Moreover, it has an additional invariance under the transformation,
\begin{equation}
  \delta A_{\mu\nu}(x) = 0 ,
\quad \delta \Lambda_\mu(x) = \partial_\mu \omega(x).
\end{equation}
Therefore, we define the BRST transformation of the field $\Lambda$ as
\begin{equation}
  \delta_B \Lambda_\mu(x) = - m C_\mu(x) + \partial_\mu C'(x) ,
\label{BRST-Lambda}
\end{equation}
by introducing additional ghost $C'$.  The BRST transformation of $C'$ and $\bar C'$ is defined by
\begin{eqnarray}
 \delta_B C'(x) &=& im d(x), 
\\
 \delta_B \bar C'(x) &=& iB'(x) ,
\\
 \delta_B B'(x) &=& 0 .
\end{eqnarray}
The first BRST transformation is determined by nilpotency of (\ref{BRST-Lambda}). Other two transformations are also suggested from nilpotency of BRST transformation.
Thus the nilpotent BRST transformation is determined for all the fields.
In order to fix the gauge freedom, we must obtain the GF+PF term ${\cal L}_{GF+FP}$.
For ${\cal L}_{tot}' := {\cal L}_0^m{}' + {\cal L}_{GF+FP}$,
the generation functional of the theory is given by
\begin{equation}
  Z := \int {\cal D}A_{\mu\nu} {\cal D}\Lambda_\mu {\cal D}B_\mu
{\cal D}C_\mu {\cal D}\bar C_\mu {\cal D}d {\cal D}\bar d {\cal D}N 
{\cal D}P {\cal D}B^{(1)}{\cal D}C'{\cal D}\bar C' {\cal D}B'
\exp \left[ i \int d^4x {\cal L}_{tot}' \right] .
\end{equation}
\par
A good choice is%
\footnote{  
The author would like to thank Atsushi Nakamura \cite{Nakamura00} for helpful discussions on this Appendix.
}
\begin{eqnarray}
  {\cal L}_{GF+FP} &=& - i \delta_B \Biggr[
\bar C^\nu \left( \partial^\mu A_{\mu\nu} - \partial_\nu N - a \Lambda_\nu 
+{\alpha_1 \over 2}B_\nu  \right)
+ \bar d \left( \partial^\mu C_\mu + b C' + \alpha_2 P \right)
\nonumber\\&&
+ \bar C' \left( \partial^\mu \Lambda_\mu+{\alpha' \over 2}B' \right) \Biggr] 
\\
&=& B^\nu \left( \partial^\mu A_{\mu\nu} - \partial_\nu N - a \Lambda_\nu 
+{\alpha_1 \over 2}B_\nu  \right)
\nonumber\\ &&
+ i \bar C^\nu [\partial^\mu (\partial_\mu C_\nu-\partial_\nu C_\mu) 
-\partial_\nu P-a(\partial_\nu C'-m C_\nu)]
\nonumber\\&&
-iB^{(1)}(\partial^\mu C_\mu+bC'+\alpha_2 P)
+ \bar d \partial^\mu \partial_\mu d + b m \bar d d 
\nonumber\\&&
+ B'(\partial^\mu \Lambda_\mu+{\alpha' \over 2}B')
+ i\bar C' \partial^\mu (\partial_\mu C'-mC_\mu) ,
\end{eqnarray}
where $\alpha_1, \alpha_2, \alpha{}'$ are gauge fixing parameters and $a,b$ are parameters specified later.
From the naive viewpoint, this corresponds to the gauge fixing condition,
$\partial^\mu A_{\mu\nu}=0$, $\partial^\mu C_\mu=0=\partial^\mu \bar C_\mu$ and
$\partial^\mu \Lambda_\mu=0$.
If we integrate over $B$ and $B'$,
the sector containing $A$ and $\Lambda$ reads
\begin{equation}
  Z := \int {\cal D}A_{\mu\nu} {\cal D}\Lambda_\mu {\cal D}N  
\exp \left[ i \int d^4x {\cal L}_{tot}'' \right] ,
\end{equation}
where
\begin{eqnarray}
 {\cal L}_{tot}'' &=& {\cal L}_0^m{}[A] 
- {1 \over 2}mA_{\mu\nu}(\partial_\mu \Lambda_\nu - \partial_\nu \Lambda_\mu) - {1 \over 4}m^2 (\partial_\mu \Lambda_\nu - \partial_\nu \Lambda_\mu)^2
\nonumber\\&&
 - {1 \over 2\alpha'}(\partial^\mu \Lambda_\mu)^2 
+ {1 \over 2\alpha_1}(\partial^\nu A_{\mu\nu}-\partial_\mu N-a\Lambda_\mu)^2  ,
\end{eqnarray}
since other fields decouple from the relevant sector.
Then we perform the integration over $N$ and obtain
\begin{eqnarray}
  Z &=& \int {\cal D}A_{\mu\nu} 
\exp \left\{ i \int d^4x \left( {\cal L}_0^m{}[A] 
+ {1 \over 2\alpha_1} \left( \partial^\nu A_{\mu\nu} \right)^2 \right) \right\}
\nonumber\\&& \times
\int {\cal D}\Lambda_\mu
 \exp \left\{ i \int d^4x {\cal L}_1^m{}[A,\Lambda] \right\},
\end{eqnarray}
where
\begin{eqnarray}
{\cal L}_1^m{}[A,\Lambda] &:=& 
- {1 \over 2}mA_{\mu\nu}(\partial_\mu \Lambda_\nu - \partial_\nu \Lambda_\mu) - {1 \over 4}m^2 (\partial_\mu \Lambda_\nu - \partial_\nu \Lambda_\mu)^2
\nonumber\\
&+&  {a^2 \over 2\alpha_1}(\Lambda_\mu)^2
+ {a \over \alpha_1}(\partial^\nu A_{\mu\nu})\Lambda_\mu 
+ {a^2 \over 2\alpha_1}(\partial^\mu \Lambda_\mu)\Delta^{-1}(\partial^\nu \Lambda_\nu)  .
\end{eqnarray}
If we choose $a=m\alpha_1$, the cross term $(\partial^\nu A_{\mu\nu})\Lambda_\mu$ cancels with 
$mA_{\mu\nu}(\partial_\mu \Lambda_\nu - \partial_\nu \Lambda_\mu)$.
Hence, ${\cal L}_1^m{}[A,\Lambda]$ becomes independent of $A$ field.
Thus the $\Lambda$ field decouples from the theory of $A$,
\begin{eqnarray}
{\cal L}_1^m{}[\Lambda] =
 - {1 \over 4}m^2 (\partial_\mu \Lambda_\nu - \partial_\nu \Lambda_\mu)^2
+  {a^2 \over 2\alpha_1}(\Lambda_\mu)^2
+ {a^2 \over 2\alpha_1}(\partial^\mu \Lambda_\mu)\Delta^{-1}(\partial^\nu \Lambda_\nu)  .
\end{eqnarray}
  By taking the Landau gauge $\alpha_1=0$, we recover the partition function, 
\begin{eqnarray}
  Z = \int {\cal D}A_{\mu\nu}\delta(\partial^\nu A_{\mu\nu}) 
\exp \left\{ i \int d^4x  {\cal L}_0^m{}[A]   \right\} .
\end{eqnarray}
\par
Moreover, it is possible to consider a simpler gauge \cite{BS96},
\begin{eqnarray}
  {\cal L}_{GF+FP} &=& - i \delta_B \left[
\bar C^\mu \Lambda_\mu + \bar d C' \right] 
\\
&=&  B^\mu \Lambda_\mu + i \bar C^\mu (\partial_\mu C' - mC_\mu)
-iB^{(1)}C' + m\bar d d .
\end{eqnarray}
This corresponds to the gauge fixing condition,
$\Lambda_\mu=0$ and $C'=0$.
The integration over $N$, $P$ and $B'$ is trivial, since ${\cal L}_{tot}'$ does not include them.  The $B$ integration leads to the constraint 
$\delta(\Lambda_\mu)$,
\begin{equation}
  Z = \int {\cal D}A_{\mu\nu} {\cal D}\Lambda_\mu \delta(\Lambda_\mu)
\int {\cal D}C_\mu {\cal D}\bar C_\mu {\cal D}d {\cal D}\bar d {\cal D}B^{(1)} {\cal D}C'{\cal D}\bar C'
\exp \left[ i \int d^4x {\cal L}_{tot}'' \right] ,
\end{equation}
where
\begin{eqnarray}
 {\cal L}_{tot}'' = {\cal L}_0^m{}'[A_{\mu\nu},\Lambda_\mu] 
  + i \bar C^\mu (\partial_\mu C' - mC_\mu)
-iB^{(1)}C' + m\bar d d    .
\end{eqnarray}
When we consider the sector of $A$ and $\Lambda$, the sector described by other fields decouples and we obtain

\begin{equation}
  Z = \int {\cal D}A_{\mu\nu} {\cal D}\Lambda_\mu \delta(\Lambda_\mu)
\exp \left\{ i \int d^4x {\cal L}_0^m{}'[A_{\mu\nu},\Lambda_\mu] \right\} 
= \int {\cal D}A_{\mu\nu} 
\exp \left\{ i \int d^4x {\cal L}_0^m[A_{\mu\nu}] \right\} . 
\end{equation}
Therefore, we recover the original theory which is written by the $A$ field only with the Lagrangian ${\cal L}_0^m$.
Note that the massive antisymmetric tensor gauge theory stands for the massive spin-1 theory.

\section{\label{sec:dGL}Renormalization of dual Abelian Higgs model}
\setcounter{equation}{0}

\subsection{Goldstone bosons remain massless in higher orders}

\par
\unitlength=0.001in
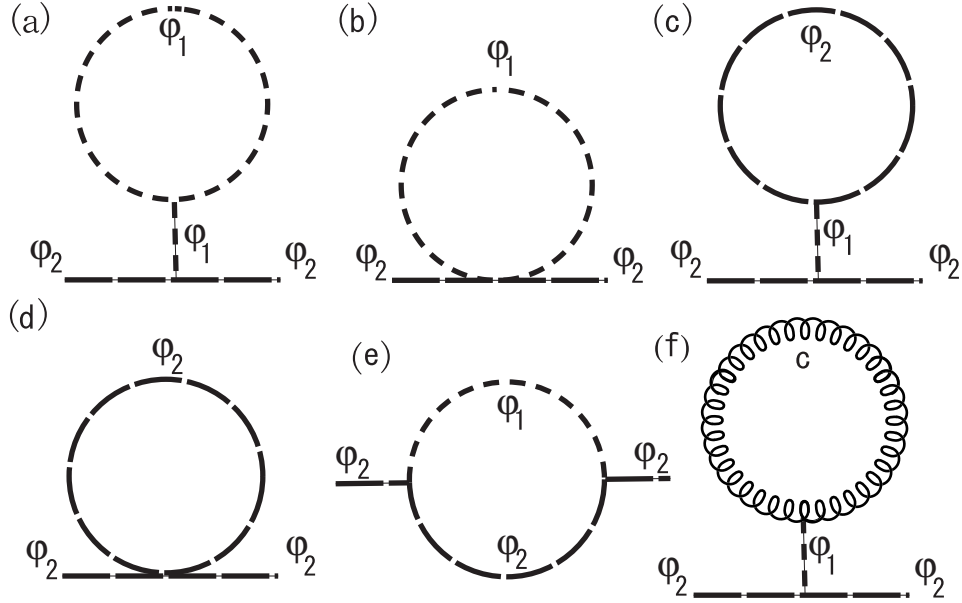
\begin{figure}[t]
\begin{center}
\begin{picture}(5000,3500)
\put(150,1500){%
   \put(0,0){\epsfbox{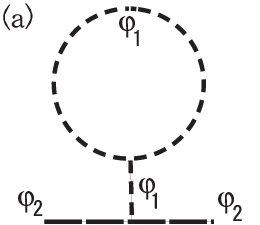}}%
   }%
\put(1850,1500){%
   \put(0,0){\epsfbox{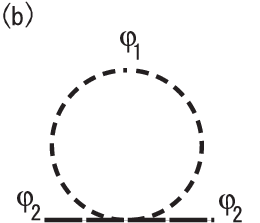}}%
   }%
\put(3500,1500){%
   \put(0,0){\epsfbox{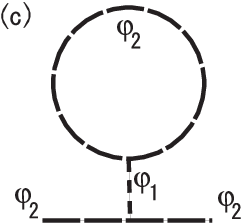}}%
   }%
\put(150,-50){%
   \put(0,0){\epsfbox{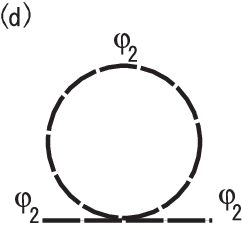}}%
   }%
\put(1850,-50){%
   \put(0,0){\epsfbox{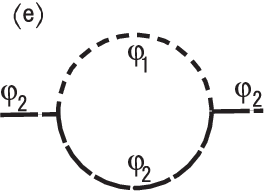}}%
   }%
\put(3500,-550){%
   \put(0,0){\epsfbox{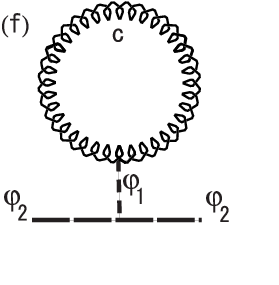}}%
   }%
\end{picture}
 \caption[]{Self-energy diagrams for the would-be Nambu-Goldstone  particle $\varphi_2$.}
 \label{fig:NG}
\end{center}
\end{figure}

We examine the self-energy diagrams for the would-be NG boson which are given in Fig.~\ref{fig:NG}.
\begin{eqnarray}
  \Sigma_a(0) &:=&  (-2i\lambda v) {i \over -m^2} (-6i\lambda v) {1 \over 2} 
\int {d^4k \over (2\pi)^4} {i \over k^2-m^2} 
= -3\lambda I(m^2) ,
\\
  \Sigma_b(0) &:=& (-2i\lambda) {1 \over 2} 
\int {d^4k \over (2\pi)^4} {i \over k^2-m^2} 
= \lambda I(m^2) ,
\\
  \Sigma_c(0) &:=& (-2i\lambda v) {i \over -m^2} (-2i\lambda v) {1 \over 2} 
\int {d^4k \over (2\pi)^4} {i \over k^2-\xi M^2} 
= -\lambda I(\xi M^2) ,
\\
  \Sigma_d(0) &:=& (-6i\lambda)  {1 \over 2} 
\int {d^4k \over (2\pi)^4} {i \over k^2-\xi M^2} 
= 3\lambda I(\xi M^2) ,
\\
  \Sigma_e(0) &:=& (-2i\lambda v)^2 
\int {d^4k \over (2\pi)^4} {i \over k^2-\xi M^2}{i \over k^2-m^2}
\nonumber\\ 
&=& {4\lambda^2 \over 2\lambda-\xi g^2}[I(m^2)-I(\xi M^2)],
\\
  \Sigma_f(0) &:=&  (-2i\lambda v) {i \over -m^2} (ig\xi M) 
\int {d^4k \over (2\pi)^4} {i \over k^2-\xi M^2} 
= \xi g^2  I(\xi M^2) ,
\end{eqnarray}

When $\xi=0$, it is easy to check that the total sum of the above self-energy parts vanishes.
This fact ensures that the Nambu-Goldstone particle remains massless even if we include the higher order corrections.

\par
\unitlength=0.001in
\begin{figure}[t]
\begin{center}
\begin{picture}(5000,3500)
\put(150,1500){%
   \put(0,0){\epsfbox{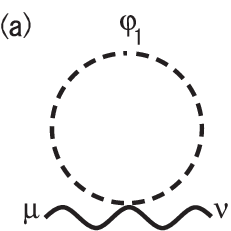}}%
   }%
\put(2000,1500){%
   \put(0,0){\epsfbox{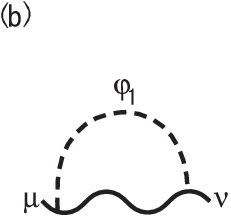}}%
   }%
\put(3450,1500){%
   \put(0,0){\epsfbox{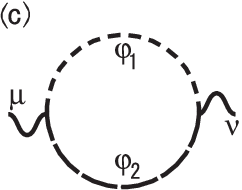}}%
   }%
\put(150,-100){%
   \put(0,0){\epsfbox{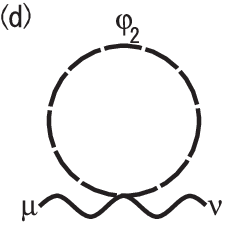}}%
   }%
\put(2000,-100){%
   \put(0,0){\epsfbox{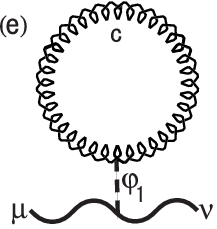}}%
   }%
\end{picture}
 \caption[]{Vacuum polarization diagrams for the $b_\mu$ particle.}
 \label{fig:dualvector}
\end{center}
\end{figure}
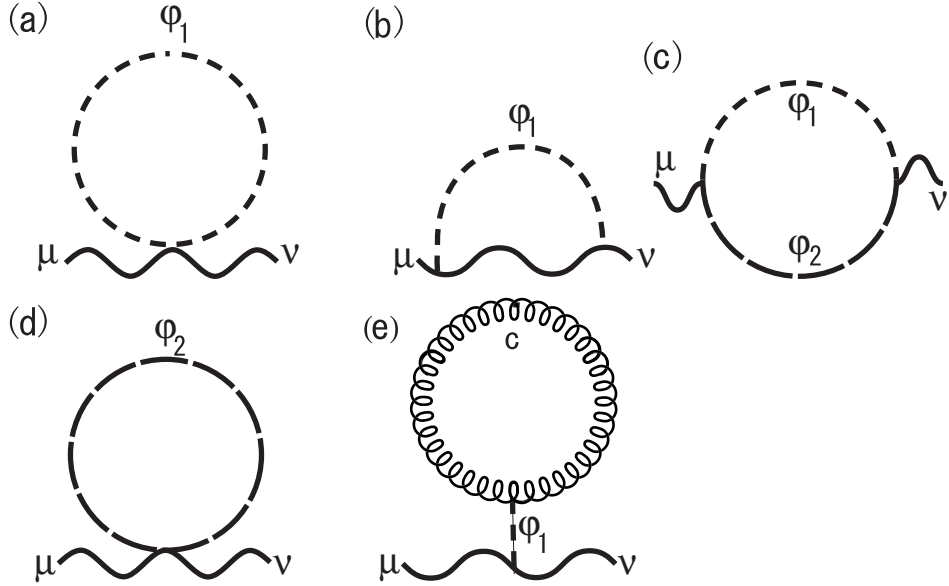

\subsection{Vacuum polarization}

Next, we consider the vacuum polarization for the $b_\mu$ field.  The relevant diagrams to one loop are enumerated in Fig.~\ref{fig:dualvector}.
The vacuum polarization tensor for the dual vector field $b_\mu$ is obtained by summing up the following contributions,
\begin{eqnarray}
 \Pi_{\mu\nu}^{(a)}(p) &:=& 2ig_m^2 g_{\mu\nu} \int {d^4k \over (2\pi)^4} {i \over k^2-m^2}, 
\quad m^2 := 2\mu^2 = 2 \lambda v^2 ,
\\
  \Pi_{\mu\nu}^{(b)}(p)  &:=& (2ig_m^2v g_{\mu\rho})(2ig_m^2v g_{\sigma\nu}) \int {d^4k \over (2\pi)^4}{i \over (p-k)^2-m^2}{-i \over k^2-M^2}
\nonumber\\&& \times
 \left[ g_{\rho\sigma} - (1-\xi){k_\rho k_\sigma \over k^2-\xi M^2} \right] ,
\\
  \Pi_{\mu\nu}^{(c)}(p)  &:=& g_m^2 \int {d^4k \over (2\pi)^4} {i \over (p-k)^2-m^2}{i \over k^2-\xi M^2} 
(2k-p)_\mu (-2k+p)_\nu ,
\\
  \Pi_{\mu\nu}^{(d)}(p)  &:=& 2ig_m^2 g_{\mu\nu} \int {d^4k \over (2\pi)^4} {i \over k^2-\xi M^2} ,
\\
 \Pi_{\mu\nu}^{(e)}(p)  &:=& 2ig_m^2 v g_{\mu\nu} ig\xi M \int  {d^4k \over (2\pi)^4} {i \over k^2-\xi M^2}{i \over -m^2}  .
\end{eqnarray}

First, the dimensional regularization yields
\begin{eqnarray}
 \Pi_{\mu\nu}^{(a)}(p) &:=& - 2 g_m^2 g_{\mu\nu} \int {d^Dk \over i(2\pi)^D} {1 \over k^2-m^2}
= i2 g_m^2 g_{\mu\nu} \int {d^Dk \over i(2\pi)^D} {1 \over -k^2+m^2}
\\
&=& i2 g_m^2 g_{\mu\nu} {\Gamma(\epsilon-1) \over (4\pi)^{2-\epsilon}}
(m^2)^{1-\epsilon}
\\
&=& i2 g_m^2 g_{\mu\nu} {1 \over (4\pi)^2} \left( -{1 \over \epsilon} + \gamma_E -1 \right) m^2 (1-\epsilon \ln m^2)(1+\epsilon \ln 4\pi)
\\
&=&
i{2g_m^2 \over (4\pi)^2} g_{\mu\nu} m^2 \left( -N_\epsilon -1 + \ln m^2 \right) ,
\end{eqnarray}
where we have defined 
$\epsilon :=2-D/2$ and
\begin{equation}
 N_\epsilon := {1 \over \epsilon} + \ln 4\pi - \gamma_E .
\end{equation}
and used the Laurent expansion of the Gamma function $\Gamma(\epsilon-1)$.

\par
Second, we want to calculate
\begin{eqnarray}
 && \Pi_{\mu\nu}^{(b)}(p)  
\nonumber\\
&:=& (2ig_m^2v g_{\mu\rho})(2ig_m^2v g_{\sigma\nu}) \int {d^4k \over (2\pi)^4}{i \over (p-k)^2-m^2}{-i \over k^2-M^2} \left[ g_{\rho\sigma} - (1-\xi){k_\rho k_\sigma \over k^2-\xi M^2} \right] 
\nonumber\\
&=& -4 g_m^4 v^2 \int {d^4k \over (2\pi)^4}{1 \over -(p-k)^2+m^2}\left[
 { g_{\mu\nu}-{k_\mu k_\nu \over M^2} \over -k^2+M^2} 
+ {{k_\mu k_\nu \over M^2} \over -k^2+\xi M^2}  \right] .
\end{eqnarray}
The Feynman parameter formula can combine two denominators into a common denominator.
Then the dimensional regularization leads to
\begin{eqnarray}
 && \Pi_{\mu\nu}^{(b)}(p)  
\nonumber\\
&=& -i 4 g_m^4 v^2 \int_0^1 dx \Biggr\{
\int {d^Dk \over i(2\pi)^D}
 {\left( g_{\mu\nu}-{k_\mu k_\nu \over M^2} \right)
 \over [-k^2+2xp \cdot k -xp^2+xm^2+(1-x) M^2]^2} 
\nonumber\\&&
\quad \quad +   \int {d^Dk \over i(2\pi)^D}
{{k_\mu k_\nu \over M^2} \over [-k^2+2xp \cdot k -xp^2+xm^2+(1-x)\xi M^2]^2} 
    \Biggr\}  
\\
&=&  {-i 4 g_m^4 v^2 \over (4\pi)^{2-\epsilon}\Gamma(2)}
\int_0^1 dx \Biggr\{ 
\Gamma(\epsilon) g_{\mu\nu} [(x^2-x)p^2 +xm^2+(1-x) M^2]^{-\epsilon}
\nonumber\\&&
 - \Gamma(\epsilon) x^2 {p_\mu p_\nu \over M^2} 
[(x^2-x)p^2 +xm^2+(1-x) M^2]^{-\epsilon}
\nonumber\\&&
+ \Gamma(\epsilon-1){1 \over 2}{g_{\mu\nu} \over M^2}
[(x^2-x)p^2 +xm^2+(1-x) M^2]^{1-\epsilon}
\nonumber\\&&
+ \Gamma(\epsilon) x^2{p_\mu p_\nu \over M^2} 
[(x^2-x)p^2 +xm^2+(1-x) \xi M^2]^{-\epsilon}
\nonumber\\&&
- \Gamma(\epsilon-1){1 \over 2}{g_{\mu\nu} \over M^2}
[(x^2-x)p^2 +xm^2+(1-x) \xi M^2]^{1-\epsilon}
    \Biggr\} .
\end{eqnarray}

Hence, using the Laurent expansion of the Gamma function $\Gamma(\epsilon)$,
we have
\begin{eqnarray}
  && \Pi_{\mu\nu}^{(b)}(p)  
\nonumber\\
&=&  {-i 4 g_m^4 v^2 \over (4\pi)^{2}\Gamma(2)}
\int_0^1 dx \Biggr[
g_{\mu\nu} \left\{ N_\epsilon - \ln [(x^2-x)p^2 +xm^2+(1-x) M^2]  \right\}
\nonumber\\&&
+ {1 \over 2}{g_{\mu\nu} \over M^2}
[(x^2-x)p^2 +xm^2+(1-x) M^2]
\nonumber\\&& \times
\left\{ -N_\epsilon -1   
  + \ln [(x^2-x)p^2 +xm^2+(1-x) M^2] \right\} 
\nonumber\\&&
+ {1 \over 2}{g_{\mu\nu} \over M^2}[(x^2-x)p^2 +xm^2+(1-x) \xi M^2]
\nonumber\\&& \times
\left\{ N_\epsilon + 1   
-  \ln [(x^2-x)p^2 +xm^2+(1-x) \xi M^2]  \right\}
\nonumber\\&&
 + x^2 {p_\mu p_\nu \over M^2} 
\Biggr\{   
 +  \ln [(x^2-x)p^2 +xm^2+(1-x) M^2]  
\nonumber\\&&
 - \ln [(x^2-x)p^2 +xm^2+(1-x) \xi M^2] \Biggr\} \Biggr] 
  + O(\epsilon) .
\end{eqnarray}
\par
Using the low-energy expansion,
\begin{eqnarray}
 && \ln [(x^2-x)p^2 +xm^2+(1-x) \xi M^2]
\nonumber\\
&=& \ln [xm^2+(1-x) \xi M^2] + {x^2-x \over xm^2+(1-x)\xi M^2}p^2 +O\left({p^4 \over M^4} \right) ,
\end{eqnarray}
we obtain
\begin{eqnarray}
 && \Pi_{\mu\nu}^{(b)}(p)  
\nonumber\\
&=& { -i 4 g_m^4 v^2 \over (4\pi)^{2}\Gamma(2)}
  \Biggr[
  g_{\mu\nu} \int_0^1 dx \left\{ N_\epsilon - 
\ln [xm^2+(1-x)  M^2] + {x^2-x \over xm^2+(1-x) M^2}p^2
  \right\}
\nonumber\\&&
+ {1 \over 2}{g_{\mu\nu} \over M^2}
\int_0^1 dx [xm^2+(1-x) M^2]
\nonumber\\&& \times
\Biggr\{ -N_\epsilon -1   
  + \ln [xm^2+(1-x)  M^2] 
 + {x^2-x \over xm^2+(1-x) M^2}p^2
 \Biggr\} 
\nonumber\\&&
+ {1 \over 2}{g_{\mu\nu} \over M^2}p^2
\int_0^1 dx (x^2-x)
\Biggr\{ -N_\epsilon -1   
  + \ln [xm^2+(1-x) M^2] 
 \Biggr\} 
\nonumber\\&&
+ {1 \over 2}{g_{\mu\nu} \over M^2}\int_0^1 dx [xm^2+(1-x) \xi M^2]
\nonumber\\&& \times
\Biggr\{ N_\epsilon + 1   
-  \ln [xm^2+(1-x) \xi M^2] 
 - {x^2-x \over xm^2+(1-x)\xi M^2}p^2
  \Biggr\}
\nonumber\\&&
+ {1 \over 2}{g_{\mu\nu} \over M^2}p^2  \int_0^1 dx (x^2-x) 
\Biggr\{ N_\epsilon + 1   
-  \ln [xm^2+(1-x) \xi M^2] 
  \Biggr\}
\nonumber\\&&
 + {p_\mu p_\nu \over M^2} \int_0^1 dx x^2  
 \Biggr\{   
 + \ln [xm^2+(1-x) M^2]  
 - \ln [xm^2+(1-x) \xi M^2]
 \Biggr\}
 \Biggr] 
\nonumber\\&&
  + O(p^4) .
\end{eqnarray}
In particular, when $\xi=0$, we obtain
\begin{eqnarray}
 \Pi_{\mu\nu}^{(b)}(p)
&=& -i{4g_m^4v^2 \over (4\pi)^{2}} g_{\mu\nu} \left[ {3 \over 4}(N + 1)
- {m^2 \ln m^2 - M^2 \ln M^2 \over m^2-M^2}  
 - {1 \over 8} {m^2 \over M^2} (-1+2 \ln m^2) \right] 
\nonumber\\&&
 + O\left( {p^2 \over M^2} \right) .
\nonumber\\
\end{eqnarray}

\par
Third, we can proceed in the similar way to (b) and obtain
\begin{eqnarray}
 &&  \Pi_{\mu\nu}^{(c)}(p) 
\\ 
 &=& i{g_m^2 \over (4\pi)^{2}} \Biggr[ g_{\mu\nu} (N_\epsilon+1) \left( -{1 \over 3}p^2 + m^2 -\xi M^2 \right) 
\nonumber\\ &&
- 2 g_{\mu\nu} \Biggr\{  \int_0^1 dx [xm^2+(1-x)\xi M^2] \ln [xm^2+(1-x)\xi M^2]
\nonumber\\&&
+ p^2 \int_0^1 dx (x^2-x)
+ p^2 \int_0^1 dx(x^2-x) \ln [xm^2+(1-x)\xi M^2] \Biggr\}
\nonumber\\ &&
 + {1 \over 3}N_\epsilon p_\mu p_\nu 
- p_\mu p_\nu \int_0^1 dx (4x^2-4x+1) \ln [xm^2+(1-x)\xi M^2] \Biggr]
 + O(p^4) .
\end{eqnarray}
In particular, when $\xi=0$, we obtain
\begin{eqnarray}
 \Pi_{\mu\nu}^{(c)}(p)
&=& i{g_m^2 \over (4\pi)^{2}} \Biggr\{ g_{\mu\nu} \left[ p^2 \left( -{N_\epsilon \over 3}-{13 \over 36} - 6 \ln m^2 \right)    + m^2 \left( N_\epsilon+{3 \over 2}-\ln m^2 \right) \right] 
\nonumber\\&&
+ p_\mu p_\nu \left[ {N_\epsilon \over 3} + {4 \over 9} -{1 \over 3} \ln m^2 \right] \Biggr\} + O(p^4) .
\end{eqnarray}
The remaining quantities, $\Pi_{\mu\nu}^{(d)}$ and $\Pi_{\mu\nu}^{(e)}$ are calculated in the same manner as $\Pi_{\mu\nu}^{(a)}$.  However, they vanish for $\xi=0$ in the dimensional regularization.
\par
Therefore, the total sum of the vacuum polarization for $\xi=0$ is given by\begin{eqnarray}
 && i \Pi_{\mu\nu}(p) 
\nonumber\\
&=&
 -{2g_m^2 \over (4\pi)^2} m^2 g_{\mu\nu}  \left( -N_\epsilon -1 + \ln {m^2 \over \mu^2} \right) 
\nonumber\\&&
+{4g_m^2 \over (4\pi)^{2}} (g_mv)^2 g_{\mu\nu} \left[ {3 \over 4}(N_\epsilon + 1)
- {m^2 \ln {m^2 \over \mu^2} - M^2 \ln {M^2 \over \mu^2} \over m^2-M^2}  
 - {1 \over 8} {m^2 \over M^2} \left(-1+2 \ln {m^2 \over \mu^2} \right) \right] 
\nonumber\\&& 
-{g_m^2 \over (4\pi)^{2}}  m^2 g_{\mu\nu} \left( N_\epsilon+{3 \over 2}-\ln {m^2 \over \mu^2} \right)   
\nonumber\\&& 
-{g_m^2 \over (4\pi)^{2}} \Biggr\{ - p^2 g_{\mu\nu}  \left( {N_\epsilon \over 3}+{13 \over 36} + 6 \ln {m^2 \over \mu^2} \right)      
+ p_\mu p_\nu \left[ {N_\epsilon \over 3} + {4 \over 9} -{1 \over 3} \ln {m^2 \over \mu^2} \right] \Biggr\} 
\nonumber\\&&
+ O\left({p^2 \over M^2}\right) ,
\end{eqnarray}
where 
$m^2=2\lambda v^2$ and $M^2=g_m^2 v^2$.

Finally, the renormalization factors are obtained as
\begin{eqnarray}
 \delta_b &=& Z_b - 1 
= -{g_m^2 \over (4\pi)^{2}}  \left( {N_\epsilon \over 3}+{13 \over 36} + 6 \ln {m^2 \over \mu^2} \right) + O(g_m^4) ,
\nonumber\\
\delta_m  &=& Z_b(M_b)^2 - (M_b^R)^2 
=  -{2g_m^2 \over (4\pi)^2} m^2  \left( -N_\epsilon -1 + \ln {m^2 \over \mu^2} \right) 
\nonumber\\&&
+{4g_m^2 \over (4\pi)^{2}} (g_mv)^2 \left[ {3 \over 4}(N_\epsilon + 1)
- {m^2 \ln {m^2 \over \mu^2} - M^2 \ln {M^2 \over \mu^2} \over m^2-M^2}  
 - {1 \over 8} {m^2 \over M^2} \left(-1+2 \ln {m^2 \over \mu^2} \right) \right] 
\nonumber\\&& 
-{g_m^2 \over (4\pi)^{2}}  m^2  \left( N_\epsilon+{3 \over 2}-\ln {m^2 \over \mu^2} \right) + O(g_m^4) .
\end{eqnarray}

\section{\label{sec:WLcalc}Calculation of the Wilson loop}
\setcounter{equation}{0}

In the following we show that the area law decay of the Wilson loop is obtained from the result,
\begin{equation}
 \langle W(C) \rangle_{YM} =
 \exp \left\{ -{1 \over 2}(2Jg\rho^{-1}K^{1/2})^2  (\tilde \Xi_\mu, D_m^{-1} \tilde \Xi^\mu) 
\right\} ,
\end{equation}
where $\Xi$ is the one-form defined by
\begin{equation}
  \Xi := *d \Theta\Delta^{-1} = \delta * \Theta \Delta^{-1} ,
\end{equation}
with the component,
\begin{eqnarray}
\Xi^\mu(x)
&=& {1 \over 2} \epsilon^{\mu\alpha\beta\gamma} \partial_\alpha^x \int d^4y \Theta_{\beta\gamma}(y) \Delta^{-1}(y-x) 
\\
&=&  {1 \over 2} \epsilon^{\mu\alpha\beta\gamma} \partial_\alpha^x \int_S d^2 S_{\beta\gamma}(x') \Delta^{-1}(x'-x) .
\label{Theta}
\end{eqnarray}
Then the argument of the exponential is cast into the following form,
\begin{eqnarray}
  (\Xi_\mu, D_m^{-1} \Xi^\mu) 
&=& (\delta * \Theta \Delta^{-1}, D_m^{-1} \delta * \Theta \Delta^{-1})
\nonumber\\
&=& (\Theta, \Delta^{-1}*d \delta * d \Delta^{-1} D_m^{-1} \Theta)
\nonumber\\
&=& (\Theta, \Delta^{-1}\delta d \Delta^{-1} D_m^{-1} \Theta)
\nonumber\\
&=&  (\Theta, \Delta^{-1}D_{m}^{-1}(\Delta) \Theta) 
- (\Theta, \Delta^{-1}d \delta \Delta^{-1} D_m^{-1} \Theta) 
\nonumber\\
&=&  (\Theta, \Delta^{-1}D_{m}^{-1}(\Delta) \Theta) 
- (\delta \Theta, \Delta^{-2} D_m^{-1} \delta \Theta) .
\label{areaperi}
\end{eqnarray}
For the rectangular loop with side lengths $T$ and $R$ in the $x_1-x_4$ plane, we take
\begin{equation}
  \Theta_{\alpha\beta}(z) = \delta_{\alpha 1}\delta_{\beta 4} \delta(z_2) \delta(z_3) \theta(z_1) \theta(R-z_1) \theta(z_4) \theta(T-z_4) .
\end{equation}
Then the Fourier transformation is given by
\begin{eqnarray}
  \Theta_{\alpha\beta}(p) &\equiv& \int d^4z \Theta_{\alpha\beta}(z) e^{-ip \cdot z}
\nonumber\\ 
&=& \delta_{\alpha 1}\delta_{\beta 4} \int_0^R dz_1 e^{-ip_1 z_1}
\int_0^T dz_4 e^{-ip_4 z_4}
\nonumber\\ 
&=& \delta_{\alpha 1}\delta_{\beta 4} {2 \over p_1} e^{-i{p_1 R \over 2}} \sin {p_1 R \over 2} {2 \over p_4} e^{-i{p_4 T \over 2}} \sin {p_4 T \over 2} .
\end{eqnarray}
In the momentum representation, we have
\begin{eqnarray}
(\Theta, \Delta^{-1}D_{m}^{-1} \Theta) 
= \int {d^4p \over (2\pi)^4} \Theta_{\alpha\beta}(p) \Theta_{\alpha\beta}(-p)
 [\Delta^{-1}D_{m}^{-1}](p) .
\end{eqnarray}
If we use the formula following \cite{maKanazawa},
\begin{equation}
  \lim_{R \rightarrow \infty}\left({\sin aR \over a}\right)^2
 = \pi R \delta(a), 
\end{equation}
for large $R$ and large $T$,
then we obtain
\begin{eqnarray}
(\Theta, \Delta^{-1}D_{m}^{-1}\Theta) 
&\cong&  \int {d^4p \over (2\pi)^4} (2\pi)^2 T R \delta(p_1)\delta(p_4) 
[\Delta^{-1}D_{m}^{-1}](p)
\nonumber\\
&=& T R  \int {d^2p \over (2\pi)^2}   
[\Delta^{-1}D_{m}^{-1}](0,p_2,p_3,0)
\nonumber\\
&=& - T R  \int {d^2p \over (2\pi)^2}   \kappa \left[ 
{1 \over p_2^2+p_3^2+m_1^2} - {1 \over p_2^2+p_3^2+m_2^2} \right] .
\end{eqnarray}
Here the logarithmic divergence of the integral is removed by introducing the ultraviolet cutoff $\Lambda$ as 
\begin{eqnarray}
(\Theta, \Delta^{-1}D_{m}^{-1}\Theta) 
&=& - T R  \int {d^2p \over (2\pi)^2}   \kappa \left[ 
{1 \over p_2^2+p_3^2+m_1^2} - {1 \over p_2^2+p_3^2+m_2^2} \right] 
\nonumber\\
&=& - TR \lim_{\Lambda \rightarrow \infty} \int_0^{\Lambda^2} {d |p|^2 \over 4\pi}  \kappa \left[ 
{1 \over |p|^2+m_1^2} - {1 \over |p|^2+m_2^2} \right]
\nonumber\\
&=& TR {\kappa \over 4\pi} \lim_{\Lambda \rightarrow \infty} \ln {m_1^2 \over \Lambda^2+m_1^2}{\Lambda^2+m_2^2 \over m_2^2}
\nonumber\\
&=& TR {\kappa \over 4\pi} \ln {m_1^2 \over m_2^2} .
\label{areacontribution}
\end{eqnarray}
So we obtain the $\Lambda$-independent result.
The last term in (\ref{areaperi}) gives a perimeter decay part, since $\delta \Theta$ is the boundary of the rectangular surface.  The dominant term in the large loop is given by the contribution (\ref{areacontribution}) which exhibits the area law.
Thus we arrive at the result (\ref{tension}).

\section{\label{sec:dexp}Derivative expansion of the string}
\setcounter{equation}{0}

In this section, we begins with the expression,
\begin{eqnarray}
   \langle W(C) \rangle_{YM} 
&=&  \exp \left[ -  
  \int_{S_C} dS^{\mu\nu}(x) \int_{S_C} dS^{\rho\sigma}(y) 
   G_{\mu\nu,\rho\sigma}(x,y) \right] ,
\end{eqnarray}
where
\begin{eqnarray}
   G_{\mu\nu,\rho\sigma}(x,y) &:=&  - 2J^2g^2 \rho^{-2} K 
\langle {\cal D}[\partial_x]h_{\mu\nu}^\xi(x) {\cal D}[\partial_y]h_{\rho\sigma}^\xi(y) \rangle_{APEGT} .
\end{eqnarray}
In this paper, we have obtained the result,
\begin{eqnarray}
   G_{\mu\nu,\rho\sigma}(x,y) 
&=&  - 2J^2g^2 \rho^{-2} K I_{\mu\nu,\rho\sigma}
  [\Delta D_m]^{-1} 
\\ 
&=&  - 2J^2g^2 \rho^{-2} K I_{\mu\nu,\rho\sigma}
  \left[ {\chi \over \Delta-m_1^2} - {\chi \over \Delta-m_2^2} \right] .
\end{eqnarray}
\par
We define the Euclidean propagator:
\begin{equation}
  G_m(x) = (-\Delta_E+m^2)^{-1}(x,0)
= \int {d^4k \over (2\pi)^4} e^{ik \cdot x} {1 \over k^2+m^2} .
\label{Gm}
\end{equation}
It is written in the form, 
\begin{equation}
  G_m(x) = {1 \over 4\pi^2}{m \over |x|} K_1(m|x|)
=  {1 \over 4\pi^2} m^2 {K_1(|x|/\xi) \over |x|/\xi},
\label{Dfunc}
\end{equation}
where $K_1(z)$ is the modified Bessel function and we have defined the correlation length $\xi$ by $\xi=m^{-1}$.
(\ref{Dfunc})  is obtained as follows.
Substituting the identity,
\begin{equation}
  {1 \over k^2+m^2} = \int_0^\infty ds e^{-s(k^2+m^2)} ,
\end{equation}
into (\ref{Gm}) and performing the Gaussian integration over the four momentum $k$, we obtain
\begin{eqnarray}
  G_m(x) &=& \int_0^\infty ds e^{-sm^2} 
\int {d^4k \over (2\pi)^4} e^{- sk^2+ik \cdot x }  
\nonumber\\
&=&  \int_0^\infty ds e^{-sm^2} 
\exp \left[ -{x^2 \over 4s} \right] {1 \over (2\pi)^4}
\left( \sqrt{{\pi \over s}} \right)^{4}
\nonumber\\
&=& {1 \over 16\pi^2} \int_0^\infty ds {1 \over s^2}
\exp \left[ -sm^2 -{x^2 \over 4s} \right] .
\end{eqnarray}
The above result (\ref{Dfunc}) is immediately obtained by applying the integration formula \cite{GR80}:
\begin{equation}
  \int_0^\infty dx x^{\nu^1} \exp \left[ -{\beta \over x} -\gamma x 
\right] 
= 2 (\beta \gamma^{-1})^{\nu/2} K_\nu(2\sqrt{\beta \gamma}) 
\quad (\Re \beta > 0, \Re \gamma > 0)),
\end{equation}
to the case $\nu=-1, \beta = x^2/4, \gamma = m^2$, since 
$K_{-\nu}(z)=K_{\nu}(z)$.
\par
In Euclidean space,
\begin{eqnarray}
 G_{\mu\nu,\rho\sigma}(x,x') 
&=&  - 2J^2g^2 \rho^{-2} K I_{\mu\nu,\rho\sigma} \kappa
  \left[ G_{m_1}(x-x') - G_{m_2}(x-x') \right]  
\\
&=& 2 I_{\mu\nu,\rho\sigma}[F_1((x-x')^2) - F_2((x-x')^2)],
\\
F_i((x-x')^2) &:=& - J^2g^2 \rho^{-2} K 
{\chi \over 4\pi^2}  {m_i  K_1(m_i|x-x'|) \over |x-x'|}  .
\label{F}
\end{eqnarray}
We define
\begin{eqnarray}
   J_i := 2 \int_{S_C} dS^{\mu\nu}(x(\sigma)) \int_{S_C} dS^{\rho\sigma}(x(\sigma{}')) 
   I_{\mu\nu,\rho\sigma} F_i((x(\sigma)-x(\sigma{}'))^2) .
\end{eqnarray}
It is shown \cite{AES96,Antonov99} that the derivative expansion in powers of 
\begin{eqnarray}
  \zeta^a := (\sigma{}'-\sigma)^a/\xi_i ,
\quad \xi_i := m_i^{-1} ,
\end{eqnarray}
leads to
\begin{eqnarray}
   J_i =  \int d^2\sigma \sqrt{g} \left[ 4 \xi_i^2 M_0^i
- {1 \over 4}\xi_i^4 M_2^i g^{ab}(\partial_a t_{\mu\nu})(\partial_b t_{\mu\nu}) \right] 
+ O(\xi_i^6 ) ,
\end{eqnarray}
with the moment,
\begin{eqnarray}
  M_n^i := \int d^2z (z^{2})^n F_i(z^2) , \quad z^a := g^{1/4}\zeta^a ,
\end{eqnarray}
where we have used the conformal gauge for the induced metric,
$g_{ab}(\sigma)=\sqrt{g(\sigma})\delta_{ab}$, hence
$\zeta^a \zeta^b g_{ab}=g^{-1/2}g_{ab}z^a z^b=z^a z^b \delta_{ab}:=z^2$.
Thus, the confining string theory derived in this paper is characterized by the parameters,

\begin{eqnarray}
  \sigma &=& 4 \int d^2z [m_1^{-2}F_1(z^2)-m_2^{-2}F_2(z^2)] ,
\label{stension}
\\
 \alpha_0^{-1} &=& - {1 \over 4} \int d^2z z^2 [m_1^{-4} F_1(z^2) - m_2^{-4} F_2(z^2)] ,
\label{alphaeq}
\\
 \kappa &=& {1 \over 6} \int d^2z z^2 [m_1^{-4} F_1(z^2) - m_2^{-4} F_2(z)] .
\end{eqnarray}
By substituting (\ref{F}) into (\ref{stension}), we obtain
\begin{eqnarray}
  \sigma 
&=& - 4J^2g^2 \rho^{-2}K {\chi \over 4\pi^2} \left[ 
\int_{{m_1 \over \Lambda}}^{\infty} 2\pi |z| d|z|  {K_1(|z|) \over |z|}
- \int_{{m_2 \over \Lambda}}^{\infty} 2\pi |z| d|z| {K_1(|z|) \over |z|} \right] 
\\
&=& - 4J^2g^2 \rho^{-2}K {\chi \over 2\pi} \left[ K_0\left({m_1 \over \Lambda} \right) - K_0\left({m_2 \over \Lambda} \right) \right]  ,
\end{eqnarray}
where we have used $K_1(x) = - K_0(x)$, $K_0(\infty)=0$ and introduced the ultraviolet cutoff $\Lambda$ ($K_0(0)=\infty$).  Incidentally, the asymptotics of $K_p(z)$ for $z>0$,
\begin{equation}
 K_p(z) \sim \sqrt{{\pi \over 2z}} e^{-z} [1+O(z^{-1})] ,
\end{equation}
means that $K_p(z)$ decreases exponentially for large $z$.
After removing the cutoff $\Lambda$, the above expression reduces to a finite value,
\begin{equation}
 \sigma_{st}  
\cong {(2Jg)^2  \over 4\pi} \rho^{-2}K \chi \ln \left( {m_1 \over m_2} \right) .
\end{equation}

The coefficient of the rigidity term is calculated as
\begin{eqnarray}
 \alpha_0^{-1} &=& J^2g^2 \rho^{-2}K {\chi \over 4\pi^2} \int_{0}^{\infty} 2\pi |z| d|z| \left[ {|z|^2 \over m_1^2}{K_1(|z|) \over |z|}
- \int_{0}^{\infty} |z| d|z| {|z|^2 \over m_2^2}{K_1(|z|) \over |z|} \right] 
\nonumber\\
&=& J^2g^2 \rho^{-2}K {\chi \over 4\pi^2} \left[ {4\pi \over m_1^2} 
-   {4\pi \over m_2^2}  \right] .
\nonumber\\
&=& - J^2g^2 \rho^{-2}K {1 \over \pi}  <0  ,
\label{alpha0}
\end{eqnarray}
where we have used the integration formula,
\begin{equation}
 \int_0^\infty dx x^{\mu-1} K_\nu(ax) = 2^{\mu-2}a^{-\mu} \Gamma\left({\mu-\nu \over 2}\right) \Gamma\left({\mu+\nu \over 2}\right) \quad (\Re \mu > \Re \nu) .
\end{equation}
Here note that the integral in (\ref{alpha0}) is finite and we don't have to introduce the cutoff.
Finally, the $\kappa$ is calculated as
\begin{eqnarray}
 \kappa = {2 \over 3} J^2g^2 \rho^{-2}K {1 \over \pi}  > 0  .
\end{eqnarray}



\baselineskip 14pt

\begin{thebibliography}{99}
\bibitem{Nambu74}
  Y. Nambu,
  Phys. Rev. D 10, 4262-4268 (1974).
\\
G. 't Hooft,
  in: High Energy Physics, edited by A. Zichichi 
(Editorice Compositori, Bologna, 1975).
\\
S. Mandelstam,
  Phys. Report  23, 245-249 (1976).

\bibitem{tHooft81}
  G. 't Hooft,
  Nucl.Phys. B 190 [FS3], 455-478 (1981).

 
\bibitem{EI82}
  Z.F. Ezawa and A. Iwazaki,
  Phys. Rev. D 25, 2681-2689 (1982).

\bibitem{review}
  A. DiGiacomo, 
  hep-lat/9802008;  
  hep-th/9603029.
\\
M.I. Polikarpov,
  hep-lat/9609020.
M.N. Chernodub and M.I. Polikarpov,
  hep-th/9710205.
  \\
G.S. Bali,
  hep-ph/9809351.

\bibitem{KLSW87}
  A. Kronfeld, M. Laursen, G. Schierholz and U.-J. Wiese,
  Phys. Lett. B 198, 516-520 (1987).  

\bibitem{SY90}
  T. Suzuki and I. Yotsuyanagi,
  Phys. Rev. D 42, 4257-4260 (1990).

\bibitem{SNW94}
S. Hioki, S. Kitahara, S. Kiura, Y. Matsubara, O. Miyamura, S. Ohno, T. Suzuki,
Phys. Lett. B 272, 326-332 (1991), Erratum-ibid. B281, 416 (1992). 
\\
J.D. Stack, S.D. Nieman and R. Wensley,
hep-lat/9404014,
Phys. Rev. D 50, 3399-3405 (1994).
\\
H. Shiba and T. Suzuki,
hep-lat/940401,
Phys. Lett. B 333, 461-466 (1994).
\\
O. Miyamura, 
  Phys. Lett. B 353, 91-95 (1995).
\\
S. Sasaki and O. Miyamura,
hep-lat/9811029,
Phys. Rev. D 59, 094507 (1999). 
\\
S. Sasaki and O. Miyamura,
hep-lat/9810039,
  Phys. Lett. B, 443, 331-337 (1998).

\bibitem{Suzuki88}
  T. Suzuki, 
  Prog. Theor. Phys. 80, 929-934 (1988).
  \\
S. Maedan and T. Suzuki,
  Prog. Theor. Phys. 81, 229-240 (1989). 
  \\
T. Suzuki, 
  Prog. Theor. Phys. 81, 752-757 (1989).
\\
S. Maedan, Y. Matsubara and T. Suzuki,
  Prog. Theor. Phys. 84, 130-141 (1990).

\bibitem{Zwanziger71}
  D. Zwanziger,
  Phys. Rev. D 3, 880-891 (1971).
\\
R.A. Brandt, F. Neri and D. Zwanziger,
  Phys. Rev. D 19, 1153-1167 (1979).


\bibitem{dualsupertype}
  G.S. Bali, V. Bornyakov, M. M\"uller-Preussker and K. Schilling,
hep-lat/9603012,
Phys. Rev. D 54, 2863 (1996).
\\
M.N. Chernodub, S. Kato, N. Nakamura, M.I. Polikarpov and T. Suzuki,
hep-lat/9902013.


\bibitem{KondoI}
  K.-I. Kondo,
  hep-th/9709109,
  Phys. Rev. D 57, 7467-7487 (1998).
  \\
K.-I. Kondo, 
  hep-th/9803063,
  Prog. Theor. Phys. Supplement, No. 131, 243-255.


\bibitem{KondoII}
  K.-I. Kondo, 
  hep-th/9801024,
  Phys. Rev. D 58, 105019 (1998).

\bibitem{KondoIII}
  K.-I. Kondo,
  hep-th/9803133,
  Phys. Rev. D 58, 085013 (1998).

\bibitem{KondoIV}
  K.-I. Kondo,
  hep-th/9805153,
  Phys. Rev. D 58, 105016 (1998).

\bibitem{KondoV}
  K.-I. Kondo,
  hep-th/9810167,
  Phys. Lett. B 455, 251-258 (1999).

\bibitem{KondoVI}
  K.-I. Kondo,
  hep-th/9904045,
  Intern. J. Mod. Phys. A, to be published.

\bibitem{QR98}
  M. Quandt and H. Reinhardt,
 hep-th/9707185,
 Int. J. Mod. Phys. A13, 4049-4076 (1998). 

\bibitem{AC75}
  T. Appelquist and J. Carazonne, 
Phys. Rev. D11, 2856-2861 (1975).

  
\bibitem{Schaden99}
  M. Schaden,
  hep-th/9909011, revised version 3.

\bibitem{KS00a}
  K.-I. Kondo and T. Shinohara,
  Abelian dominance in low-energy Gluodynamics due to dynamical mass generation,
  CHIBA-EP-120,
  hep-th/0004158,
  Phys. Lett. B, to be published.

\bibitem{MLP85}
  H. Min, T. Lee and P.Y. Pac,
  Phys. Rev. D 32, 440-449 (1985).

\bibitem{BRST}
  C. Becchi, A. Rouet and R. Stora,
  Commun. Math. Phys. 42, 127 (1975);
  Ann. Phys. 98, 287 (1976).
\\
I.V. Tyutin,
  Lebedev preprint, FIAN No.39 (in Russian) (1975).

\bibitem{AS99}
  K. Amemiya and H. Suganuma,
  Phys. Rev. D 60, 114509 (1999).

\bibitem{maKanazawa}
  S. Fujimoto, S. Kato, M. Murata and T. Suzuki,
hep-lat/9909103.
\\
S. Fujimoto, S. Kato and T. Suzuki,
hep-lat/0002006.
\\
M.N. Chernodub, S. Fujimoto, S. Kato, M. Murata, M.I. Polikarpov and T. Suzuki,
hep-lat/0006025.
\\
S. Kitahara, K. Yamagishi, T. Suzuki,
hep-lat/0002011.

\bibitem{Polyakov96}
  A.M. Polyakov,
hep-th/9607049,
  Nucl. Phys. B 486, 23-33 (1997). 


\bibitem{DP89}
  D.I. Diakonov and V.Yu. Petrov, 
  Phys. Lett. B 224, 131-135 (1989).
\\
D. Diakonov and V. Petrov,
  hep-th/9606104.

\bibitem{KT99}
  K.-I. Kondo and Y. Taira,
  hep-th/9906129,
Mod. Phys. Lett. A 15, 367-377 (2000); 
\\
K.-I. Kondo and Y. Taira,
  hep-th/9911242.
 
\bibitem{HU99}
  M. Hirayama and M. Ueno,
  hep-th/9907063,
  Prog. Theor. Phys. 103, 151-159 (2000).

\bibitem{Dosch87}
  H.G. Dosch,
Phys. Lett. B 190, 177-181 (1987).
\\
H.G. Dosch and Yu.A. Simonov,
Phys. Lett. B 205, 339-344 (1988).

\bibitem{DDM97}
  M. D'Elia, A. Di Giacomo and E. Meggiolaro,
hep-lat/970503,
Phys. Lett. B 408, 315-319 (1997).
\\
A. Di Giacomo, M. D'Elia, H. Panagopoulos and E. Meggiolaro,
hep-lat/9808056.

\bibitem{BBDV98}
  M. Baker, N. Brambilla, H.G. Dosch and A. Vairo,
hep-ph/9802273,
Phys. Rev. D 58, 034010 (1998).

\bibitem{HN93}
  H. Hata and I. Niigata,
  Nucl. Phys. B 389, 133-152 (1993).

\bibitem{NS83}
  C. Nash and S. Sen,
  Topology and Geometry for Physicists 
(Academic Press, New York, 1983).

\bibitem{Nakahara90}
  M. Nakahara, 
  Geometry, Topology and Physics
(IOP, Bristol, 1990).

\bibitem{KS00b}
  K.-I. Kondo and T. Shinohara,  Renormalizable Abelian-projected effective gauge theory derived from Quantum Chromodynamics,
hep-th/0005125, and papers in preparation.


\bibitem{KS00c}
  K.-I. Kondo and T. Shinohara, in preparation.

\bibitem{KM00}
  K.-I. Kondo and T. Murakami,
in preparation.


\bibitem{Townsend79}
  P.K. Townsend,
Phys. Lett. B 88, 97-101 (1979).

\bibitem{Kimura80}
  T. Kimura,
  Prog. Theor. Phys. 64, 357-360 (1980).
\\
T. Kimura,
  Prog. Theor. Phys. 65, 338-350 (1981).

\bibitem{HKO81}
  H. Hata, T. Kugo and N. Ohta,
  Nucl. Phys. B 178, 527-544 (1981).



\bibitem{tHooft71}
  G. 't Hooft,
  Nucl. Phys. B 35, 167-188 (1971).

\bibitem{FLS72}
K. Fujikawa, B.W. Lee and A.I. Sanda,
  Phys. Rev. D 6, 2923-2943 (1972).
\\
K. Fujikawa,
  Phys. Rev. D 7, 393-398 (1973).

\bibitem{SST95}
  S. Sasaki, H. Suganuma and H. Toki,
  Prog. Theor. Phys. 94, 373-384 (1995).
\\
H. Suganuma, S. Sasaki and H. Toki,
  Nucl. Phys. B 435, 207-240 (1995).

\bibitem{Polchinski92}
  J. Polchinski,
Strings and QCD, 
hep-th/9210045.

\bibitem{tHooft74a}
  G. 't Hooft,
  Nucl. Phys. B 72, 461-473 (1974).


\bibitem{oldstring}
  M. Kalb and P. Ramond,
Phys. Rev. D 9, 2273-2284 (1974).
\\
D. F\"orster,
Nucl. Phys. B 81, 84-92 (1974).
\\
J.L. Gervais and B. Sakita,
Nucl. Phys. B 91, 301-316 (1975).

\bibitem{Polyakov86}
A.M. Polyakov,
  Nucl. Phys. B 268, 406-412 (1986).

\bibitem{Kleinert86}
H. Kleinert,
  Phys. Lett. B 174, 335-338 (1986). 
\\
H. Kleinert,
  Phys. Lett. B211, 151-155 (1988). 

\bibitem{David89}
  F. David,
Phys. Reports 184, 221-227 (1989).

\bibitem{PWZ93}
  M.I. Polikarpov, U.-J. Wiese and M.A. Zubkov,
  hep-lat/9303007,
  Phys. Lett. B 309, 133-138 (1993).

\bibitem{Lee93}
  K. Lee, 
  Phys. Rev. D 48, 2493-2498 (1993).

\bibitem{Orland94}
  P. Orland,
  hep-th/9404140,
  Nucl. Phys. B 428, 221-232 (1994).

\bibitem{SY95}
  M. Sato and S. Yahikozawa,
  hep-th/9406208,
  Nucl. Phys. B 436, 100-128 (1995).


\bibitem{ACPZ96}
  E.T. Akhmedov, M.N. Cherndub, M.I. Polikarpov and M.A. Zubkov,
  hep-th/9505070,
  Phys. Rev. D 53, 2087-2095 (1996).
\\
E.T. Akhmedov,
hep-th/9605214.

\bibitem{BS00}
M. Baker and R. Steinke,
  hep-ph/9905375,
  Phys. Lett. B 474, 67-72 (2000).
\\
M. Baker and R. Steinke,
  hep-ph/0006069.


\bibitem{Alvarez81}
O. Alvarez,
Phys. Rev. D 24, 440-449 (1981).

\bibitem{Arvis83}
J.F. Arvis,
Phys. Lett. B 127, 106-108 (1983).

\bibitem{Olesen85}
P. Olesen,
Phys. Lett. B 160, 144-148 (1985).

\bibitem{LSW80}
  M. L\"uscher, K. Symanzik and P. Weiz,
Nucl. Phys. B 173, 365 (1980).

\bibitem{PS91}
J. Polchinski and A. Strominger,
Phys. Rev. Lett. 67, 1681-1684 (1991).

\bibitem{KC96}
  H. Kleinert and A.M. Chervyakov,
  hep-th/9601030.

\bibitem{PY92}
  J. Polchinski and Z. Yang,
Phys. Rev. D 46, 3667-3669 (1992).


\bibitem{DQT96}
  M.C. Diamantini, F. Quevedo and C.A. Trugenberger,
hep-th/9612103,
  Phys. Lett. B 396, 115-121 (1997).

\bibitem{DT97}
M.C. Diamantini and C.A. Trugenberger, 
  hep-th/9712008, 
  Phys. Lett. B 421, 196-202 (1998).
\\
M.C. Diamantini and C.A. Trugenberger,
  hep-th/9803046, 
  Nucl. Phys. B531, 151-167 (1998). 
 

\bibitem{AES96}
D.V. Antonov, D. Ebert and Yu.A. Simonov,
hep-th/9605086,
  Mod. Phys. Lett. A 11, 1905-1918 (1996).

\bibitem{AE98}
  D. Antonov and D. Ebert,
  hep-th/9806153,
  Eur. Phys. J. C 8, 343-351 (1999).

\bibitem{AE99}
  D. Antonov and D. Ebert,
hep-th/9902177.

\bibitem{Antonov99}
  D. Antonov,
hep-th/9909209,
  Surveys High Energ. Phys. 14, 265-355 (2000). 




\bibitem{Polyakov97}
  A.M. Polyakov, 
hep-th/9711002.


\bibitem{AGO98}
  E. Alvarez, C. Gomez and T. Ortin,
hep-th/9806075.

\bibitem{GR80}
  I.S. Gradstheyn and I.M. Ryzhik,
  {\it Table of Integrals, Series and Products}
(Academic Press, Bosoton, 1980).

\bibitem{preparation}
  T. Imai, K.-I. Kondo, T.-S. Lee, T. Murakami and T. Shinohara, 
in preparation.


\bibitem{KU82}
  T. Kugo and S. Uehara,
  Nucl. Phys. B 197, 378 (1982).

\bibitem{BFV}
E.S. Fradkin and G.A. Vilkovisky,
  Phys. Lett. B 55, 224 (1975).
\\
I.A. Batalin and G.A. Vilkovisky,
  Phys. Lett. B 69, 309 (1977).
\\
Y. Igarashi,
  Soryushiron-Kenkyu (Kyoto) (in Japanese) 82, 414-436 (1991). 



\bibitem{Nakamura00}
  A. Nakamura, private communication and a paper in preparation.

\bibitem{BS96}
  C. Bizdadea and S.O. Saliu,
  Phys. Lett. B 368, 202-208 (1996).

\end{thebibliography}
\end{document}